\documentclass[12pt]{article}
\usepackage{jheppub}
\usepackage{amsmath}
\usepackage{array}

\usepackage{arydshln}
\usepackage{multirow}
\usepackage{subfigure}

\allowdisplaybreaks %Let LaTeX automatically break large equations

\def\NeqFive{{\cal N} = 5}
\def\NeqFour{{\cal N} = 4}
\def\NeqEight{{\cal N} = 8}
\def\fig#1{Fig.~\ref{fig:#1}}
\def\figs#1#2{Figs.~\ref{fig:#1} and~\ref{fig:#2}}
\def\sect#1{Sect.~\ref{sec:#1}}
\def\eqn#1{Eq.~(\ref{eqn:#1})}
\def\eqns#1#2{Eqs.~(\ref{eqn:#1}) and~(\ref{eqn:#2})}

\def\tab#1{Table~\ref{tab:#1}}
\def\tabs#1#2{Table~\ref{tab:#1} and~\ref{tab:#2}}
\def\tree{{\rm tree}}
\def\eps{\epsilon}
\def\I{{\cal I}}
\def\dI{d\hskip0.3mm{\cal I}}

\def\fourloop{{\rm 4\hbox{-}loop}}
\def\fiveloop{{\rm 5\hbox{-}loop}}
\def\Ord{{\cal O}}
\def\dlog{d {\rm log}}

\def\res#1{\underset{#1}{\textrm{Res}\ }}
\def\la{\langle}
\def\ra{\rangle}
\def\K{\mathcal{K}}
\def\cut#1{ #1 \bigr|_{\rm cut}}
\def\maxcut#1{\left. #1 \right|_{\substack{\rm max \\ \rm cut}}}

\newcommand{\ab}[1]{\langle #1 \rangle}     % for angle brackets
\newcommand\labellast{\addtocounter{equation}{1}\tag{\theequation}}

\title{Logarithmic singularities and maximally supersymmetric amplitudes}
\author{Zvi Bern$^1$,}
\author{Enrico Herrmann$^2$,}
\author{Sean Litsey$^1$,}
\author{James Stankowicz$^1$,}
\author{Jaroslav Trnka$^2$}

\affiliation{$^1$ Department of Physics and Astronomy, UCLA, Los
Angeles, CA 90095, USA}
\affiliation{$^2$ Walter Burke Institute for Theoretical Physics, \\
California Institute of Technology, Pasadena, CA 91125, USA}

\emailAdd{bern@physics.ucla.edu, eherrmann@caltech.edu, slitsey@ucla.edu,
jjstankowicz@ucla.edu, trnka@caltech.edu}

\abstract{The dual formulation of planar $\NeqFour$ super-Yang-Mills
  scattering amplitudes makes manifest that the integrand has only
  logarithmic singularities and no poles at infinity.  Recently,
  Arkani-Hamed, Bourjaily, Cachazo and Trnka conjectured the same
  singularity properties hold to all loop orders in the nonplanar
  sector as well.  Here we conjecture that to all loop orders these
  constraints give us the key integrand level analytic information
  contained in dual conformal symmetry.  We also conjecture that to
  all loop orders, while $\NeqEight$ supergravity has poles at
  infinity, at least at four points it has only logarithmic
  singularities at finite locations.  We provide nontrivial evidence
  for these conjectures.  For the three-loop four-point $\NeqFour$
  super-Yang-Mills amplitude, we explicitly construct a complete basis
  of diagram integrands that has only logarithmic singularities and no
  poles at infinity.  We then express the complete amplitude in terms
  of the basis diagrams, with the coefficients determined by
  unitarity.  We also give examples at three loops showing how to make
  the logarithmic singularity properties manifest via $\dlog$ forms.
  We give additional evidence at four and five loops supporting the
  nonplanar logarithmic singularity conjecture.  Furthermore, we
  present a variety of examples illustrating that these constraints
  are more restrictive than dual conformal symmetry.  Our
  investigations show that the singularity structures of planar and
  nonplanar integrands in $\NeqFour$ super-Yang-Mills are strikingly
  similar.  While it is not clear how to extend either dual conformal
  symmetry or a dual formulation to the nonplanar sector, these
  results suggest that related concepts might exist and await
  discovery.  Finally, we describe the singularity structure of
  $\NeqEight$ supergravity at three loops and beyond.  } \preprint{
\begin{flushright}UCLA/14/TEP/109 \hfill CALT-TH-2014-172\end{flushright}
}

\begin{document}
\maketitle

%===========================================================================  
\section{Introduction}
\label{sec:IntroductionSection}
%============================================================================  

Recent years have seen remarkable progress in our understanding of the
structure of scattering amplitudes in planar $\NeqFour$
super-Yang-Mills (sYM) theory~\cite{sYM1,sYM2}.  (See reviews
e.g.~Refs.~\cite{BDKReview, Cachazo:2005ga, Beisert:2010jr,
  Drummond:2011ic, Elvang:2013cua, Henn:2014yza}.)  Along with the
conceptual progress have come significant computational advances,
including new explicit results for amplitudes after integration up to
high loop orders (see
e.g.~Refs.~\cite{Dixon:2011nj,Dixon:2014xca,Dixon:2014voa,DixonThreeLoop}),
as well as some all-loop order predictions~\cite{BDS}. Among the
theoretical advances are connections to twistor string
theory~\cite{Witten:2003nn,RSV}, on-shell recursion
relations~\cite{BCF, BCFW, ArkaniHamed:2010kv}, unveiling of hidden
dual conformal symmetry~\cite{DualConformalMagic, Alday:2007hr,
  Drummond:2008vq}, momentum twistors~\cite{Hodges}, a dual
interpretation of scattering amplitudes as supersymmetric Wilson
loops~\cite{Mason:2010yk,CaronHuot:2010ek,Alday:2010zy} and a duality
to correlation functions~\cite{CorrelatorDualityI,
  CorrelatorDualityII}. More recently, four-dimensional planar
integrands in $\NeqFour$ sYM were reformulated using on-shell diagrams
and the positive Grassmannian \cite{ArkaniHamed:2012nw,
  ArkaniHamed:2009dn, ArkaniHamed:2009vw, Mason:2009qx,
  ArkaniHamed:2009dg, ArkaniHamed:2009sx} (see related work in
Ref.~\cite{Huang:2013owa, Huang:2014xza, Kim:2014hva,
  ElvangGrassmannian, NonPlanarOnShell}).  This reformulation fits
nicely into the concept of the
amplituhedron~\cite{Arkani-Hamed:2013jha, Arkani-Hamed:2013kca,
  Franco:2014csa, Lam:2014jda,Bai:2014cna}, and makes an extremely
interesting connection to active areas of research in algebraic
geometry and combinatorics (see
e.g.~\cite{Lusztig,postnikov,postnikov2,lauren,goncharov,knutson}).
This picture also makes certain properties of amplitudes completely
manifest, including properties like Yangian
invariance~\cite{Drummond:2009fd} that are obscure in standard
field-theory descriptions.

A special feature of $\NeqFour$ sYM scattering amplitudes that appears after 
integration is uniform transcendentality
\cite{LipatovTranscendentality,Eden:2006rx,BeisertEdenStaudacher}, a
property closely related to the $\dlog$-structure of the integrand in
the dual formulation~\cite{ArkaniHamed:2012nw} (for recent discussion
on integrating $\dlog$ forms see Ref.~\cite{Lipstein:2013xra}). The
dual formulation can perhaps also be extended to integrated results
via special functions that are motivated by the positive
Grassmannian~\cite{Goncharov:2010jf, Golden:2013xva, Golden:2014xqa,
  Golden:2014pua}.  Such an extension might naturally incorporate the 
integrability of $\NeqFour$ sYM theory~\cite{Beisert:2003yb}. So far this 
has not played a major role in the dual formulation, but is very useful 
in the flux tube $S$-matrix approach~\cite{Basso:2013vsa,Basso:2013aha,
Basso:2014koa,Basso:2014jfa,Basso:2014nra},
leading to some predictions at finite coupling. Integrability should
be present in the dual formulation of the planar theory through
Yangian symmetry. Therefore, it is natural to attempt to search for either a
Yangian-preserving regulator of infrared divergences of amplitudes
\cite{Ferro:2012xw,Ferro:2013dga,Beisert:2014qba,Broedel:2014hca}, or
directly for Yangian-preserving deformations of the Grassmannian
integral~\cite{Ferro:2014gca,Bargheer:2014mxa}.

In this paper we are interested in understanding how to carry these many
advances and promising directions over to the nonplanar sector of
$\NeqFour$ sYM theory.  Unfortunately much less is known
about nonplanar $\NeqFour$, in part because of the difficulty of
carrying out loop integrations. In addition, lore suggests that we lose
integrability and thereby many nice features of planar amplitudes
believed to be associated with it. (We do not use a $1/N$ expansion.)  
Even at the integrand level, the absence of a unique integrand makes it difficult
to study nonplanar amplitudes globally, rather than in some particular
expansion.  One approach to extending planar properties to the
nonplanar sector is to search for the dual formulation of the theory
using on-shell diagrams. Despite the fact that these are well-defined
objects beyond the planar limit with many interesting
properties~\cite{NonPlanarOnshellToAppear}, yet it is still not known how to
expand scattering amplitudes in terms of these objects.

Nevertheless, there are strong hints that at least some of the
properties of the planar theory survive the extension to the nonplanar
sector.  In particular, the Bern--Carrasco--Johansson (BCJ) duality
between color and kinematics~\cite{BCJ,BCJLoop} shows that the
nonplanar sector of $\NeqFour$ sYM theory is intimately linked to the
planar one, so we should expect that some of the properties carry
over.  BCJ duality can be used to derive $\NeqEight$ supergravity
integrands starting from $\NeqFour$ sYM ones, suggesting that some
properties of the gauge theory should extend to $\NeqEight$
supergravity as well. An encouraging observation is that the two-loop
four-point amplitude of both $\NeqFour$ sYM theory and $\NeqEight$
supergravity have a uniform transcendental
weight~\cite{KLOV,LipatovTranscendentality, SchnitzerN8UniformTrans,
  QueenMaryN8UniformTrans,GehrmannFormFactor,Log}.  Related to the leading
transcendentality properties is the recent conjecture by Arkani-Hamed,
Bourjaily, Cachazo and one of the authors~\cite{Log} that, to all loop
orders, the full $\NeqFour$ sYM amplitudes, including the nonplanar
sector, have only logarithmic singularities and no poles at infinity.
This is motivated by the possibility of dual formulation that would
make these properties manifest~\cite{ArkaniHamed:2012nw}.  As evidence
for their conjecture, they rewrote the two-loop four-point
amplitude~\cite{BRY} in a format that makes these properties hold term
by term.

In this paper we follow this line of reasoning, showing that key
features of planar $\NeqFour$ sYM amplitudes carry over to the
nonplanar sector.  In particular, we demonstrate that the three-loop
four-point amplitude of $\NeqFour$ sYM theory has only logarithmic
singularities and no poles at infinity. We find a diagrammatic
representation of the amplitude, using standard Feynman propagators,
where these properties hold diagram by diagram. While we do not expect
that these properties can be made manifest in each diagram to all loop
orders, for the amplitudes studied in this paper this strategy works
well.  We proceed here by analyzing all singularities of the
integrand; this includes singularities both from propagators and from
Jacobians of cuts.  We then construct numerators to cancel unwanted
singularities, where we take ``unwanted singularities'' to mean double
or higher poles and poles at infinity. In the planar case, subsets of
these types of constraints have been used in
Refs.~\cite{Drummond:2007aua,Bourjaily:2011hi}.  As a shorthand, we
call numerators with the desired properties ``$\dlog$ numerators''
(and analogously for ``$\dlog$ integrands'' and ``$\dlog$ forms'').
Once we have found all such objects, we use unitarity constraints to
determine the coefficients in front of each contribution.  To verify
that the amplitude so deduced is complete and correct we evaluate a
complete set of unitarity cuts.  The representation of the three-loop
four-point amplitude that we find in this way differs from the
previously found ones~\cite{GravityThreeLoop,BCJLoop} by contact terms
that have been nontrivially rearranged via the color Jacobi identity.
While all forms of this amplitude have only logarithmic singularities,
it is not at all obvious in earlier representations that this is true,
because of nontrivial cancellations between different diagrams.

After constructing the three-loop basis of $\dlog$ integrands, we address
some interesting questions.  One is whether there is a simple pattern
dictating the coefficients with which the basis integrands appear in
the amplitude.  Indeed, we show that many of the coefficients follow
the rung rule~\cite{BRY}, suggesting that a new structure remains to
be uncovered.  Another question is whether it is possible to
explicitly write the integrands we construct as $\dlog$ forms. In
general, this requires a nontrivial change of variables, but we have
succeeded in writing all but one type of basis integrand form as $\dlog$
forms.  We present three explicit examples at three loops showing how
this is done.  These $\dlog$ forms make manifest that the integrand
basis elements have only logarithmic singularities, although the 
singularity structure at infinity is not manifest.

The requirement of only logarithmic singularities and no poles at
infinity strongly restricts the integrands.  In fact, we conjecture
that in the planar sector logarithmic singularities and absence of
poles at infinity imply dual conformal invariance in the integrand.
We check this for all contributions at four loops and give a five-loop
example illustrating that these singularity conditions impose even
stronger constraints on the integrand than dual conformal symmetry.

Related to the $\dlog$ forms, the results presented in this paper
offer a useful bridge between integrands and integrated results.  The
objects we construct here are a subset of the uniform transcendental
integrals needed in the Henn and Smirnov procedure~\cite{Henn,
  HennSmirnov, HennSmirnov2, Henn:2014qga} to find a relatively simple
set of differential equations for them. The importance of uniform
transcendental weight was first realized in Ref.~\cite{KLOV}. It was
noted that through three loops the $\NeqFour$ sYM anomalous dimensions
of Wilson twist 2 operator match the terms in the corresponding QCD
anomalous dimension that have maximal transcendental weight.  The
ideas of uniform transcendental weight were expanded on in a variety
of subsequent papers and include examples with nonplanar
contributions~\cite{LipatovTranscendentality, SchnitzerN8UniformTrans,
  QueenMaryN8UniformTrans,GehrmannFormFactor,Log}.  In this paper we
focus mainly on integrands relevant to $\NeqFour$ sYM theory, which
correspond to the subset of integrands with no poles at infinity.  In
any case, a side benefit of the methods described here is that it should
offer an efficient means for identifying integrals of uniform
transcendental weight.  Ref.~\cite{Log} noted a simple relation
between the singularity structure of the two-loop four-point amplitude
of $\NeqEight$ supergravity and the one of $\NeqFour$ sYM.  How much
of this continues at higher loops?  Starting at three loops, the
situation is more complicated because the integrals appearing in the
two theories are different.  Nevertheless, by making use of the BCJ
construction~\cite{BCJ,BCJLoop}, we can obtain the corresponding
$\NeqEight$ amplitude in a way that makes its analytic properties
relatively transparent. In particular, it allows us to immediately
demonstrate that away from infinity, $\NeqEight$ supergravity has only
logarithmic singularities.  We also find that starting at three loops,
$\NeqEight$ supergravity amplitudes have poles at infinity whose
degree grow with the number of loops.

This paper is organized as follows: In \sect{TwoLoopSection} we will
briefly discuss logarithmic singularities and poles at infinity in
loop integrands. In \sect{StrategySection} we outline our strategy for
studying nonplanar amplitudes and illustrate it using the two-loop
four-point amplitude. In \sect{ThreeLoopBasisSection} we construct a
basis of three-loop four-point integrands that have only logarithmic
singularities and no poles at infinity. We then express the three-loop
four-point amplitude in this basis and show that the rung rule
determines a large subset of the coefficients.  Then in
\sect{DlogSection} we discuss $\dlog$ forms in some detail.  In
\sect{HigherLoops}, we give a variety of multiloop examples corroborating
that only logarithmic singularities are present in $\NeqFour$
sYM theory.  In \sect{PlanarIntegrand}, we present
evidence that the constraints of only logarithmic singularities and no
poles at infinity incorporate the constraints from dual conformal
symmetry.  In \sect{GravitySection}, we comment on the singularity
structure of the ${\cal N}=8$ supergravity four-point amplitude.
In \sect{ConclusionsSection}, we present our conclusions
and future directions.

%==============================================================================
\section{Singularities of the integrand}
\label{sec:TwoLoopSection}
%==============================================================================

Integrands offer enormous insight into the structure of scattering
amplitudes.  This includes the discovery of dual conformal
symmetry~\cite{DualConformalMagic}, the Grassmannian
structure~\cite{ArkaniHamed:2009dn,
  ArkaniHamed:2009vw,Mason:2009qx,ArkaniHamed:2012nw}, the geometric
structures~\cite{Arkani-Hamed:2013jha}, and ultraviolet
properties~\cite{BDDPR,Manifest3,ColorKinematics}.  The singularity
structure of integrands, along with the integration contours, determine
the properties of integrated expressions.  In particular, the uniform
transcendentality property is determined by the singularity structure
of the integrand.  The nonplanar sector of $\NeqFour$ sYM theory is much less
developed than the planar one.  Studying integrands offers a means of
making progress in this direction, especially at high loop orders
where it is difficult to obtain integrated expressions.

It would be ideal to study the amplitude as a single object and not to
rely on an expansion using diagrams as building blocks which carry
their own labels. In the planar sector, we can avoid such an expansion
by using globally defined dual variables to obtain a unique rational
function called the {\it integrand} of the amplitude. Unfortunately,
it is unclear how to define such a unique object in the nonplanar
case.  In this paper we sidestep the lack of global variables by
focusing on smaller pieces of the amplitude, organized through
covariant, local diagrams with only three-point vertices and Feynman
propagators.  Such diagrams have also proved useful in the generalized
unitarity method. Diagrams with only cubic vertices are sufficient in
gauge and gravity theories, because it is possible to express diagrams
containing higher point vertices in terms of ones with only cubic
vertices by multiplying and dividing by appropriate Feynman
propagators.  For future reference, when not stated otherwise, this is
what we mean by a ``diagrammatic representation'' or an ``expansion in terms
of diagrams''.

For a given diagram, there is no difficulty in having a well-defined
notion of an integrand, at least for a given set of momentum labels.
For this to be useful, we need to be able to expose the desired
singularity properties one diagram at a time, or at worst for a
collection of a small subset of diagrams at a time.  In general, there
is no guarantee that this can be done, but in cases where it can be,
it offers a useful guiding principle for making progress in the
nonplanar sector.  A similar strategy proved successful for BCJ
duality.  In that case, at most three diagrams at a time need to share
common labels in order to define the duality, bypassing the need for
global labels.

At three loops we will explicitly construct a basis of integrands
that have only logarithmic singularities and
no poles at infinity. We also discuss some higher-loop examples.
Before doing so, we summarize the dual formulation of planar amplitudes,
in order to point out the properties that we wish to carry over to 
the nonplanar case.

\subsection{Dual formulation of planar theory}

Here we summarize the dual formulation of planar $\NeqFour$
sYM theory, with a focus on our approach to extending this
formulation to the nonplanar sector.  For details beyond what appear
here, we refer the reader to Refs.~\cite{ArkaniHamed:2009dn,ArkaniHamed:2009vw,ArkaniHamed:2012nw}.

As mentioned in the previous subsection, for planar amplitudes we can define an {\it
  integrand} based on a global set of variables valid for all terms in
the amplitude~\cite{ArkaniHamed:2010kv}.  Up to terms that vanish
under integration, the integrand of a planar amplitude is a unique
rational function constrained by the requirement that all unitarity
cuts of the function are correct.  Methods based on unitary and factorization
construct the integrand using only on-shell input information.
On-shell diagrams capitalize on this efficiency by
representing integrands as graphs where all internal lines are
implicitly on shell, and all vertices are three-point amplitudes.

An important further step is to promote on-shell diagrams from being
only reference data to being actual building blocks for the
amplitude. This idea was exploited in Ref.~\cite{ArkaniHamed:2009dn}
where loop-level~\cite{ArkaniHamed:2010kv} recursion relations for
integrands were interpreted directly in terms of higher-loop on-shell
diagrams.  A preliminary version of this notion is already visible in the
early version of the BCFW recursion relations~\cite{BCF}, where the
tree-level amplitudes are expressed in terms of leading singularities
of one-loop amplitudes.  

More recently, the construction of amplitudes from on-shell diagrams has been
connected~\cite{ArkaniHamed:2012nw} to modern developments in
algebraic geometry and combinatorics~\cite{Lusztig, postnikov,
  postnikov2, lauren, goncharov, knutson} where the same type of
diagrams appeared in a very different context. Each on-shell diagram
can be labeled using variables associated with edges $e_i$ or faces
$f_j$, from which one can build a $k\times n$ matrix $C$, where $n$
is the number of external particles, and $k$ is related to the number
of negative helicity particles.  This matrix has a ${\rm GL}(k)$
symmetry and therefore belongs to a Grassmannian $C \in G(k,n)$. If
the edge and face variables are taken to be real and to have fixed sign based on
certain rules, all the maximal minors of the matrix $C$ are positive
and produce cells in the positive Grassmannian $G_{+}(k,n)$. This is
more than just a mathematical curiosity, as this viewpoint can be used
to evaluate on-shell diagrams independently of the notion of the notion of gluing together
three-point on-shell amplitudes.

After parametrizing the on-shell diagram as described, the diagram
takes the value~\cite{ArkaniHamed:2009dn}
\begin{equation}
\Omega = \frac{df_1}{f_1}\wedge \frac{df_2}{f_2}\wedge \cdots \wedge
\frac{df_m}{f_m} \, \delta (C(f_j)\cdot {\cal W})\,,
\label{eqn:LogForm}
\end{equation}
where we collectively encode all external data, both bosonic and
fermionic, in ${\cal W}$. The delta function implies a set of equations
that can be solved for the $f_j$ in terms of external data.  Any
on-shell diagrams that have an interpretation as building blocks for
tree-level amplitudes exactly determine all variables $f_j$ so that $\Omega$
becomes a function of external data only, and $\Omega$ gives exactly
the tree amplitude.  Likewise, any on-shell diagrams that have an
interpretation as building blocks of an $L$-loop integrand leave $4L$
variables $f_j$ unfixed, and $\Omega$ is the $4L$-form giving exactly
the unique $L$-loop integrand. Even on-shell diagrams that do not
directly correspond to tree amplitudes or loop integrands have some
meaning as cuts or factorizations of the amplitude. This construction
is often referred to as the {\it dual formulation} of planar
amplitudes.  One of our motivations is to look for possible extensions
of this formulation to the nonplanar sector.

A crucial feature of $\Omega$ is that it has only logarithmic
singularities in $f_j$, inherited from the structure of
\eqn{LogForm}. As written there, these singularities are in the
abstract Grassmannian space, or equivalently in the extended bosonic
variables within the amplituhedron construction of the integrand.
When translated back to momentum (or twistor or momentum twistor)
space, the logarithmic property is lost due to the supersymmetric-part
of the delta function in \eqn{LogForm}. However, for both MHV ($k=2$)
and NMHV ($k=3$) on-shell diagrams, the supersymmetric-part of delta
functions can be separated from the bosonic parts, resulting in a
logarithmic form in momentum space~\cite{Arkani-Hamed:2013jha,
  Arkani-Hamed:2013kca}.  The other property that is completely
manifest when forms are written in momentum twistor space is the
absence of poles at infinity.  Both these properties are preserved for
all on-shell diagrams and so are true for all tree-level amplitudes
and integrands for planar loop amplitudes.

We can also compute nonplanar on-shell diagrams, either by gluing
together three-point on-shell amplitudes or by using the relation to
the Grassmannian. The relation to the positive part of the
Grassmannian is naively lost, but reappears under more careful
scrutiny~\cite{NonPlanarOnshellToAppear}.  We can still associate a
logarithmic form to diagrams as in \eqn{LogForm}.  Using the same
arguments as in the planar sector, all MHV and NMHV on-shell diagrams
have logarithmic singularities in momentum space. However, it is not
known at present how to construct complete $\NeqFour$ sYM amplitudes,
including the nonplanar parts, using recursion relations of these
nonplanar on-shell diagrams.  Unlike in the planar sector, a major
obstacle in the nonplanar sector is the absence of a unique
integrand. If this problem can be solved so that the amplitude is
expressible in terms of on-shell diagrams, then the same arguments as
used in the planar sector would prove that the full nonplanar ${\cal
  N}=4$ sYM amplitudes have logarithmic singularities.  In any case,
even if the existence of a dual formulation for the nonplanar sector
cannot be straightforwardly established, we can still test the key
consequences: only logarithmic singularities and no poles at infinity.
This is what we turn to now.

\subsection{Logarithmic singularities}

Before discussing the basis of integrands for $\NeqFour$ sYM amplitudes, 
we consider some simple toy cases that display the 
properties relevant for subsequent sections. It is natural to define an 
integrand form $\Omega(x_1,\dots,x_m)$ of the integral $F$ by stripping 
off the integration symbol
\begin{equation}
F = \int \Omega(x_1,\dots,x_m)\,,
\end{equation}
and to study its singularity structure. There is a special
class of forms that we are interested in here: those that have only
{\it logarithmic singularities}. A form has only logarithmic
singularities if near any pole $x_i\rightarrow a$ it behaves as
\begin{equation}
{\Omega}(x_1,\dots,x_m) \rightarrow \frac{dx_i}{x_i-a}\,
\wedge  \hat\Omega(x_1,\dots, \hat{x}_i,\dots,x_m)\,,
\end{equation}
where $\hat \Omega(x_1,\dots,\hat{x}_i,\dots x_m)$ is an $(m-1)$-form\footnote{The signs from the
  wedge products will not play a role because at the end we will
  construct basis elements whose normalization in the amplitude is
  fixed from unitarity cuts.}
in all variables except $\hat{x}_i$.  An equivalent terminology is 
that there are only simple poles.
  That is, we are interested in
integrands where we can change variables $x_i\rightarrow g_i(x_j)$
such that the form becomes
\begin{equation}
\Omega = \dlog\,g_1\wedge \dlog\,g_2 \wedge \dots \wedge \dlog\,g_m \,,
\end{equation}
where we denote
\begin{equation}
\dlog\,x \equiv \frac{dx}{x}\,.
\end{equation}
We refer to this representation as a ``$\dlog$ form''. 

A simple example of such a form is
$\Omega(x) = dx/x\equiv \dlog\, x$, while $\Omega(x) = dx$ or
$\Omega(x) = dx/x^2$ are examples of forms which do not have this
property. A trivial two-form with logarithmic singularities is $\Omega(x,y)
= dx\wedge dy/(xy) = \dlog\,x\wedge \dlog\,y$.  A less trivial
example of a $\dlog$ form is
\begin{equation}
\Omega(x,y) = \frac{dx\wedge dy}{xy(x+y+1)} =
 \dlog \frac{x}{x+y+1} \wedge
 \dlog \frac{y}{x+y+1} \, .
\label{eqn:ExampledLog}
\end{equation}
In this case, the property of only logarithmic singularities is not
obvious from the first expression, but a change of variables resulting
in the second expression makes the fact that $\Omega$ is a $\dlog$
form manifest. This may be contrasted with the form 
\begin{equation}
\Omega(x,y) = \frac{dx \wedge dy}{xy(x+y)} \,,
\end{equation}
which is not logarithmic because near the pole $x=0$
it behaves as $dy/y^2$; this form cannot be written as a $\dlog$ form.
In general, the nontrivial changes of variables required can make it
difficult to find the explicit $\dlog$ forms even where they exist.

In a bit more detail, consider the behavior of a form near $x=0$.  If
the integrand scales as $dx/x^m$ for integer $m$, we consider two
different regimes where integrands can fail to have logarithmic
singularities. The first is when $m \ge 2$, which results in double or
higher poles at $x = 0$. The second is when $m\le 0$, which results
in a pole at infinity.  Avoiding unwanted
singularities, either at finite or infinite values of $x$, leads to
tight constraints on the integrand of each diagram. Since we
take the denominators to be the standard Feynman propagators
associated to a given diagram, in our expansion of the amplitude 
the only available freedom is to adjust the kinematic numerators. 
As a simple toy example, consider the form
\begin{equation}
\Omega(x,y) = \frac{dx\wedge dy\,\,N(x,y)}{xy(x+y)}\,.
\end{equation}
As noted above, for a constant numerator $N(x,y)=1$ the form develops
a double pole at $x=0$. Similarly, for $N(x,y)=x^2+y^2$ the form
behaves like $dy$ for $x=0$ and again it is not logarithmic. There is
only one class of numerators that make the form logarithmic near $x=0$
and $y=0$ : $N(x,y) = a_1 x + a_2 y$ for arbitrary $a_1$ and $a_2$.

Our discussion of loop integrands will be similar: constant numerators
(i.e.~those independent of loop momenta) are dangerous for they may allow
double or higher poles located at finite values of loop momenta,
while a numerator with too many powers of loop momentum can develop
higher poles at infinity. It turns out that the first case is generally 
the problem in gauge  theory, whereas the second case usually arises for gravity 
amplitudes, because the power counting of numerators is boosted relative 
to the gauge-theory case. For sYM integrands, we will carefully tune 
numerators so that only logarithmic singularities are present.  
The desired numerators live exactly on the boundary between too many
and too few powers of loop momenta.
%
%%%%%%%%%%%%%%%%%%%%%%%%%%%%%%%%%%%%%%%%%%%%%%%%%%%%%%%%%%%%%%%%%%%
%
\subsection{Loop integrands and poles at infinity}
\label{subsec:LoopIntegralsPoleInfty}
%
%%%%%%%%%%%%%%%%%%%%%%%%%%%%%%%%%%%%%%%%%%%%%%%%%%%%%%%%%%%%%%%%%%%
%
%
%%%%%%%%% FIGURE %%%%%%%%%%%%%%%
\begin{figure}[tb]
\begin{center}
\begin{tabular}
{>{\centering\arraybackslash}m{0.30\textwidth}
>{\centering\arraybackslash}m{0.30\textwidth}
>{\centering\arraybackslash}m{0.30\textwidth}
}
\includegraphics{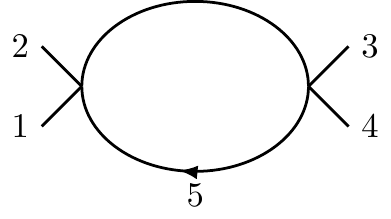} &
\includegraphics{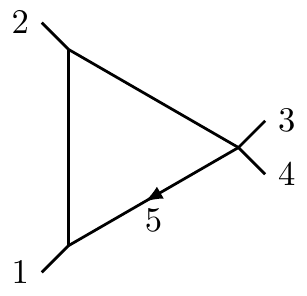} &
\includegraphics{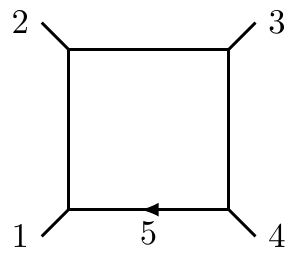} \\
(a) & (b) & (c)\\
\end{tabular}
\caption[]{
The (a) bubble, (b) triangle and (c) box one-loop diagrams.
     }
\label{fig:OneLoopBubTriBox}
\end{center}
\end{figure}
%%%%%%%%%%%%%%%%%%%%%%%%%%%%

Now consider the special class of four-forms that correspond to
one-loop integrands. Standard integral reduction
methods~\cite{PassarinoVeltman,BDKIntegrals} reduce any massless
one-loop amplitude to a linear combination of box, triangle and bubble
integrals.  In nonsupersymmetric theories there are additional
rational terms arising from loop momenta outside of $D=4$; these are
not relevant for our discussion of $\NeqFour$ sYM theory.  While it
will eventually be necessary to include the $(-2 \eps)$ dimensional
components of loop momenta, since these are in general required by
dimensional regularization, for the purposes of studying the
singularities of the integrand we simply put this matter aside.  In
any case, direct checks reveal that these $(-2 \eps)$ dimensional
pieces do not lead to extra contributions through at least six loops
in $\NeqFour$ sYM four-point amplitudes~\cite{DDimCheck}.  That is,
the naive continuation of the four-dimensional integrand into $D$
dimensions yields the correct results. As usual, infrared
singularities are regularized using dimensional regularization.
(See for example,
Refs.~\cite{GravityThreeLoop,Bern:2010tq,ColorKinematics}.)  We focus
here on the four-point case, but a similar analysis can be performed
for larger numbers of external legs as well, although in this case we
expect nontrivial corrections from $(-2 \eps)$ components of the loop
momenta.

Consider the bubble, triangle and box integrals in
\fig{OneLoopBubTriBox}.  In these and all following diagrams, we take
all external legs as outgoing.  The explicit forms in $D=4$ are
\begin{align}
d \I_2 & = {d^4\ell_5}  \frac{1} {\ell_5^2(\ell_5-k_1-k_2)^2} \,,  \nonumber\\
d \I_3 & = {d^4\ell_5}  \frac{s} {\ell_5^2(\ell_5-k_1)^2(\ell_5-k_1-k_2)^2} \,,	\\
d \I_4 & = {d^4\ell_5}  \frac{st} {\ell_5^2(\ell_5-k_1)^2(\ell_5-k_1-k_2)^2(\ell_5+k_4)^2} \,, \nonumber
\end{align}
where we have chosen a convenient normalization.  
The variables $s = (k_1+k_2)^2$ and $t = (k_2 + k_3)^2$ are the usual 
Mandelstam invariants, depending only on external momenta. 
Under integration, these forms are infrared or ultraviolet divergent
and need to be regularized, but as mentioned about we set this 
aside and work directly in four dimensions.

In $D=4$, we can parametrize the loop momentum in terms of four
independent vectors constructed from the spinor-helicity variables
associated with the external momenta $k_1 = \lambda_1 \tilde
\lambda_1$ and $k_2 = \lambda_2 \tilde \lambda_2$.  A clean choice for
the four degrees of freedom of the loop momentum is
\begin{equation}
\ell_5 = \alpha_1\lambda_1\widetilde{\lambda}_1 +
 	 \alpha_2\lambda_2\widetilde{\lambda}_2 +
	 \alpha_3\lambda_1\widetilde{\lambda}_2 +
	 \alpha_4\lambda_2\widetilde{\lambda}_1\,,
\label{eqn:OneLoopParametrization}
\end{equation}
where the $\alpha_i$ are now the independent variables.
Writing $d \I_2$ in this parametrization we obtain
\begin{equation}
d \I_2 = \frac{d\alpha_1\wedge d\alpha_2\wedge d\alpha_3\wedge d\alpha_4}
{(\alpha_1\alpha_2-\alpha_3\alpha_4)
  (\alpha_1\alpha_2-\alpha_3\alpha_4-\alpha_1-\alpha_2+1)} \,.
\end{equation}
In general, since we are not integrating the expressions, we
ignore Feynman's $i \epsilon$ prescription and any factors of $i$
from Wick rotation.

To study the singularity structure, we can focus on subregions of the
integrand by imposing on-shell or cut conditions.  As an example, the cut
condition $\ell_5^2=0$ can be computed in these variables by setting
\begin{equation}
0=\ell_5^2 = (\alpha_1\alpha_2-\alpha_3\alpha_4)s\,.
\end{equation}
We can then eliminate one of the $\alpha_i$, say $\alpha_4$, by computing
the residue on the pole located at 
$\alpha_4=\alpha_1\alpha_2/\alpha_3$. This results in a residue,
\begin{equation}
\underset{\ell_5^2 = 0}{\textrm{Res}} \ d \I_2 =
\frac{d\alpha_3}{\alpha_3}\wedge
\frac{d\alpha_2}{(\alpha_2+\alpha_1-1)} \wedge d\alpha_1 \,.
\end{equation}
Changing variables to $\alpha_\pm  = \alpha_1 \pm \alpha_2$, this becomes
\begin{equation}
\underset{\ell_5^2 = 0}{\textrm{Res}} \ d \I_2 =
 \frac{d\alpha_3}{\alpha_3}\wedge \frac{d\alpha_+}{(\alpha_+ -1)}
  \wedge d\alpha_- \,.
\end{equation}
We can immediately see that the form $d\I_2$ is non-logarithmic in $\alpha_-$,
and thus the bubble integrand has a nonlogarithmic singularity in this
region.

Carrying out a similar exercise for the triangle $d\I_3$
using the parametrization in \eqn{OneLoopParametrization}, we obtain
\begin{equation}
d \I_3 =  \frac{d\alpha_1\wedge d\alpha_2\wedge d\alpha_3\wedge d\alpha_4}
{(\alpha_1\alpha_2-\alpha_3\alpha_4)
(\alpha_1\alpha_2-\alpha_3\alpha_4-\alpha_2)
(\alpha_1\alpha_2-\alpha_3\alpha_4-\alpha_1-\alpha_2+1)} \,.
\end{equation}
We can make a change of variables and rewrite it in the manifest $\dlog$ form,
\begin{equation}
d \I_3 = \dlog\,(\alpha_1\alpha_2-\alpha_3\alpha_4)\wedge
\dlog\,(\alpha_1\alpha_2-\alpha_3\alpha_4-\alpha_2)\wedge
\,\dlog(\alpha_1\alpha_2-\alpha_3\alpha_4-\alpha_1-\alpha_2+1)\wedge
\,\dlog\,\alpha_3 \,.
\label{eqn:dLogTriAlphas}
\end{equation}
Translating this back into momentum space:
\begin{equation}
d \I_3 = \dlog\ell_5^2
\wedge \dlog(\ell_5-k_1)^2
\wedge \dlog(\ell_5-k_1-k_2)^2
\wedge \dlog(\ell_5 - k_1)\cdot(\ell_5^\ast-k_1)\,,
\label{eqn:dLogTriangle}
\end{equation}
where $\ell_5^\ast \equiv \beta \lambda_2 \widetilde\lambda_1 + 
   \lambda_1 \widetilde{\lambda}_1$ is one of the two solutions 
to the triple cut.  The parameter $\beta$ is arbitrary in the 
triple cut solution, and the $\dlog$ form is independent of it. 
For the box integral, a similar
process followed in Ref.~\cite{ArkaniHamed:2012nw} results in
\begin{equation}
d \I_4 =  \dlog \frac{\ell_5^2}{(\ell_5-\ell_5^\ast)^2} \wedge
				\,\dlog \frac{(\ell_5-k_1)^2}{(\ell_5-\ell_5^\ast)^2}\wedge
				\,\dlog \frac{(\ell_5-k_1-k_2)^2}{(\ell_5-\ell_5^\ast)^2} \wedge
				\,\dlog \frac{(\ell_5+k_4)^2}{(\ell_5-\ell_5^\ast)^2} \,,
\label{eqn:dLogBox}				
\end{equation}
where $\ell_5^\ast \equiv -\frac{\la 1 4\ra}{\la 2 4\ra} \lambda_2
\widetilde \lambda_1 + \lambda_1 \widetilde\lambda_1$; see also our
discussion in subsection~\ref{subsec:dlogOneLoop}. 

While both triangle and box integrands can be written in $\dlog$ form,
there is an important distinction between the triangle form $d \I_3$
and the box form $d \I_4$. On the cut $\alpha_4 = \alpha_1
\alpha_2/\alpha_3$, only one $\dlog$-factor in $d \I_3$ depends on
$\alpha_3$ and develops a singularity in the limit
$\alpha_3\rightarrow\infty$ (which implies $\ell_5\rightarrow\infty$),
while $d \I_4$ does not. We refer to any singularity that develops as
a loop momentum approaches infinity (in our example, $\ell_5
\rightarrow \infty$) at any step in the cut structure as a {\it pole
  at infinity}.  To be more specific, even if a $\dlog$ form has no
pole at infinity before imposing any cut conditions, it is possible to
generate such poles upon taking residues, as we saw in the example of
the triangle integrand above.  In this sense, the pole at infinity
property is more refined than simple power counting, which only
considers the overall scaling of an integrand before taking any cuts.

The issue of poles at infinity will be important for our discussion of
$\NeqFour$ sYM theory as well as $\NeqEight$ supergravity amplitudes.
While a lack of poles at infinity implies ultraviolet finiteness,
having poles at infinity does not necessarily mean that there are
divergences.  For example, the triangle integral contains such a pole
in the cut structure but is ultraviolet finite.  In principle, there
can also be nontrivial cancellations between different contributions.

%%%%%%%%%%%%%%%%%%%%% FIGURE %%%%%%%%%%%%%%%
\begin{figure}[tb]
\centering
\includegraphics{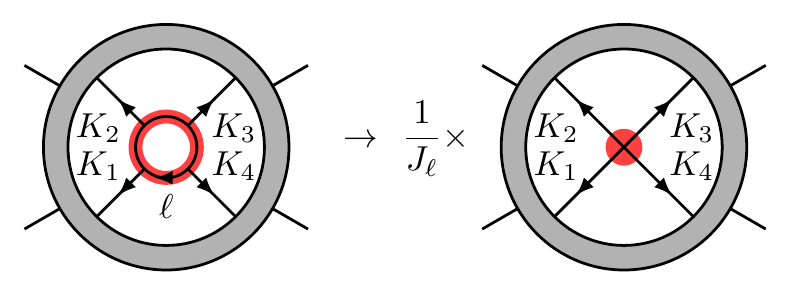}
\caption{ The left diagram is a generic $L$-loop contribution to the
  four-point $\NeqFour$ sYM amplitude.  The thick (red) highlighting
  indicates propagators replaced by on-shell conditions. After this
  replacement, the highlighted propagators leave behind the simplified
  diagram on the right multiplied by an inverse Jacobian,
  \eqn{CutBoxJacobian}.  The four momenta $K_1, \ldots , K_4$ 
  can correspond either to external legs or propagators of the higher-loop diagram.  }
\label{fig:JacobianRules}
\end{figure}
%%%%%%%%%%%%%%%%%%%%%%%%%%%%%%%%%%%%%%%%

To find numerators that do not allow these poles
at infinity and also ensure only logarithmic singularities, it is not necessary to
compute every residue of the integrand.  This is because cutting box
subdiagrams from a higher loop diagram, as on the left in
\fig{JacobianRules}, can only increase the order of remaining poles in
the integrand.  To see this, consider computing the four residues that
correspond to cutting the four highlighted propagators in
\fig{JacobianRules},
\begin{equation}
\ell^2  = (\ell - K_1)^2 = (\ell - K_1 - K_2)^2 = (\ell + K_4)^2 = 0 \,.
\label{eqn:OnShellBox}
\end{equation}
This residue is equivalent to computing the Jacobian
obtained by replacing the box propagator with on-shell 
delta functions.  This Jacobian is then
\begin{equation}
J_{\ell} = |\partial P_i / \partial \ell^\mu| \,,
\label{eqn:GeneralJacobian}
\end{equation}
where the $P_i$ correspond to the four inverse propagators placed
on shell in \eqn{OnShellBox}. See, for example, 
Ref.~\cite{CachazoSharpening} for more details.
Another way to obtain this Jacobian by reading off the rational
factors appearing in front of the box integrals---see
appendix I of Ref.~\cite{BDKIntegrals}.

For the generic case $J_\ell$ contains square roots making, it
difficult to work with.  In special cases it simplifies. For example
for $K_1 = k_1$ massless, the three-mass normalization is
\begin{equation}
J_{\ell} = (k_1 + K_2)^2(K_4+k_1)^2 - K_2^2 K_4^2 \,. 
\label{eqn:CutBoxJacobian}
\end{equation}
If in addition $K_3=k_3$ is massless, the so called ``two-mass-easy''
case, the numerator factorizes into a product of two factors, a
feature that is important in many calculations.  This gives,
\begin{equation}
J_\ell=(K_2+k_1)^2(K_4+k_1)^2-K_2^2K_4^2=(K_2\cdot q)(K_2\cdot \overline{q})\,,
\label{eqn:TwoMassEasyJacobian}
\end{equation}
where $q=\lambda_1\widetilde{\lambda}_3$ and $\overline{q}=\lambda_3\widetilde{\lambda}_1$.
If instead both $K_1 = k_1$ and $K_2=k_2$ are massless we get so called two-mass-hard normalization 
\begin{equation}
J_\ell = (k_1+k_2)^2(K_3+k_2)^2\,.
\end{equation}
These formula are useful at higher loops, where the $K_j$ 
depend on other loop momenta.

These Jacobians go into the denominator of the integrand after a
box-cut is applied. It therefore can only raise the order of the
remaining poles in the integrand.  Our basic approach utilizes this
fact: we cut embedded box subdiagrams from diagrams of interest and
update the integrand by dividing by the obtained Jacobian
(\ref{eqn:GeneralJacobian}).  Then we identify all kinematic regions
that can result in a double pole in the integrand.

It would be cumbersome to write out all cut equations for every such
sequence of cuts, so we introduce a compact notation:
\begin{equation}
{\rm cut} = \{ \ldots,\,  (\ell - K_i)^2 \,, \ldots, B(\ell) \, , \ldots ,\, 
B(\ell',(\ell' -Q) ) \, , \ldots \} \,.
\label{eqn:CutSequenceDef}
\end{equation}
Here:
\begin{itemize}
\item Cuts are applied in the order listed.
\item A propagator listed by itself, as $(\ell - K_i)^2$ is, means: ``Cut just this propagator.''
\item $B(\ell)$ means: ``Cut the four propagators that depend on
  $\ell$.''  This exactly corresponds to cutting the box propagators
  as in \eqn{OnShellBox} and \fig{JacobianRules}.
\item $B(\ell',(\ell' -Q) )$ means: ``Cut the three standard propagators
  depending on $\ell'$, as well as a fourth $1/(\ell' -Q)^2$ resulting
  from a previously obtained Jacobian.''
  The momentum $Q$ depends on other momenta besides $\ell'$.  The four
  cut propagators form a box.
\end{itemize}
We use this notation in subsequent sections.

%%%%%%%%%%%%%%%%%%%%%%%%%%%%%%%%%%%%%%%%%%%%%
%
\subsection{Singularities and maximum transcendental weight}
%
%%%%%%%%%%%%%%%%%%%%%%%%%%%%%%%%%%%%%%%%%%%%%

There is an important link between the singularity structure of the 
integrand and the transcendental weight of an integral, as straightforwardly seen at one loop. 
If we evaluate the bubble, triangle and box integrals displayed in 
\fig{OneLoopBubTriBox} in dimensional
regularization~\cite{DimensionalRegularization}, through ${\cal
  O}(\epsilon^0)$ in the dimensional regularization parameter
$\epsilon$, we have~\cite{BDKIntegrals,Factorization}
\begin{align}
I_2 & =  \frac{1}{\epsilon} + \log(-s/\mu^2) + 2 \,, 	\nonumber \\
	\label{eqn:NormalizedOneLoopInts}
I_3 & = \frac{1}{\epsilon^2} - \frac{\log(-s/\mu^2)}{\epsilon} +
			\frac{\log^2(-s/\mu^2) -\zeta_2}{2}\,, 	\\
I_4 & = \frac{4}{\epsilon^2} - 2 \, \frac{\log(-s/\mu^2) 
       + \log(-t/\mu^2)}{\epsilon} +\log^2(-s/\mu^2) 
   + \log^2(-t/\mu^2) - \log^2(s/t) - 8 \zeta_2\,.	\nonumber
\end{align}
Here $\mu$ is the dimensional regularization scale parameter, and the
integrals are normalized by an overall multiplicative factor of
\begin{equation}
- i \frac{e^{\gamma \eps}(4 \pi)^{2-\eps}} {(2\pi)^{4-2\eps} } \,,
\end{equation}
where $s,t<0$.
In the bubble integral the $1/\eps$ singularity originates from
the ultraviolet, while in the triangle and box integrals all
$1/\eps$ singularities originate from the infrared.

Following the usual rules for counting transcendental weight in the
normalized expressions of \eqn{NormalizedOneLoopInts}, we count
logarithms and factors of $1/\eps$ to have weight 1 and $\zeta_2 =
\pi^2/6$ to have weight 2. Integers have weight 0.  With this counting
we see that the bubble integral, which has nonlogarithmic
singularities as explained in the previous
subsection~\ref{subsec:LoopIntegralsPoleInfty}, is not of uniform
transcendental weight, and has maximum weight 1.  On the other hand
the triangle and box, which both have only logarithmic singularities,
are of uniform weight 2.

Building on the one-loop examples, a natural conjecture is that the
uniform transcendentality property of integrated expressions noted by
Kotikov and Lipatov~\cite{LipatovTranscendentality} is directly linked
to the appearance of only logarithmic singularities.  In fact,
experience shows that after integration the $L$-loop planar ${\cal
  N}=4$ sYM amplitudes have transcendental weight $2L$. Various
examples are found in
Refs.~\cite{ABDK,BDS,Dixon:2014voa,Dixon:2014xca}.  One of our
motivations is to make the connection between logarithmic forms and
transcendental functions more precise. It is clearly an important
connection that deserves further study.

Recently, Henn et al.~observed~\cite{Henn, HennSmirnov, HennSmirnov2,
  Henn:2014qga} that integrals with uniform transcendental weight lead
to simple differential equations.  An interesting connection is that
the integrands we construct do appear to correspond to integrals of
uniform transcendental weight.\footnote{We thank Johannes Henn for
  comparisons with his available results showing that after
  integration our integrands are of uniform transcendental weight.}
Here we mainly focus on the particular subset with no poles at
infinity, since they are the ones relevant for $\NeqFour$ sYM theory.

%==========================================================================
\section{Strategy for nonplanar amplitudes}
\label{sec:StrategySection}
%==========================================================================

As introduced in \sect{TwoLoopSection}, instead of trying to define a
nonplanar global integrand, we subdivide the amplitude into diagrams
with their own momentum labels and analyze them one by one.  In
Ref.~\cite{Log}, the $\NeqFour$ sYM four-point two-loop amplitude was
rewritten in a form with no logarithmic singularities and no
poles at infinity.  In this section, we develop a strategy for doing
the same at higher loop orders.  We emphasize that we are working at
the level of the amplitude integrand prior to integration. In
particular we do not allow for any manipulations that involve the
integration symbol (e.g.~integration-by-parts identities) to shuffle
singularities between contributions.

Our general procedure has four steps:
\begin{enumerate}
\item Define a set of parent diagrams whose propagators are the
	standard Feynman ones.  The parent diagrams are defined to have only
	cubic vertices and loop momentum flowing through all propagators.
\item Construct $\dlog$ {\it numerators}.  These are a basis set of
  numerators constructed so that each diagram has only logarithmic
  singularities and no poles at infinity. These numerators also
  respect diagram symmetries, including color signs. Each $\dlog$
  numerator, together with the diagram propagators, forms a basis
  diagram.
\item Use simple unitarity cuts to determine the linear combination
of basis diagrams that gives the amplitude. 
\item Confirm that the amplitude so constructed is correct and complete.
We use the method of maximal cuts~\cite{MaximalCuts} for this task.
\end{enumerate}
There is no a priori guarantee that this will succeed.
In principle, requiring $\dlog$ numerators could make it
impossible to expand the amplitude in terms of
independent diagrams with Feynman propagators.  Indeed, at a
sufficiently high loop order we expect that even in the planar sector
it may not be possible to find a covariant diagrammatic representation
manifesting the desired properties; in such circumstances we would expect that
unwanted singularities cancel between diagrams.  This may
happen even earlier in the nonplanar sector.  As in many
amplitude calculations, we simply assume that we can construct a basis with
the desired properties, and then, once we have an ansatz, we
check that it is correct by computing a complete set of cuts.  

In this section, we illustrate the process of finding diagram
integrands with the desired properties and explain the steps in some
detail. For simplicity, we focus on the four-point amplitude, but we
expect that a similar strategy is applicable for higher-point
amplitudes in the MHV and NMHV sectors as well. 

We use the one- and two-loop contributions to the four-point amplitudes 
to illustrate the procedure, before turning to three loops
in \sect{ThreeLoopBasisSection}.  We find that the
canonical one-loop numerator is already a $\dlog$ numerator,
while the two-loop result illustrates the issues that we face at
higher loops.  The two-loop amplitude was first obtained in
\cite{BRY}, but in a form that does not make clear the singularity
structure.  In Ref.~\cite{Log}, the two-loop amplitude was rewritten in
a form that makes these properties manifest by rearranging contact
terms in the amplitude by using the color-Jacobi identity.  In
this section we replicate this result by following our strategy of
systematically constructing a basis of integrands with the desired
properties.  In subsequent sections we apply our strategy to higher
loops.

%%%%%%%%%%%%%%%%%%%%%%%%%%%%%%%%%%%%%%%%%%%%%%%%%
\subsection{Constructing a basis}
\label{sec:ConstraintsSubsection}
%%%%%%%%%%%%%%%%%%%%%%%%%%%%%%%%%%%%%%%%%%%%%%%%%

The construction of a basis of integrands starts from a set of parent
diagrams. As mentioned in the introduction to \sect{TwoLoopSection},
we focus on graphs with only cubic vertices.  Furthermore we restrict
to diagrams that do not have triangle or bubble subdiagrams, since
these are not necessary for $\NeqFour$ amplitudes that we study.  We
also exclude any diagrams in which a propagator does not carry loop
momentum, because such contributions can be absorbed into diagrams
in which all propagators contain loop momenta. At the end, we confirm this
basis of diagrams is sufficient by verifying a set of unitarity cuts
that fully determine the amplitude.

%%%%%%%%% FIGURE %%%%%%%%%%%%%%%
\begin{figure}[tb]
\begin{center}
\begin{tabular}
{
>{\centering\arraybackslash}m{0.35\textwidth}
>{\centering\arraybackslash}m{0.35\textwidth}
}
\includegraphics{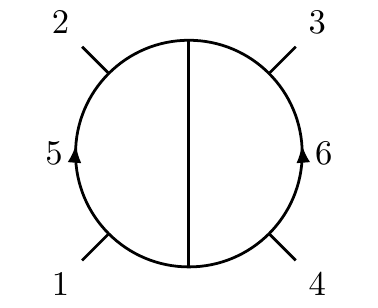} &
\includegraphics{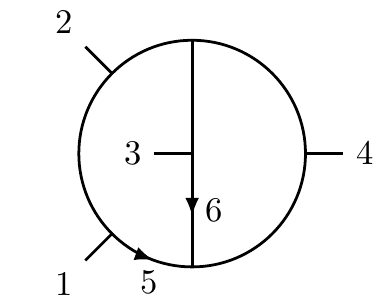} \\
(a) & (b) \\
\end{tabular}
\end{center}
\caption{Two-loop four-point parent diagrams for $\NeqFour$
  sYM theory. }
\label{fig:TwoLoopParent}
\end{figure}
%%%%%%%%%%%%%%%%%%%%%%%%%%%%

At one loop the parent diagrams are the three independent box
integrals, one of which is displayed in \fig{OneLoopBubTriBox}(c), and the
other two of which are cyclic permutations of the external legs $k_2$,
$k_3$ and $k_4$ of this one.  At two loops the four-point
amplitude of $\NeqFour$ sYM theory has twelve parent
diagrams, two of which are displayed in \fig{TwoLoopParent}; the
others are again just given by relabelings of external legs.

Unlike the planar case, there is no global, canonical way to label loop
momenta in all diagrams. In each parent diagram, we label $L$
independent loop momenta as $\ell_5,\ldots,\ell_{4+L}$. By conserving
momentum at each vertex, all other propagators have sums of the loop
and external momenta flowing in them.
We define the $L$-loop {\it integrand}, $\I^{(x)}$, of a diagram
labeled by $(x)$ by combining the kinematic part of the numerator with
the Feynman propagators of the diagram as
\begin{equation}
\I^{(x)} \equiv \frac{N^{(x)}} {\prod_{\alpha_{(x)}}{p^2_{\alpha_{(x)}}}} \,.
\label{eqn:DiagramIntegrand}
\end{equation}
The product in the denominator in \eqn{DiagramIntegrand} runs over all
propagators $p_{\alpha_{(x)}}^2$ of diagram $(x)$, and the kinematic
numerator $N^{(x)}$ generally depends on loop momenta.  From this 
we define an {\it integrand form}
\begin{equation}
\dI^{(x)} \equiv \prod_{l=5}^{4+L} d^{4} \ell_{l} \, \I^{(x)}\,.
\label{eqn:DiagramIntegrandForm}
\end{equation}
This integrand form is a $4L$ form in the $L$
independent loop momenta $\ell_5,\ldots,\ell_{4+L}$.
We have passed factors of $i$, $2\pi$, and coupling constants
into the definition of the amplitude, \eqn{LoopGauge}.
As mentioned previously, we focus on $D=4$.

We define an expansion of the numerator
\begin{equation}
N^{(x)} = \sum_i a_{i}^{(x)} N^{(x)}_i \,,
\label{eqn:BasisExpansion}
\end{equation}
where the $N^{(x)}_i$ are the $\dlog$ numerators we aim to construct,
and the $a_{i}^{(x)}$ are coefficients.  We put off a detailed
discussion of how to fix these coefficients until
subsection~\ref{subsec:expAmpl}, and here just mention that the
coefficients can be obtained by matching an expansion of the amplitude
in $\dlog$ numerators to unitarity cuts or other physical constraints,
such as leading singularities.

Starting from a generic numerator $N_i^{(x)}$, we impose the following
constraints:
\begin{itemize}
\item {\it Overall dimensionality.} $N_i^{(x)}$ must be a local
  polynomial of momentum invariants
  ({i.e.}~$k_a\cdot k_b$, $k_a \cdot \ell_b$, or $\ell_a \cdot \ell_b$)
  with dimensionality $N^{(x)}_{i}\sim
  (p^2)^K$, where $K=P-2L-2$, and $P$ is the number of propagators in the
  integrand. We forbid numerators with $K<0$.
\item {\it Asymptotic scaling.} For each loop momentum $\ell_l$, the
  integrand $\I^{(x)}$ must not scale less than $1/(\ell_l^2)^4$ for
  $\ell_l\rightarrow\infty$ in all possible labellings.
\item {\it No double/higher poles.} The integrand $\I^{(x)}$ must be
  free of poles of order two or more in all kinematic regions.
\item {\it No poles at infinity.} The integrand $\I^{(x)}$ must be
  free of poles of any order at infinity in all kinematic regions.
\end{itemize}
The overall dimensionality and asymptotic scaling give us power
counting constraints on the subdiagrams.  In practice, these two
constraints dictate the initial form of an ansatz for the numerator,
while the last two conditions of no higher degree poles and no poles
at infinity constrain that ansatz to select ``$\dlog$ numerators''.
The constraint on overall dimensionality is the requirement that the
overall mass dimension of the integrand is $-4L-4$;%
\footnote{
The $-4$ in the mass dimension originates from factoring out a 
dimensionful quantity from the final amplitude in \eqn{LoopGauge}.
}
in $D=4$ this matches the dimensionality of gauge-theory integrands.
The asymptotic scaling constraint
includes a generalization of the absence of bubble and triangle
integrals at one-loop order in $\NeqFour$ sYM theory and $\NeqEight$
supergravity \cite{BjerrumBohr:2008ji,ArkaniHamed:2008gz}.  This
constraint is a necessary, but not a sufficient, condition for
having only logarithmic singularities and no poles at infinity.

At one loop, the asymptotic scaling constraint implies that only the
box diagram, \fig{OneLoopBubTriBox}(c), appears; coupling that with
the overall dimensionality constraint implies that the numerator is
independent of loop momentum.  The box numerator must then be a single
basis element which we can normalize to be unity:
\begin{equation}
    N_1^{\rm(B)} = 1\,.
\end{equation}
In the one-loop integrand, neither higher degree poles nor poles at infinity arise.
Thus everything at one loop is consistent and manifestly exhibits only
logarithmic singularities. A more thorough treatment of the one-loop box,
including the sense in which logarithmic singularities are manifest in a box,
is found in the context of $\dlog$ forms in subsection~\ref{subsec:dlogOneLoop}

Next consider two loops.  Here the asymptotic scaling constraint
implies that only the planar and nonplanar double box diagrams in
\fig{TwoLoopParent} appear, since the constraint forbids triangle or
bubble subdiagrams.  We now wish to construct the numerators $N^{\rm
  (P)}$ and $N^{\rm (NP)}$ for the planar (\fig{TwoLoopParent}(a))
and nonplanar (\fig{TwoLoopParent}(b)) diagrams respectively.  There
are different ways of labeling the two graphs.  As already mentioned,
we prefer labels in \fig{TwoLoopParent}, where the individual loop
momenta appear in the fewest possible number of propagators.  This
leads to the tightest power counting constraints in the sense of our
general strategy outlined above. We consider the
planar and nonplanar diagrams separately.

For the planar diagram in \fig{TwoLoopParent}, only four propagators 
contain either loop momentum $\ell_5$ or $\ell_6$.  By the asymptotic 
scaling constraint, the numerator must be independent of both loop 
momenta: $N^{\rm (P)} \sim \mathcal{O}((\ell_5)^0,(\ell_6)^0) $.
Since overall dimensionality restricts $N^{\rm (P)}$ to be quadratic
in momentum, we can write down two independent numerator basis
elements:
\begin{equation}
N_1^{\rm (P)} =  s\,, \hskip 2 cm
N_2^{\rm (P)} =  t\,.
\end{equation}
The resulting numerator is then a linear combination of these
two basis elements:
\begin{equation}
N^{\rm (P)} = a^{\rm (P)}_{1} s + a^{\rm (P) }_{2} t\,,
\label{eqn:TwoLoopPNum}
\end{equation}
where the $a^{\rm (P) }_{j}$ are constants, labeled as discussed after
\eqn{BasisExpansion}.
Again, as in the one-loop case, there are no hidden double poles 
or poles at infinity from which nontrivial constraints could arise.

The nonplanar two-loop integrand $\I^{\rm (NP)}$ is the first instance
where nontrivial constraints result from requiring logarithmic
singularities and the absence of poles at infinity, so we discuss this
example in more detail.  The choice of labels in
\fig{TwoLoopParent}(b) results in five propagators with momentum
$\ell_5$ but only four with momentum $\ell_6$, so $N^{\rm (NP) }$ is
at most quadratic in $\ell_5$ and independent of $\ell_6$: $N^{\rm
  (NP)} \sim \mathcal{O}((\ell_5)^2,(\ell_6)^0)$.  Overall
dimensionality again restricts $N^{\rm (NP)}$ to be quadratic in
momentum.  This dictates the form of the numerator to be
\begin{equation}
N^{\rm (NP)} = c_1 \ell_5^2 + c_2 (\ell_5\cdot Q ) + c_3 s + c_4 t \,,
\label{eqn:NumeratorNP}
\end{equation}
where $Q$ is some vector and the $c_i$ are coefficients independent of loop momenta.  

Now we search the integrand
\begin{equation}
\I^{\rm (NP)} =
	\frac{c_1 \ell_5^2 + c_2(\ell_5\cdot Q) + c_3 s + c_4 t}
	{\ell_5^2(\ell_5+k_1)^2(\ell_5-k_3-k_4)^2 
	\ell_6^2 (\ell_5+\ell_6)^2 (\ell_5+\ell_6-k_4)^2 (\ell_6+k_3)^2 } %\,, A comma here doesn't make sense
\end{equation}
for double poles as well as poles at infinity, and
impose conditions on the $c_i$ and $Q$ such that any such poles vanish.
For the nonplanar double box, we apply this cut on
the four propagators carrying momentum $\ell_6$,
\begin{equation}
\ell_6^2=(\ell_5+\ell_6)^2 = (\ell_5+\ell_6-k_4)^2=(\ell_6+k_3)^2 = 0 \,.
\label{eqn:OnShellBox6}
\end{equation}
The Jacobian for this cut is
\begin{equation}
	J_6 = (\ell_5-k_3)^2(\ell_5-k_4)^2 - (\ell_5-k_3-k_4)^2\ell_5^2
	    = (\ell_5\cdot q)(\ell_5\cdot \overline{q}) \,,
\label{eqn:JacobianNonPlanarDoubleBox}
\end{equation}
where $q=\lambda_3\widetilde{\lambda}_4$, $\overline{q}=
\lambda_4\widetilde{\lambda}_3$. 

After imposing the quadruple cut conditions in \eqn{OnShellBox6} the remaining
integrand, including the Jacobian (\ref{eqn:JacobianNonPlanarDoubleBox}), is
\begin{equation}
\res{\ell_6\text{-}{\rm cut} }\left[ \I^{\rm (NP)} \right] \equiv \widetilde\I^{\rm (NP)} =
\frac{c_1 \ell_5^2 + c_2(\ell_5\cdot Q) + c_3 s + c_4 t}
{\ell_5^2(\ell_5+k_1)^2(\ell_5-k_3-k_4)^2(\ell_5\cdot q)
      (\ell_5\cdot \overline{q})} \,,
\end{equation}
where the integrand evaluated on the cut is denoted by a new symbol $\widetilde \I^{(NP)}$ for brevity.

To make the potentially problematic singularities visible, we
parametrize the four-dimensional part of the remaining loop momentum as
\begin{equation}
\ell_5 = 	\alpha\lambda_3\widetilde{\lambda}_3
			+ \beta \lambda_4 \widetilde{\lambda}_4
			+ \gamma \lambda_3 \widetilde{\lambda}_4
			+ \delta \lambda_4 \widetilde{\lambda}_3 \,.
\label{eqn:LoopMomentumParametrization}
\end{equation}
This gives us
\begin{align}
\widetilde\I^{\rm (NP)} =&
		\Bigl(c_1 (\alpha\beta-\gamma\delta)s
		+ c_2\left[\alpha (Q\cdot k_3)
		+ \beta (Q\cdot k_4) + \gamma \langle 3|Q|4] 
    + \delta \langle 4|Q|3]\right] + c_3s + c_4 t\Bigr)	\nonumber \\
			& \null \times
\Bigl[s^2(\alpha\beta-\gamma\delta)(\alpha\beta-\gamma\delta-\alpha-\beta+1) \nonumber \\
	&  \null \hskip 2 cm \times
		\Bigl((\alpha\beta-\gamma\delta)s+\alpha u+\beta t
			 - \gamma \langle 13\rangle[14]
	- \delta \langle 14\rangle[13]\Bigr)\gamma\delta \Bigr]^{-1} ,
 \end{align}
where we use the convention $2 k_i \cdot k_j = \ab{ij}[ij]$ and $\la i
| k_m | j ] \equiv \ab{im}[m j]$.  Our goal is to identify double- or
higher-order poles. To expose these, we take residues in a
certain order. For example, taking consecutive residues at
$\gamma=0$ and $\delta=0$ followed by $\beta = 0$ gives
\begin{equation}
\res{\substack{\gamma=\delta=0 \\ \beta=0}} \left[\widetilde \I^{\rm (NP)}\right]
	= \frac{c_2 \alpha (Q\cdot k_3) + c_3 s + c_4 t}{s^2u\alpha^2(1-\alpha)}\,.
 \label{eqn:ResidueOne}
\end{equation}
Similarly taking consecutive residues first at $\gamma=\delta=0$ 
followed by $\beta = 1$, we get
\begin{equation}
\res{\substack{\gamma=\delta=0 \\ \beta=1}} \left[\widetilde \I^{\rm (NP)} \right]
	 = 	- \frac{c_1 \alpha s + c_2\left[\alpha (Q\cdot k_3)
		+ (Q\cdot k_4)\right] + c_3 s + c_4 t}{s^2t\alpha(1-\alpha)^2} \,.
\label{eqn:ResidueTwo}
\end{equation}
In both cases we see that there are unwanted double poles in $\alpha$.
The absence of double poles forces us to choose the $c_i$
in the numerator such that the integrand reduces to at most a single pole in $\alpha$.
Canceling the double pole at $\alpha = 0$ in \eqn{ResidueOne}
requires $c_3 = c_4 =0$.  Similarly, the second residue in
\eqn{ResidueTwo} enforces $c_1 s + c_2 (Q\cdot
(k_3+k_4)) = 0$ to cancel the double pole at $\alpha=1$.
The solution that ensures $N^{\rm (NP)}$ is a $\dlog$ numerator is
\begin{equation}
N^{\rm (NP)} = \frac{c_1}{s} [\ell_5^2 (Q\cdot(k_3+k_4))
                      - (k_3+k_4)^2(\ell_5\cdot Q)] \,.
\end{equation}
The integrand is now free of the uncovered double poles, but requiring
the absence of poles at infinity imposes further constraints on the
numerator.  If any of the parameters $\alpha$, $\beta$, $\gamma$ or
$\delta$ grow large, the loop momentum $\ell_5$
\eqn{LoopMomentumParametrization} also becomes large.  Indeed, such a
pole can be accessed by first taking the residue at $\delta=0$,
followed by taking the residues at $\alpha=0$ and $\beta=0$:
\begin{equation}
\res{\substack{\delta=0 \\ \alpha=\beta=0}} \left[\widetilde \I^{\rm (NP)} \right]
		= \frac{\langle 3|Q|4]}{\gamma s^2\langle 13\rangle[14]} \,.
\end{equation}
The resulting form $d \gamma/\gamma$ has a pole for
$\gamma\rightarrow\infty$. Similarly, taking a residue at
$\gamma=0$, followed by residues at
$\alpha=0$ and $\beta=0$ results in a single pole for
$\delta\rightarrow\infty$. To prevent such poles at infinity from appearing
requires $\langle 3|Q|4] = \langle 4|Q|3] = 0$, which in turn requires
that  $Q=\sigma_1 k_3 + \sigma_2 k_4$ with the $\sigma_i$ arbitrary constants.
This is enough to determine the numerator, up to two arbitrary coefficients.

As an exercise in the notation outlined in the beginning of the
section, as well as to illustrate a second approach, we could also
consider the cut sequence $ \{ B(\ell_6) \, \} $, following the
notation defined at the end of
Sect.~\ref{subsec:LoopIntegralsPoleInfty}. The resulting Jacobian is
\begin{equation}
J_{6} = (\ell_5 - k_4)^2 (\ell_5 - k_3)^2 - (\ell_5 + k_1 +k_2)^2 \ell_5^2 \,.
\end{equation}
The two terms on the right already appear as propagators in the integrand,
and so to avoid double poles, the $\dlog$ numerator must scale as
$N^{\rm (NP)} \sim (\ell_5 + k_1 +k_2)^2 \ell_5^2$ in the kinematic regions
where $(\ell_5 - k_4)^2 (\ell_5 - k_3)^2 = 0$. This constraint is sufficient to
fix the ansatz \eqn{NumeratorNP} for $N^{\rm (NP)}$.

In both approaches, the constraints of having only logarithmic singularities and no
poles at infinity results in a numerator for the nonplanar double box
of the form,
\begin{equation}
N^{\rm (NP)} = a^{\rm (NP)}_{1} (\ell_5 -k_3)^2 + a^{\rm (NP)}_{2} 
  (\ell_5-k_4)^2 \,,
\label{eqn:TwoLoopNPNum}
\end{equation}
where $a^{\rm (NP)}_{1}$ and $a^{\rm (NP)}_{2}$ are numerical
coefficients.  Finally, we impose that the numerator should respect the 
symmetries of the diagram.  Because the nonplanar double box is
symmetric under $k_3 \leftrightarrow k_4$ this forces $a^{\rm (NP)}_2
= a^{\rm (NP)}_1$, resulting in a unique numerator up to an overall constant
\begin{equation}
 N^{\rm (NP)}_1 = (\ell_5 -k_3)^2 + (\ell_5-k_4)^2 \,.
\end{equation}
%
%%%%%%%%%%%%%%%%%%%%%%%%%%%%%%%%%
%
\subsection{Expansion of the amplitude}
\label{subsec:expAmpl}
%
%%%%%%%%%%%%%%%%%%%%%%%%%%%%%%%%%

In the previous subsection we outlined a procedure to 
construct a basis of integrands where each element has 
only logarithmic singularities and no pole at infinity.
The next step is to actually expand the amplitude 
in terms of this basis. As mentioned before, we primarily 
focus on the $L$-loop contribution to the $\NeqFour$ sYM theory,
four-point amplitudes. Following the normalization
conventions of Ref.~\cite{ColorKinematics}. these can be written in a diagrammatic representation
\begin{equation}
 {\cal A}^{L-\rm loop}_4 = {g^{2 + 2L}} \frac{i^L \K} {(2\pi)^{DL}}
	\sum_{{\cal S}_4} \sum_{x}
	\frac{1}{S^{(x)}} c^{(x)} \int {\dI ^{(x)} (\ell_5,\ldots,\ell_{4+L})} \,,
\label{eqn:LoopGauge}
\end{equation}
where $\dI ^{(x)}$ is
the integrand form defined in \eqn{DiagramIntegrandForm}, and we have
implicitly analytically continued the expression to $D$ dimensions to be
consistent with dimensional regularization.  
In \eqn{LoopGauge} the sum labeled by $x$ runs over the set of 
distinct, non-isomorphic diagrams with only
cubic vertices, and the sum over ${\cal S}_4$ is over all $4!$
permutations of external legs.  The symmetry factor $S^{(x)}$ then
removes overcounting that arises from automorphisms of the diagrams.
The color factor $c^{(x)}$ of diagram $(x)$ is given by dressing every
three-vertex with a group-theory structure constant, $\tilde f^{abc} =
i \sqrt{2} f^{abc}$. In the sum over permutations in \eqn{LoopGauge},
any given $\dI^{(x')}$ is a momentum relabeling of $\dI^{(x)}$ in
\eqn{DiagramIntegrandForm}.

For the cases we
consider, the prefactor is proportional to the color-ordered tree amplitude,
\begin{equation}
\K = s t A_4^{\tree}(1,2,3,4) \,.
\label{eqn:KappaDef}
\end{equation}
Furthermore, $\K$ has a crossing symmetry so it can also be expressed in terms of
the tree amplitude with different color orderings,
\begin{equation}
\K  = s u A_4^{\tree}(1,2,4,3) = t u A_4^{\tree}(1,3,2,4)\,.
\label{eqn:KappaSymmetry}
\end{equation}
The explicit values of the tree amplitudes are 
\begin{align}
&A_4^{\tree}(1,2,3,4) = i\frac{\delta^8({\cal Q})}{\la12\ra\la23\ra\la34\ra\la41\ra}\,,
\label{eqn:SuperTrees}
\end{align}
where the other two orderings are just relabelings of the first.
The factor $\delta^8({\cal Q})$ is the supermomentum conservation
$\delta$ function, as described in e.g. Ref.~\cite{Elvang:2013cua}.  The
details of this factor are not important for our discussion. For
external gluons with helicities $1^-, 2^-, 3^+, 4^+$ it is just $\la 1
2 \ra^4$, up to Grassmann parameters.

A simple method for expanding the amplitude in terms of $\dlog$
numerators is to use previously constructed representations of the
amplitude as reference data, rather than sew together lower-loop
amplitudes directly.  Especially at higher loops, this drastically
simplifies the process of determining the coefficients $a_{i}^{(x)}$.
To ensure that the constructed amplitude is complete and correct,  we
also check a complete set of unitarity cuts via the method of maximal
cuts \cite{FiveLoopNonPlanar}.

As an illustration of the procedure for determining the coefficients, consider
the two-loop amplitude.  A representation of the two-loop four-point
amplitude is~\cite{BRY} \eqns{LoopGauge}{BasisExpansion} with numerators
\begin{equation}
		N^{\rm(P)}_{\rm old} = s\,, \hskip 2 cm
		N^{\rm (NP)}_{\rm old} = s\,,
\label{eqn:TwoLoopNumeratorsOld}
\end{equation}
where we follow the normalization conventions of Ref.~\cite{ColorKinematics}.
Following our strategy, we demand that the numerators are linear combinations
of the basis elements constructed in \eqns{TwoLoopPNum}{TwoLoopNPNum}:
\begin{align}
N^{\rm (P)} = a^{\rm (P)}_{1} s + a^{\rm (P)}_{2} t\,,  \hskip 2 cm 
N^{\rm (NP)}  = a^{\rm (NP)}( (\ell_5 -k_3)^2 + (\ell_5 -k_4)^2 ) \,,
\label{eqn:TwoLoopNumeratorsUnfixed}
\end{align}
where, for comparison to $N^{\rm (NP)}_{\rm old}$, it is useful to rewrite the nonplanar
numerator as
\begin{equation}
N^{\rm (NP)} =  a^{\rm (NP)} ( -s + (\ell_5 - k_3 - k_4)^2 + \ell_5^2 )\,.
\label{eqn:TwoLoopNumeratorsAlt}
\end{equation}
We can determine the coefficients by comparing the new and old
expressions on the maximal cuts.  By maximal cuts we mean replacing
all propagators with on-shell conditions, $p^2_{\alpha_{(x)}} = 0$, defined in~\eqn{DiagramIntegrand}. The
planar double-box numerator is unchanged on the maximal cut,
since it is independent of all loop momenta. Comparing the two expressions in
\eqns{TwoLoopNumeratorsOld}{TwoLoopNumeratorsUnfixed} gives
\begin{equation}
a^{\rm (P)}_{1} = 1\,, \hskip 2 cm a^{\rm (P)}_{2} = 0 \,.
\end{equation}
For the nonplanar numerator we note that under the maximal cut
conditions $\ell_5^2 = (\ell_5 -k_3 - k_4)^2 = 0$.  Comparing 
the two forms of the nonplanar numerator in
\eqns{TwoLoopNumeratorsOld}{TwoLoopNumeratorsAlt} after imposing these
conditions means
\begin{equation}
		a^{\rm (NP)}_{1} = -1\,,
\label{eqn:NonplanarDoubleBoxSign}
\end{equation}
so that the final numerators are
\begin{align}
		N^{\rm (P)} = s \,,  \hskip 2.5 cm 
		N^{\rm (NP)}  = - ( (\ell_5 -k_3)^2 + (\ell_5 -k_4)^2 ) \,.
\label{eqn:TwoLoopNumerators}
\end{align}
Although this fixes all coefficients in our basis, it does not prove that
our construction gives the correct sYM amplitude.  At two loops
this was already proven in Ref.~\cite{Log}, where the difference
between amplitudes in the old and the new representation was shown to vanish
via the color Jacobi identity.  More generally, we
can appeal to the method of maximal cuts since it offers a systematic and
complete means of ensuring that our constructed nonplanar amplitudes
are correct.

\subsection{Amplitudes and sums of $\dlog$ forms}
\label{subsec:AmpsandDlogs}

At any loop order, assuming the four-point $\NeqFour$ sYM
amplitudes have only logarithmic singularities then 
we can write integrand forms as a sum of $\dlog$ forms.
At the relatively low loop orders that we are working, 
we can do this diagram by diagram, using the expansion of 
the diagrams given in \eqn{LoopGauge}.  We then take each 
diagram form in \eqn{LoopGauge} and expand it as a linear
combination of $\dlog$ forms,
\begin{equation}
d\I^{(x)} = \sum_{j=1}^3 C_j\, d{\I}_j^{(x), \dlog} \,,
\label{eqn:AmpDlog}
\end{equation}
where the $d{\cal I}^{(x),\dlog}_j$ are (potentially sums of) $\dlog$ $4L$ forms.  
As discussed in Ref.~\cite{NonPlanarOnshellToAppear},
for MHV amplitudes the coefficients $C_j$ are Park-Taylor
factors with different orderings.
This follows from super-conformal symmetry of $\NeqFour$ sYM theory,
which fixes the coefficients $C_j$ to be holomorphic functions of spinor variables $\lambda$
and normalizes $d\I^{(x)}$ to be a $\dlog$ form. 
In the four-point nonplanar case this
means that there are only three different coefficients we can get,
\begin{align}
C_1 = A_4^{\tree}(1,2,3,4) \,,
\hskip 1. cm 
C_2 = A_4^{\tree}(1,2,4,3) \,,
\hskip 1. cm 
 C_3 = A_4^{\tree}(1,3,2,4) \,,
\label{eqn:CoefficientDef}
\end{align}
where the explicit form of the tree amplitudes are given in 
\eqn{SuperTrees}. 
The three coefficient are not
independent, as they satisfy $C_1+C_2+C_3=0$. Suppose that the
basis elements in \eqn{LoopGauge} are chosen such that they have only
logarithmic singularities.   We will show, in \sect{DlogSection}, that we can
indeed write the diagram as $\dlog$ forms with coefficients given by the $C_j$.

%========================================================================
\section{Three-loop amplitude}
\label{sec:ThreeLoopBasisSection}
%=======================================================================

In this section we follow the recipe of the previous section to
find a basis of three-loop diagram integrands that have only
logarithmic singularities and no poles at infinity.
%
%%%%%%%%% FIGURE %%%%%%%%%%%%%%%
\begin{figure}[tb]
\begin{center}
\begin{tabular}
{
>{\centering\arraybackslash}m{0.30\textwidth}
>{\centering\arraybackslash}m{0.30\textwidth}
>{\centering\arraybackslash}m{0.30\textwidth}
}
\includegraphics{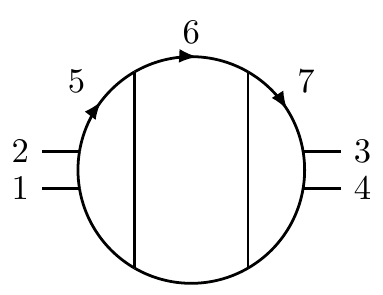} &
\includegraphics{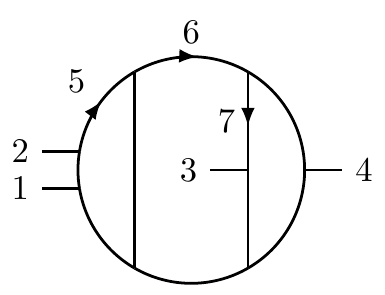} &
\includegraphics{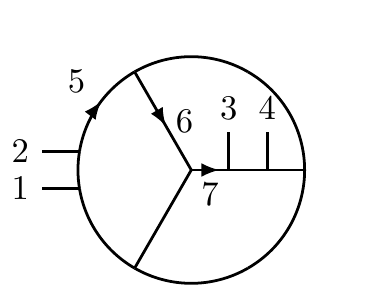} \\
(a) & (b) & (c) \\
\includegraphics{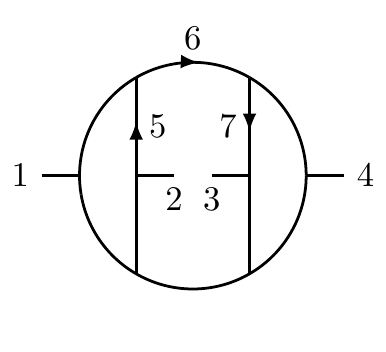} &
\includegraphics{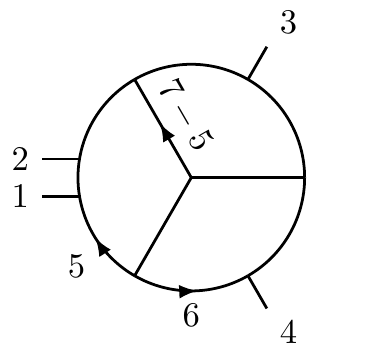} &
\includegraphics{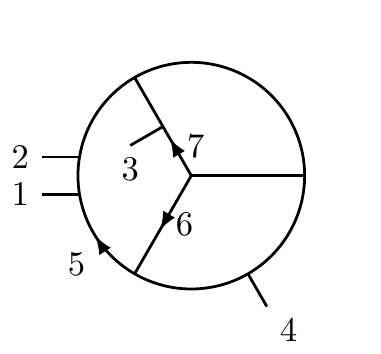} \\
(d) & (e) & (f) \\
\includegraphics{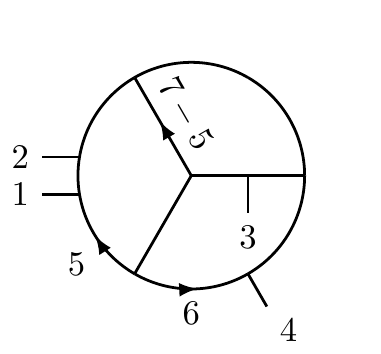} &
\includegraphics{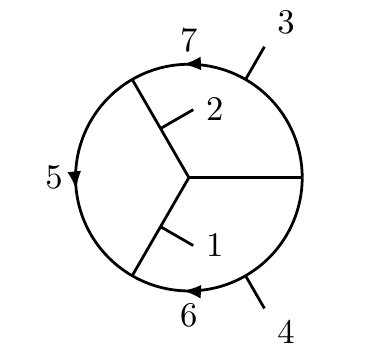} &
\includegraphics{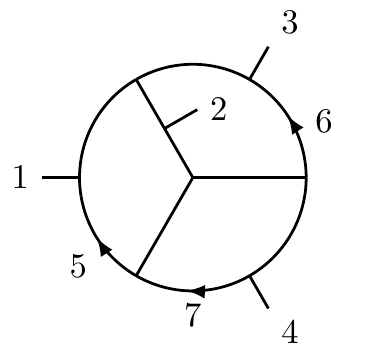} \\
(g) & (h) & (i) \\
\end{tabular}
\end{center}
\caption{ The distinct parent diagrams for three-loop four-point
  amplitudes.  The remaining parent diagrams are obtained by relabeling
  external legs.  }
\label{fig:ParentTenProp}
\end{figure}
%%%%%%%%%%%%%%%%%%%%%%%%%%%%
%
The three-loop four-point parent diagrams are shown in
\fig{ParentTenProp}.  These have been classified in
Ref.~\cite{GravityThreeLoop,Manifest3}, where an unintegrated
representation of the three-loop four-point amplitude of ${\cal N}=4$
sYM theory including nonplanar contributions was first
obtained.  As mentioned in \sect{ConstraintsSubsection}, we restrict to parent diagrams
where no bubble or triangle diagrams appear as subdiagrams;
otherwise we would find a pole at infinity that cannot be removed.
Diagrams with contact terms can be
incorporated into a parent diagram by including inverse propagators in the
numerator that cancel propagators.

Next we assign power counting of the numerator for each parent
diagram.  Applying the power-counting rules in
\sect{ConstraintsSubsection}, we find that the maximum powers of
allowed loop momenta for each parent diagram are
\begin{align}
N^{\rm (a)} &= {\cal O}(1)\,,\hskip 1.5 cm
N^{\rm (b)} = {\cal O}(\ell_6^2)\,, \hskip 1.5cm
N^{\rm (c)} = {\cal O}(\ell_5^2\,,(\ell_5\cdot\ell_7),\ell_7^2)\,,\nonumber\\
N^{\rm (d)} &= {\cal O}(\ell_6^4)\,, \hskip 1.4cm
N^{\rm (e)} = {\cal O}(\ell_5^2)\,, \hskip 1.55 cm
N^{\rm (f)} = {\cal O}(\ell_5^4)\,,\hskip 1 cm
N^{\rm (g)} = {\cal O}(\ell_5^2\ell_6^2)\,, \hskip .5 cm \nonumber\\
N^{\rm (h)} &= {\cal O}(\ell_5^2\ell_6^2, \ell_5^2\ell_7^2,\ \ell_5^2(\ell_6\cdot\ell_7)) \,, \hskip 2.4 cm
N^{\rm (i)} = {\cal O}(\ell_5^2\ell_6^2) \,,
\label{eqn:PowerCounting}
\end{align}
where we use the labels in \fig{ParentTenProp}, since these give the
most stringent power counts. For diagram (h) we need to combine
restrictions from a variety of labellings to arrive at this
stringent power count.  Ignoring the overall prefactor of $\K$, the
overall dimension of each numerator is $\Ord(p^4)$, including external
momenta.

\subsection{Diagram numerators}

The next step is to write down the most general diagram numerators
that are consistent with the power count in \eqn{PowerCounting},
respect diagram symmetry, are built only from Lorentz dot products of the loop and external momenta,
have only logarithmic singularities and have no poles at
infinity. Although the construction is straightforward, the complete
list of conditions is lengthy, so here we only present a few examples
and then write down a table of numerators satisfying the constraints.

We start with diagram (a) in \fig{ParentTenProp}.  The required
numerators are simple to write down if we follow the same logic as in
the two-loop example in \sect{ConstraintsSubsection}.  Since the
numerator of diagram (a) is independent of all loop momenta as noted
in \eqn{PowerCounting}, we can only write numerators that depend on
the Mandelstam invariants $s$ and $t$.  There are three numerators
that are consistent with the overall dimension,
\begin{equation}
  N^{\rm (a)}_1 = s^2 \,, \hskip 1.5 cm 
  N^{\rm (a)}_2 = s t \,, \hskip 1.5 cm 
  N^{\rm (a)}_3 = t^2 \,.
\end{equation}
Following similar logic as at two loops, it is straightforward 
to check that there are no double poles or poles at infinity.

The numerator for diagram (b) is also easy to obtain, this time by
following the logic of the two-loop nonplanar diagram. From
\eqn{PowerCounting}, we see that the only momentum
dependence of the numerator must be on $\ell_6$.  The two-loop subdiagram on the right
side of diagram (b) in \fig{ParentTenProp} containing $\ell_6$ is just
the two-loop nonplanar double box we already analyzed in
\sect{ConstraintsSubsection}.  Repeating the earlier nonplanar box
procedure for this subdiagram gives us the most general possible
numerator for diagram (b),
\begin{equation}
N^{(\rm b)}_1 =  s \bigl( (\ell_6-k_3)^2 + (\ell_6-k_4)^2 \bigr) \,.
\end{equation}
This is just the two-loop nonplanar numerator with an extra factor of
$s$.  A factor of $t$ instead of $s$ is disallowed because it
violates the $k_3 \leftrightarrow k_4$ symmetry of diagram (b).

As a somewhat more complicated example, consider diagram (e) in
\fig{ParentTenProp}.  Because this diagram is planar we could use dual
conformal invariance to find the desired numerator. Instead, for
illustrative purposes we choose to obtain it only from the
requirements of having logarithmic singularities and no pole at
infinity, without invoking dual conformal invariance.  We discuss
the relation to dual conformal symmetry further in
\sect{PlanarIntegrand}.

From \eqn{PowerCounting} we see that the numerator depends on the loop
momentum $\ell_5$ at most quadratically. Therefore we may start with 
the ansatz
\begin{equation}
N^{\rm (e)} = (c_1 s + c_2 t)
\bigl(\ell_5^2 + d_1 (\ell_5\cdot Q) + d_2 s + d_3 t\bigr) \,,
\label{eqn:NumeratorAnsatzE}
\end{equation}
where $Q$ is a vector independent of all loop momenta and the
$c_i$ and $d_i$ are numerical constants.  We have included an
overall factor depending on $s$ and $t$ so that the numerator
has the correct overall dimensions, but this factor does not play a role in
canceling unwanted singularities of the integrand.

In order to extract conditions on the numerator ansatz
\eqn{NumeratorAnsatzE}, we need to find any hidden double
poles or poles at infinity in the integrand.  The 
starting integrand is 
\begin{align*}
\I^{\rm (e)} = & \frac{N^{\rm (e)}}
{\ell_6^2 (\ell_6+\ell_5)^2 (\ell_6+\ell_7)^2 (\ell_6+k_4)^2
(\ell_7-\ell_5)^2 (\ell_7-k_1-k_2)^2  (\ell_7+k_4)^2 }
\\
\null & 
 \times
\frac{1}{\ell_5^2 (\ell_5 - k_1)^2 (\ell_5 - k_1-k_2)^2 }\,.
\labellast \label{eqn:DiagEIntStart}
\end{align*}
Since our numerator ansatz (\ref{eqn:NumeratorAnsatzE}) is a function
of $\ell_5$, we seek double poles only in the regions of momentum
space that we can reach by choosing convenient on-shell values for
$\ell_6$ and $\ell_7$.  This leaves the numerator ansatz unaltered, making it
straightforward to determine all coefficients.

To locate a double pole, consider the cut sequence%
\begin{equation}
{\rm cut} = \{ B(\ell_6) , B(\ell_7, \ell_7) \} \,,
\end{equation}
where we follow the notation defined at the end of
Sect.~\ref{subsec:LoopIntegralsPoleInfty}.  Here $B(\ell_7,\ell_7)$
indicates that we cut the $1/\ell_7^2$ propagator produced by the
$B(\ell_6)$ cut.  This produces an overall Jacobian
\begin{equation}
J_{6,7} = s \left[(\ell_5+k_4)^2\right]^2 \,.
\label{eqn:DiagramEJacobian}
\end{equation}
After this sequence of cuts, the integrand of \eqn{DiagEIntStart} becomes:
\begin{equation}
\res{\substack{\ell_6\text{--}{\rm cut} \\ \ell_7\text{--}{\rm cut}}}
\left[\I^{\rm (e)}\right] = \frac{N^{\rm (e)}}
{\ell_5^2(\ell_5-k_1)^2(\ell_5-k_1-k_2)^2
\left[(\ell_5+k_4)^2\right]^2 s}\,,
\end{equation}
exposing a double pole at $(\ell_5 + k_4)^2 = 0$.

To impose our desired constraints on the integrand, we need to cancel
the double pole in the denominator with an appropriate numerator.
We see that choosing the ansatz in
\eqn{NumeratorAnsatzE} to have $Q = k_4$, $d_1 = 2$, $d_2 = 0$, $d_3 =
0$ gives us the final form of the allowed numerator,
\begin{equation}
N^{\rm (e)} = (c_1 s + c_2 t) (\ell_5+k_4)^2 \,,
\end{equation}
so we have two basis numerators,
\begin{equation}
N^{\rm (e)}_1 =  s (\ell_5+k_4)^2 \,, \hskip 2 cm 
N^{\rm (e)}_2 =  t (\ell_5+k_4)^2 \,.
\end{equation}
We have also checked that this numerator passes all other double-pole
constraints coming from different regions of momentum space.  In
addition, we have checked that it has no poles at infinity.  It is
interesting that, up to a factor depending only on external momenta,
these are precisely the numerators consistent with dual conformal
symmetry. As we discuss in \sect{PlanarIntegrand}, this is no accident.

Next consider diagram (d) in \fig{ParentTenProp}.  From
the power counting arguments summarized in
\eqn{PowerCounting}, we see that the numerator for this diagram is a
quartic function of momentum $\ell_6$, but that it depends on neither
$\ell_5$ nor $\ell_7$. When constructing numerators algorithmically
we begin with a general ansatz, but to more easily illustrate the role
of contact terms we start from the natural guess that
diagram (d) is closely related to a product of two two-loop
nonplanar double boxes. Thus our initial guess is that the desired numerator is the
product of numerators corresponding to the two-loop nonplanar subdiagrams:
\begin{align}
\widetilde N^{\rm (d)} = & \left[(\ell_6+k_1)^2+(\ell_6+k_2)^2\right]
\left[(\ell_6-k_3)^2+(\ell_6-k_4)^2\right] \,.
\end{align}
We label this numerator $\widetilde N^{\rm (d)}$ because, as we see below, it is
not quite the numerator $N^{\rm (d)}$ that satisfies our pole constraints.
As always, note that we have required the numerator to satisfy the symmetries of
the diagram.

Although we do not do so here, one can show that this ansatz
satisfies nearly all constraints on double poles and poles at infinity.
The double pole not removed by the numerator
is in the kinematic region:
\begin{equation}
{\rm cut} = \{ \ell_5^2, (\ell_5+k_2)^2, \ell_7^2, (\ell_7-k_3)^2, B(\ell_6)\} \,.
\end{equation}
Before imposing the final box cut, we solve the first four 
cut conditions in terms of two parameters $\alpha$ and $\beta$:
\begin{equation}
\ell_5 = \alpha k_2 \,, \hskip 1 cm \ell_7 = -\beta k_3 \,.
\label{eqn:DEllFiveEllSixSolutions}
\end{equation}
The final $B(\ell_6)$ represents a box-cut of four
of the six remaining propagators that depend on $\alpha$ and $\beta$:
\begin{equation}
(\ell_6-\alpha k_2)^2=
(\ell_6-\alpha k_2+k_1)^2 =
(\ell_6+\beta k_3)^2 = 
(\ell_6+\beta k_3-k_4)^2=0\,.
\label{eqn:QuadCutD}
\end{equation}
Before cutting the $B(\ell_6)$ propagators,
the integrand is
\begin{equation}
\res{\substack{\ell_5\text{--}{\rm cut} \\ \ell_7\text{--}{\rm cut}}}
\widetilde{\I}^{\rm (d)} = 
\frac{\widetilde N^{\rm (d)}} {\ell_6^2(\ell_6+k_1+k_2)^2(\ell_6-\alpha k_2)^2
(\ell_6-\alpha k_2+k_1)^2(\ell_6+\beta k_3)^2(\ell_6+\beta k_3-k_4)^2} \,.
\label{eqn:DEllSixEllSevenResidue}
\end{equation}
Localizing further to the $B(\ell_6)$ cuts produces a Jacobian
\begin{equation}
J_6=su(\alpha-\beta)^2 \,,
\end{equation}
while a solution to the box-cut conditions of \eqn{QuadCutD}
\begin{equation}
\ell_6^* =\alpha\lambda_4\widetilde{\lambda}_2
   \frac{\langle12\rangle}{\langle14\rangle}
 - \beta\lambda_1\widetilde{\lambda}_3
\frac{\langle34\rangle}{\langle14\rangle} \,,
\label{eqn:CutSolutionD}
\end{equation}
turns the remaining uncut propagators of \eqn{DEllSixEllSevenResidue}
into:
\begin{equation}
\ell_6^2 = s\alpha\beta \,, \hskip 2 cm
(\ell_6+k_1+k_2)^2 = s(1+\alpha)(1+\beta)\,.
\end{equation}
The result of completely localizing all momenta in this way is:
\begin{equation}
\res{\rm cuts \ } \widetilde{\I}^{\rm (d)} =
-\frac{s^2 (\alpha(1+\beta) + \beta(1+\alpha))^2}
{s^3u \alpha\beta(1+\alpha)(1+\beta)(\alpha-\beta)^2} \,.
\label{eqn:DoublePole}
\end{equation}
We see that there is a double pole located at $\alpha - \beta = 0$.
To cancel this double pole, we are forced to add an extra term to the
numerator. A natural choice is a term that collapses
both propagators connecting the two two-loop nonplanar subdiagrams:
$\ell_6^2 (\ell_6 + k_1 + k_2)^2$.
On the support of the cut solutions \eqn{CutSolutionD}, this becomes
$s^2 \alpha \beta (\alpha+1) (\beta + 1)$.  We can cancel
the double pole at $\alpha - \beta = 0$ in \eqn{DoublePole} by choosing the linear
combination
\begin{equation}
N_1^{\rm (d)} =  \left[(\ell_6+k_1)^2+(\ell_6+k_2)^2\right]
\, \left[(\ell_6-k_3)^2+(\ell_6-k_4)^2\right]
  - 4 \ell_6^2  (\ell_6 + k_1 + k_2)^2 \,.
\label{eqn:NumeratorD}
\end{equation}
Indeed, with this numerator the diagram lacks even a single
pole at $\alpha - \beta = 0$.

It is interesting to note that if we relax the condition that the
numerator respects the diagram symmetry $k_1 \leftrightarrow k_2$ and $k_3
\leftrightarrow k_4$, there are four independent numerators with no
double pole.  For example,
\begin{equation}
\tilde N^{\rm (d)} = (\ell_6+k_1)^2 (\ell_6-k_3)^2 - 
   \ell_6^2  (\ell_6 + k_1 + k_2)^2 \,,
\end{equation}
is a $\dlog$ numerator.  When we require that
$N^{\rm (d)}$ respect diagram symmetry, we need
the first four terms in \eqn{NumeratorD}, each with
its own ``correction'' term $- \ell_6^2 (\ell_6 + k_1 + k_2)^2$.
This accounts for the factor of four on the last term
in \eqn{NumeratorD}.

We have carried out detailed checks of all potentially dangerous regions of the integrand
of diagram (d) showing that the numerator of \eqn{NumeratorD}
results in a diagram with only logarithmic singularities and no
poles at infinity.  In fact, the numerator (\ref{eqn:NumeratorD}) is
the only one respecting the symmetries of diagram (d) with these
properties.  We showed this by starting with a general ansatz subject to
the power counting constraint in \eqn{PowerCounting} and showing that no other 
solution exists other than the one in \eqn{NumeratorD}.

We have gone through the diagrams in 
\fig{ParentTenProp} in great detail, finding the numerators that respect diagram
symmetry (including color signs), and that have only logarithmic singularities
and no poles at infinity.  This gives us a set of basis $\dlog$
numerators associated with each diagram.  For the diagrams where numerator
factors do not cancel any propagators, the set of numerators is collected
in \tab{BasisTenProp}.  In addition, there are also diagrams where 
numerators do cancel propagators.  For the purpose of constructing amplitudes,
it is convenient to absorb these contact contributions into the parent
diagrams of \fig{ParentTenProp} to make color assignments
manifest. This allows us to treat all contributions on an equal
footing, such that we can read off the color factors directly from the
associated parent diagram by dressing each three vertex with an
$\tilde f^{abc}$.  This distributes the contact term diagrams in
\tab{BasisEightandNine} among the parent diagrams, listed in
\tab{BasisParentContactEightNine}.  When distributing the contact
terms to the parent diagrams, we change the momentum labels to those of each parent
diagram and then multiply and divide by the missing propagator(s). The
reason the numerators in \tab{BasisParentContactEightNine} appear more
complicated than those in \tab{BasisEightandNine} is that a single term from
\tab{BasisEightandNine} can appear with multiple momentum relabellings in order 
to enforce the symmetries of the parent diagrams on the numerators.

As an example of the correspondence between the numerators in
\tab{BasisParentContactEightNine} and \tab{BasisEightandNine},
consider diagram (j) and the associated numerators, $N_1^{\rm (j)}$
and $N_2^{\rm (j)},$ in \tab{BasisEightandNine}.  To convert this into a
contribution to diagram (i) in \tab{BasisParentContactEightNine}, we
multiply and divide by the missing propagator $1/(\ell_5 + \ell_6 +
k_4)^2$.  Then we need to take the appropriate linear combination so
that the diagram (i) antisymmetry (including the color sign) under $\{
k_1 \leftrightarrow k_3, \ell_5 \leftrightarrow \ell_6, 
\ell_7 \leftrightarrow -\ell_7 \}$ is satisfied.  This gives,
\begin{align}
% B92
N^{{\rm (i)}}_2 & = \frac{1}{3}
          (\ell_5 +\ell_6 +k_4)^2  \left[t - s\right] \,.
\label{eqn:BasisNinePropA}
\end{align}
In fact, there are three alternative propagators that can be inserted
instead of $1/(\ell_5 +\ell_6 +k_4)^2$
which are all equivalent to the three relabelings of external lines for
diagram (i).  We have absorbed a combinatorial factor of $\frac{1}{3}$ into
the definition of the numerator because of the differing symmetries between
diagram (i) in \tab{BasisParentContactEightNine} and diagram (j) in
\tab{BasisEightandNine}.

%%%%%%%%%%%%%%%%% TABLE %%%%%%%%%%%%%%%%%%%%%%%%%%%%%%%%% 
\begin{table}[t]
\begin{center}
\begin{tabular}[t]
{ >{\centering\arraybackslash} m{0.02\textwidth} 
 >{\centering\arraybackslash} m{0.12\textwidth}  %Horizontally and vertically center column 1
 >{$}l<{$} %Automatically put column two in math mode, left aligned
}
 & \hskip -2. cm  Diagram & \textrm{Numerators} \\
\hline \hline
%A
(a) & \includegraphics{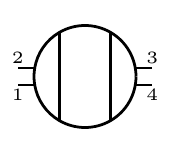} & 
\begin{array}{r l}
N^{{\rm(a)}}_{1} & = 
%%%%% begin : Num[a,1]
 s^2
%%%%% end : Num[a,1]
 \,, 
\hskip .6 cm 
N^{{\rm(a)}}_{2} =
%%%%% begin : Num[a,2]
 s t
%%%%% end : Num[a,2]
\,,
\hskip .6 cm 
N^{{\rm(a)}}_{3} =
%%%%% begin : Num[a,3]
  t^2
%%%%% end : Num[a,3]
\,, \\ 
\end{array}
\\ [-.3cm]
%\hline
%B
(b)& \includegraphics{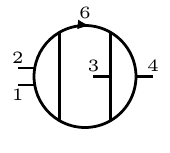} & 
\begin{array}{r l}
 N^{{\rm(b)}}_{1} & =
%%%%% begin : Num[b,1]
 s {} \left[ (\ell_{6}-k_{3})^{2}
		+ (\ell_{6}-k_{4})^{2} \right]
%%%%% end : Num[b,1]
 \,, \\
\end{array}
\\ [-.3cm]
%\hline
%C
(c) & \includegraphics{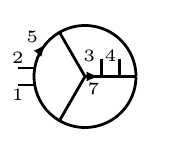} & 
\begin{array}{r l}
N^{{\rm(c)}}_{1} & =
%%%%% begin : Num[c,1]
	s {} \left[ (\ell_{5}-\ell_{7})^{2}
	+ (\ell_{5}+\ell_{7}+k_{1}+k_{2})^{2} \right]
%%%%% end : Num[c,1]
 \,, \\
\end{array}
\\ [-.3cm]
%\hline
%D
(d) & \includegraphics{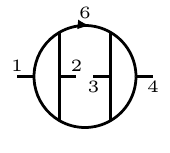} & 
\begin{array}{r l}
N^{{\rm(d)}}_{1} & =
%%%%% begin : Num[d,1]
	\left[ (\ell_{6}+k_{1})^{2}+(\ell_{6}+k_{2})^{2}\right]^{2}
     -4\ell_{6}^{2}(\ell_{6}+k_{1}+k_{2})^{2} 
%%%%% end : Num[d,1]
\,,  \\
\end{array}
\\[-.3cm]
%\hline
%E
(e) & \includegraphics{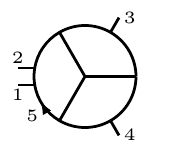} & 
\begin{array}{r l}
N^{{\rm(e)}}_{1} & = 
%%%%% begin : Num[e,1]
s{}(\ell_{5}+k_{4})^{2}
%%%%% end : Num[e,1]
 \,, \hskip .6 cm 
N^{{\rm(e)}}_{2}  = 
%%%%% begin : Num[e,2]
t{}(\ell_{5}+k_{4})^{2}
%%%%% end : Num[e,2]
 \,, \\
\end{array}
\\[-.3cm]
%\hline
%F
(f) & \includegraphics{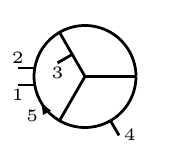} & 
\begin{array}{r l}
N^{{\rm(f)}}_{1}  =
%%%%% begin : Num[f,1]
	(\ell_{5}+k_{4})^{2}
    \left[(\ell_{5}+k_{3})^{2} + (\ell_{5}+k_{4})^{2} \right]
%%%%% end : Num[f,1]
 \,, \\[-.3cm]
\end{array}
\\
%\hline
%G
(g) & \includegraphics{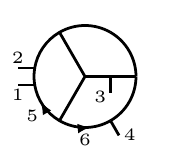} & 
\begin{array}{r l}
N^{{\rm(g)}}_{1} & = 
%%%%% begin : Num[g,1]
s{}(\ell_{5}+\ell_{6}+k_{3})^{2}
%%%%% end : Num[g,1]
 \,, \vphantom{\Big|} \hskip 2.3 cm 
 N^{{\rm(g)}}_{2}  = 
%%%%% begin : Num[g,2]
t{}(\ell_{5}+\ell_{6}+k_{3})^{2}
%%%%% end : Num[g,2]
 \,, \vphantom{\Big|} \\
N^{{\rm(g)}}_{3} & = 
%%%%% begin : Num[g,3]
(\ell_{5}+k_{3})^{2}(\ell_{6}+k_{1}+k_{2})^{2}
%%%%% end : Num[g,3]
 \,, \hskip .7 cm
 N^{{\rm(g)}}_{4}  = 
%%%%% begin : Num[g,4]
 (\ell_{5}+k_{4})^{2}(\ell_{6}+k_{1}+k_{2})^{2}
%%%%% end : Num[g,4]
 \,, \\
\end{array}
\\
& & \\ [-.3cm]
%\hline
%H
(h) & \includegraphics{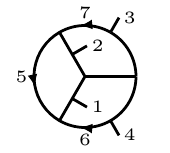} & 
\begin{array}{r l}
N^{{\rm(h)}}_{1} & =
%%%%% begin : Num[h,1]
  \Big[(\ell_6+\ell_7)^2(\ell_5+k_2+k_3)^2 -\ell^2_5 (\ell_6+\ell_7-k_1-k_2)^2  \\
											  & \qquad -(\ell_5+\ell_6)^2(\ell_7+k_2+k_3)^2
																 -(\ell_5+\ell_6+k_2+k_3)^2 \ell^2_7											  \\
												& \qquad -(\ell_6+k_1+k_4)^2(\ell_5-\ell_7)^2
																 -(\ell_5-\ell_7+k_2+k_3)^2 \ell^2_6
									 \Big]																																		\\
			&-\Big[
			[(\ell_5-k_1)^2+(\ell_5-k_4)^2][(\ell_6+\ell_7-k_1)^2+(\ell_6+\ell_7-k_2)^2] \\
			& \qquad -4\times \ell^2_5(\ell_6+\ell_7-k_1-k_2)^2													 \\
			&	\qquad  -(\ell_7+k_4)^2(\ell_5+\ell_6-k_1)^2																			
							  -(\ell_7+k_3)^2(\ell_5+\ell_6-k_2)^2																\\
			&	\qquad	-(\ell_6+k_4)^2(\ell_5-\ell_7+k_1)^2																		
								-(\ell_6+k_3)^2(\ell_5-\ell_7+k_2)^2
		\Big]
%%%%% end : Num[h,1]
% \,,\\
%
%N^{{\rm(h)}}_{2} & =
%%%%% begin : Num[h,2]
%	\Big[
%			[(\ell_5-k_1)^2+(\ell_5-k_4)^2][(\ell_6+\ell_7-k_1)^2+(\ell_6+\ell_7-k_2)^2] \\
%			& \qquad -4\times \ell^2_5(\ell_6+\ell_7-k_1-k_2)^2													 \\
%			&	\qquad  -(\ell_7+k_4)^2(\ell_5+\ell_6-k_1)^2																			
%							  -(\ell_7+k_3)^2(\ell_5+\ell_6-k_2)^2																\\
%			&	\qquad	-(\ell_6+k_4)^2(\ell_5-\ell_7+k_1)^2																		
%								-(\ell_6+k_3)^2(\ell_5-\ell_7+k_2)^2
%		\Big]
% N^{{\rm(h)}}_{1} & =
%%%%% begin : Num[h,1]
%   (\ell_{5}+k_{2}+k_{3})^{2} (\ell_{6}+\ell_{7})^{2}
%	-\ell_{5}^{2}(\ell_{6}+\ell_{7}-k_{1}-k_{2})^{2}
%%%%% end : Num[h,1]
% \,,\\
%
%N^{{\rm(h)}}_{2} & =
%%%%% begin : Num[h,2]
%\left[ (\ell_{6}+\ell_{7}-k_{1})^{2}+(\ell_{6}+\ell_{7}-k_{2})^{2}\right]
%\left[ (\ell_{5}-k_{1})^{2}
%   + (\ell_{5}-k_{4})^{2} \right] 
%   	\\
%      	    \null & 
%	\hphantom{=[(} %Puts in horizontal space equal to "=({" to match preceding line
%	-4\, \ell_{5}^{2}(\ell_{6}+\ell_{7}-k_{1}-k_{2})^{2}
%%%%% end : Num[h,2]
\,, 
      \\
\end{array}
\\
& & \\ [-.3cm]
%\hline
%I
%\vskip -2cm
(i) & \includegraphics{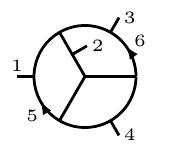} & 
\begin{array}{r l}
N^{{\rm(i)}}_{1} & =
%%%%% begin : Num[i,1]
	(\ell_{6}+k_{4})^{2}
	\left[( \ell_{5}-k_{1}-k_{2})^{2}
		+ ( \ell_{5}-k_{1}-k_{3})^{2}
	\right]\\ 
\null & 
	\hphantom{=(} %Puts in horizontal space equal to "=(" to match preceding line
	\vphantom{\Bigl|} 
        - (\ell_{5}+k_{4})^{2}
	\left[	(\ell_{6}+k_{1}+k_{4})^{2}
		+ (\ell_{6}+k_{2}+k_{4})^{2} \right] \\
\null & 
	\hphantom{=(}
	\vphantom{\Bigl|}
	  -\ell^2_5 (\ell_6-k_2)^2 + \ell^2_6(\ell_5-k_2)^2
%%%%% end : Num[i,1]
 \,.  \\
\end{array}
\\
%\hline
\end{tabular}
\vspace{-.5cm}
\caption{\label{tab:BasisTenProp}
The parent numerator basis elements corresponding to the labels of the diagrams in
  \fig{ParentTenProp}. The basis elements respect the symmetries of the diagrams.
}
$\null$
\end{center}
\vskip -1 cm 
\end{table}
%%%%%%%%%%%%%%%%% TABLE %%%%%%%%%%%%%%%%%%%%%%%%%%%%%%%%%

%%%%%%%%%%%%%%%%% TABLE %%%%%%%%%%%%%%%%%%%%%%%%%%%%%%%%%
\begin{table}[ht]
\begin{center}
\begin{tabular}[t]
{>{\centering\arraybackslash} m{0.02\textwidth} %Horizontally and vertically center column 1
 >{\centering\arraybackslash} m{0.12\textwidth}  %Horizontally and vertically center column 1
 >{$}l<{$} %Automatically put column two in math mode, left aligned
}
&\hskip -1 cm Diagram & \textrm{Numerator} \\
\hline \hline
%C8
(c) & \includegraphics{./Diagrams/SmallDiagramC}
	&
\begin{array}{r l}
% 8  propagator
N^{\rm (c)}_{2} & = 
%%%%% begin : Num[c,2]
                \left( \ell_{5} \right)^{2} \left( \ell_{7} \right)^{2} +
                ( \ell_{5} + k_{1} + k_{2} )^{2}\left( \ell_{7} \right)^{2}
         +    \left( \ell_{5} \right)^{2} ( \ell_{7} + k_{1} + k_{2})^{2} 
        \\
	\null & \hphantom{=(}
        +       ( \ell_{5} + k_{1} + k_{2})^{2}
                ( \ell_{7} + k_{1} + k_{2})^{2}
%%%%% end : Num[c,2]
 \,, 
\\
\end{array}
\\
% 8 propagator
(f) & \includegraphics{./Diagrams/SmallDiagramF}
	&
\begin{array}{r l}
N^{\rm (f)}_{2} & = 
%%%%% begin : Num[f,2]
        \ell_{5}^{2}( \ell_{5} - k_{1} - k_{2} )^{2}
%%%%% end : Num[f,2]
\,, 
	\\
\end{array}
\\
(g) & \includegraphics{./Diagrams/SmallDiagramG}
 &
\begin{array}{r l}
% 9 propagator
%N^{{\rm (g)}}_{5}  & =
%%%%% begin : Num[g,5]
% \left( \ell_{5} - k_{1} -k_{2} \right)^2 
%        \left[\left(\ell_{6}+k_{3}\right)^{2}-\ell_{6}^{2}\right]
%%%%% end : Num[g,5]
%\,, \\
% 8 propagator
N^{\rm (g)}_{5} & = 
%%%%% begin : Num[g,6]
 ( \ell_{5} - k_{1} - k_{2} )^{2}
	( \ell_{6} - k_{4})^{2} 
%%%%% end : Num[g,6]
\,,
	\\
% 8 propagator
N^{\rm (g)}_{6} & = 
%%%%% begin : Num[g,7]
  \ell_{5}^{2} ( \ell_{6} - k_{4})^{2}
%%%%% end : Num[g,7]
 \,, 
        \\
\end{array}
\\
& &\\
%H9GHI
(h) & \includegraphics{./Diagrams/SmallDiagramH}
 &
\begin{array}{r l}
% 9 propagator
%N^{{\rm (h)}}_{3}  & = 
%%%%% begin : Num[h,3]
%        \left( \ell_{5} -\ell_{7} \right)^2  
%        \left[\left(\ell_{6}+k_{4}+k_{1}\right)^{2}-\left( \ell_{6} +k_{4}\right)^{2}\right]
%        \\
%	\null & \hphantom{=(}
%        - \left( \ell_{6} +k_{4} \right)^2  
%        \left[\left(\ell_{5}-\ell_{7}+k_{1}\right)^{2}-\left( \ell_{5} -\ell_{7}\right)^{2}\right] 
%        \\
%	\null & \hphantom{=(}
%        + \left( \ell_{5} + \ell_{6} \right)^2  
%        \left[\left(\ell_{7}+k_{3}+k_{2}\right)^{2}-\left(\ell_{7}+k_{3}\right)^{2}\right]
%        \\
%	\null & \hphantom{=(}
%        + \left( \ell_{7}+k_{3} \right)^2  
%\left[\left(\ell_{5} + \ell_{6} + k_{2}\right)^{2}-\left(\ell_{5} + \ell_{6}\right)^{2}\right] 
%        \\
%	\null & \hphantom{=(}
%        + \left( \ell_{6} \right)^2  
% \left[\left(\ell_{5} - \ell_{7} + k_{2} + k_{3} \right)^{2}-
%              \left(\ell_{5} - \ell_{7} + k_{2}\right)^{2}\right] 
%      \\
%	\null & \hphantom{=(}
%        - \left( \ell_{5} - \ell_{7} + k_{2} \right)^2  
%        \left[\left( \ell_{6} + k_{3} \right)^{2}-\left(\ell_{6}\right)^{2}\right]
%        \\
%	\null & \hphantom{=(}
%        - \left( \ell_{5} + \ell_{6} - k_{1} \right)^2  
%        \left[\left( \ell_{7} + k_{4} \right)^{2}-\left(\ell_{7}\right)^{2}\right] 
%        \\
%	\null & \hphantom{=(}
%        - \left( \ell_{7} \right)^2  
%        \left[\left( \ell_{5} + \ell_{6} - k_{1} + k_{4} \right)^{2}
%        -\left(\ell_{5} + \ell_{6} - k_{1}\right)^{2}\right] \,,
%%%%% end : Num[h,3]
%        \\
% 8 propagator
N^{\rm (h)}_{2} & = 
%%%%% begin : Num[h,4]
%         \vphantom{\Bigr|}
                \ell_{6}^{2} ( \ell_{5} - \ell_{7})^{2} +
                 \ell_{7}^{2} ( \ell_{5} + \ell_{6})^{2}
        +       ( \ell_{6} + k_{4})^{2}
                ( \ell_{5} - \ell_{7} + k_{2} )^{2} 
        \\
	\null & \hphantom{=(}
        +     ( \ell_{5} + \ell_{6} - k_{1} )^{2}
                ( \ell_{7} + k_{3})^{2}
%%%%% end : Num[h,4]
 \,, 
        \\
\end{array}
\\
& & \\
(i) & \includegraphics{./Diagrams/SmallDiagramI}
&
\begin{array}{r l}
% 9 propagator
N^{{\rm (i)}}_2  & = 
%%%%% begin : Num[i,2]
\frac{1}{3}
   (\ell_5 +\ell_6 +k_4)^2  \left[t - s\right]
%%%%% end : Num[i,2]
 \,,
   	\\
% 9 propagator
% N^{\rm (i)}_{3} & = 
%%%%% begin : Num[i,3]
%                 \left( \ell_6 \right)^2 
%        \left[\left(\ell_{5}+k_{2}\right)^{2}-\ell_{5}^{2}\right]
%                -\left( \ell_5 \right)^2  
%        \left[\left(\ell_{6}+k_{2}\right)^{2}-\ell_{6}^{2}\right]
%%%%% end : Num[i,3]
% \,,
%\\
% 8 propagator  
N^{\rm (i)}_{3} & = 
%%%%% begin : Num[i,4]
                \left( \ell_{6} \right)^{2}( \ell_{5} - k_{1})^{2} -
                \left( \ell_{5} \right)^{2} ( \ell_{6} - k_{3})^{2}
%%%%% end : Num[i,4]
 \,. 
        \\
\end{array}
\\
\end{tabular}
\caption{ The parent diagram numerator basis elements where a
  numerator factor cancels a propagator.  Each term in
  brackets does not cancel a propagator, while the remaining factors each
  cancel a propagator.  Each basis numerator maintains the symmetries
  of the associated diagram, including color signs.  The associated
  color factor can be read off from each diagram.  }
\label{tab:BasisParentContactEightNine}
\end{center}
\end{table}
%%%%%%%%%%%%%%%%% TABLE %%%%%%%%%%%%%%%%%%%%%%%%%%%%%%%%%

%%%%%%%%%%%%%%%%% TABLE %%%%%%%%%%%%%%%%%%%%%%%%%%%%%%%%%
\begin{table}[ht]
\begin{center}
\begin{tabular}[t]
{>{\centering\arraybackslash} m{0.02\textwidth} %Horizontally and vertically center column 1
 >{\centering\arraybackslash} m{0.12\textwidth}  %Horizontally and vertically center column 2
 >{$}l<{$} %Automatically put column two in math mode, left aligned
}
& \hskip -1 cm Diagram & \textrm{Numerator} \\
\hline \hline
%III
(j) & \includegraphics{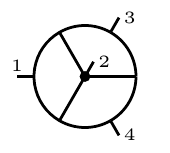}
 &
\begin{array}{r l}
N^{\rm (j)}_1 & =
%%%%% begin : Num[j,1]
 s
%%%%% end : Num[j,1]
 \,, \hskip .6 cm 
N^{\rm (j)}_2 =
%%%%% begin : Num[j,2]
 t
%%%%% end : Num[j,2]
 \,,
\end{array}
\\
%CFGGHI % updated
(k) & \includegraphics{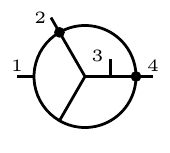}
 &
\begin{array}{r l}
N^{\rm (k)}_1 & =
%%%%% begin : Num[l,1]
 1
%%%%% end : Num[l,1]
\,. \\
\end{array}
\\
\end{tabular}
\caption{The numerator basis elements corresponding to 
the contact term diagrams. Black dots indicate contact terms.  
Written this way, the numerators are simple, but the color factors
cannot be read off from the diagrams. 
}
 \label{tab:BasisEightandNine}
\end{center}
\end{table}
%%%%%%%%%%%%%%%%% TABLE %%%%%%%%%%%%%%%%%%%%%%%%%%%%%%%%%

% updated 
As a second example, consider diagram (k) in \tab{BasisEightandNine},
corresponding to the basis element $N^{\rm (k)}_1$.  If we put back
the two missing propagators by multiplying and dividing by the
appropriate inverse propagators, the contribution from diagram (k) in
\tab{BasisEightandNine}, corresponds to numerators $N^{\rm (c)}_{2}$,
$N^{\rm (f)}_{2}$, $N^{\rm (g)}_{5}$, $N^{\rm (g)}_{6}$, $N^{\rm (h)}_{2}$ and $N^{\rm (i)}_{3}$ 
in \tab{BasisParentContactEightNine}.

In summary, the diagrams along with the numerators in
\tabs{BasisTenProp}{BasisParentContactEightNine} are a complete set
with the desired power counting, have only logarithmic singularities
and no poles at infinity.  They are also constructed to satisfy
diagram symmetries, including color signs.

\subsection{Determining the coefficients}

We now express the three-loop four-point $\NeqFour$ sYM amplitude in
terms of our constructed basis.  We express the numerator in
\eqn{LoopGauge} directly in terms of our basis via
\eqn{BasisExpansion}. Because we have required each basis numerator to
reflect diagram symmetry, we need only specify one numerator of each
diagram topology and can obtain the remaining ones simple by
relabeling of external legs.

The coefficients in front of all basis elements are straightforward to
determine using simple unitarity cuts, together with previously determined
representations of the three-loop amplitude.  We start from the $\NeqFour$ sYM
numerators as originally determined in Ref.~\cite{GravityThreeLoop},
since they happen to be in a particularly compact form. Rewriting 
these numerators using our choice of momentum labels gives
\begin{align}
N_{\rm old}^{\rm (a)} & = N_{\rm old}^{\rm (b)} = N_{\rm old}^{\rm (c)} =
N_{\rm old}^{\rm (d)} = s^2\,, \nonumber \\
N_{\rm old}^{\rm (e)} & = N_{\rm old}^{\rm (f)} = N_{\rm old}^{\rm (g)} =
  s (\ell_5 + k_4)^2 \,, \nonumber \\
N_{\rm old}^{\rm (h)} & = - s t  +  2 s (k_2 + k_3) \cdot \ell_5
     + 2 t (\ell_6 + \ell_7) \cdot (k_1 + k_2)\,,  \nonumber \\
N_{\rm old}^{\rm (i)} & = s (k_4 + \ell_5)^2 - t (k_4 + \ell_6)^2
     - \frac{1}{3} (s - t) (k_4 + \ell_5 + \ell_6)^2\,.
\label{eqn:OldForm}
\end{align}
To match to our basis we start by considering the maximal cuts, where
all propagators of each diagram are placed on shell. The complete set
of maximal cut solutions are unique to each diagram, so we can match
coefficients by considering only a single diagram at a time.  We start
with diagram (a) in \tab{BasisTenProp}. Here the numerator is a linear 
combination of three basis elements
\begin{equation}
N^{\rm (a)} =  a_1^{\rm (a)} N^{\rm (a)}_1 + a_2^{\rm (a)} N^{\rm (a)}_2 
  + a_3^{\rm (a)}  N^{\rm (a)}_3 \,,
\end{equation}
corresponding to $N_j^{\rm (a)}$ in \tab{BasisTenProp}. The 
$a_j^{\rm (a)}$ are numerical parameters to be determined.
This is to be compared to the old form
of the numerator in \eqn{OldForm}.  Here the maximal cuts have no
effect because both the new and old numerators are independent of loop
momentum.  Matching the two numerators, the coefficients in
front of the numerator basis are $a^{{\rm(a)}}_1 = 1$,
$a^{{\rm(a)}}_2 = 0$ and $a^{{\rm(a)}}_3 = 0$.

Now consider diagram (b) in \fig{ParentTenProp}.  Here the basis
element is of a different form compared to the old version of the
numerator in \eqn{OldForm}. The new form of the numerator is
\begin{equation}
N^{\rm (b)} = a_1^{\rm (b)} N_1^{\rm (b)} =
a_1^{\rm (b)}  s \left[ (\ell_{6}-k_{3})^{2}
		+ (\ell_{6}-k_{4})^{2} \right]
\,. 
\end{equation}
In order to make the comparison to the old version we impose
the maximal cut conditions involving only $\ell_6$:
\begin{equation}
\ell_6^2 = 0\,, \hskip 1 cm  (\ell_6 - k_2 - k_3)^2 = 0 \,.
\end{equation}
Applying these conditions:
\begin{equation}
\left[ (\ell_{6}-k_{3})^{2} + (\ell_{6}-k_{4})^{2} \right] \rightarrow -s \,.
\end{equation}
Comparing to $N_{\rm old}^{\rm (b)}$ in \eqn{OldForm} gives us the
coefficient $a_1^{{\rm(b)}}  = -1$.

As a more complicated example, consider diagram (i).  In this case the
numerators depend only on $\ell_5$ and $\ell_6$.  The relevant cut
conditions read off from \fig{ParentTenProp}(i) are
\begin{align}
& \ell_5^2 = 
\ell_6^2 =
(\ell_5 - k_1)^2 = 
(\ell_6 - k_3)^2 = 
 (\ell_5+\ell_6 - k_3 - k_1)^2 = 
(\ell_5+\ell_6 + k_4)^2 = 0 \,.
\label{eqn:NiCutConditions}
\end{align}
With these cut conditions, the old numerator in \eqn{OldForm} becomes
\begin{equation}
N^{\rm (i)}_{\rm old} \bigr|_{\rm cut} =
2 s \left( k_4 \cdot \ell_5 \right) - 2 t \left( k_4 \cdot \ell_6 \right) \,.
\label{eqn:NiOldCut}
\end{equation}
The full numerator for diagram (i) is a linear combination of the three
basis elements for diagram (i) in 
\tabs{BasisTenProp}{BasisParentContactEightNine}, 
\begin{equation}
  N^{\rm (i)} = a^{\rm (i)}_1 N^{\rm (i)}_1 + a^{\rm (i)}_2 N^{\rm (i)}_2
+ a^{\rm (i)}_3 N^{\rm (i)}_3 \,.
\end{equation}
The maximal cut conditions immediately set to zero the last two
of these numerators because they contain inverse propagators.
Applying the cut conditions \eqn{NiCutConditions}
to the nonvanishing term results in
\begin{align}
 N^{\rm (i)} \bigr|_{\rm cut} = &
a^{\rm (i)}_1[ -2 (\ell_6 \cdot k_4)t + 2 (\ell_5 \cdot k_4) s ] \,.
\label{eqn:NiNewCut}
\end{align}
Comparing \eqn{NiOldCut} to \eqn{NiNewCut} fixes $a^{\rm (i)}_{1} = 1$.
The two other coefficients for diagram (i), $a^{\rm (i)}_2$ and $a^{\rm (i)}_3$
cannot be fixed from the maximal cuts.

In order to determine all coefficients and to prove that the answer is
complete and correct, we need to evaluate next-to-maximal and
next-to-next-to-maximal cuts.  We need only evaluate the cuts
through this level because of the especially good 
power counting of $\NeqFour$ sYM.
We do not describe this procedure
in detail here.  Details of how this is done may be found in 
Ref.~\cite{MaximalCuts}. Using these cuts we have the 
solution of the numerators in terms of the basis elements as
%%%%%%%%%%%%%%%%%%%%
% scaled a_f -> 2 a_f and all other a_x -> 8 a_x
% then a_c -> d_1, a_f -> d_2, a_h -> d_3, a_9 -> d_4
\newcommand{\ptp}[1]{ \left( \textrm{#1} \right) }
\begin{align}
N^{\rm{(a)}} & =
%%%%% begin : Num[a]
 N^{\rm{(a)}}_{1}
%%%%% end : Num[a]
 \,, \nonumber
	\\
N^{\rm{(b)}} & =
%%%%% begin : Num[b]
 - N^{\rm{(b)}}_{1} 
%%%%% end : Num[b]
\,, \nonumber
	\\
N^{\rm{(c)}} & =
%%%%% begin : Num[c]
 - N^{\rm{(c)}}_{1} + 2  d_1 N^{\rm{(c)}}_{2} 
%%%%% end : Num[c]
\,, \nonumber
	\\
N^{\rm{(d)}} & = 
%%%%% begin : Num[d]
N^{\rm{(d)}}_{1}
%%%%% end : Num[d]
 \,, \nonumber
	\\
N^{\rm{(e)}} & =
%%%%% begin : Num[e]
 N^{\rm{(e)}}_{1} 
%%%%% end : Num[e]
\,,  \label{eqn:Solution}
	\\
N^{\rm{(f)}} & = 
%%%%% begin : Num[f]
- N^{\rm{(f)}}_{1} + 2 d_2 \, N^{\rm{(f)}}_{2} 
%%%%% end : Num[f]
\,, 	\nonumber
	\\
N^{\rm{(g)}} & = 
%%%%% begin : Num[g]
 - N^{\rm{(g)}}_{1} + N^{\rm{(g)}}_{3}
           + N^{\rm{(g)}}_{4} 
	+ (d_1 + d_3 - 1) N^{\rm{(g)}}_{5}
	+ (d_1 - d_2) N^{\rm{(g)}}_{6} 
%%%%% end : Num[g]
\,, \nonumber
	\\
N^{\rm{(h)}} & = 
%%%%% begin : Num[h]
  N^{\rm{(h)}}_{1} %- N^{\rm{(h)}}_{2}
%        - N^{\rm{(h)}}_{3} 
  + 2 d_3\, N^{\rm{(h)}}_{2} 
%%%%% end : Num[h]
\,, \nonumber
	\\
N^{\rm{(i)}} & =
%%%%% begin : Num[i]
         N^{\rm{(i)}}_{1} 
	+ N^{\rm{(i)}}_{2}
%	- \, N^{\rm{(i)}}_{3}
	+ (d_3 - d_2) N^{\rm{(i)}}_{3}
%%%%% end : Num[i]
 \,, \nonumber
\end{align}
%%%%%%%%%%%%%%%%%%%%
where the three $d_{i}$ are free parameters not fixed by any physical
constraint.

The ambiguity represented by the three free parameters, $d_i$ in
\eqn{Solution}, derives from color factors not being independent but
instead related via the color Jacobi identity.  This allows us to move
contact terms between different diagrams without altering the
amplitude.  Different choices of $d_1, d_2, d_3$ correspond to three
degrees of freedom from color Jacobi identities. These allow us to
move contact contributions of diagram (k), where two propagators are
collapsed, between different parent diagrams.  The contact term in
diagram (j) of \tab{BasisEightandNine} does not generate a fourth
degree of freedom because the three resulting parent diagrams are all
the same topology, corresponding to relabelings of the external legs
of diagram (i).  The potential freedom then cancels within a single
diagram.  We have explicitly checked that the $d_i$ parameters in
\eqn{Solution} drop out of the full amplitude after using appropriate
color Jacobi identities.  One choice of free parameters is to take
them to all vanish
\begin{equation}
d_1  = 0\,, \hskip 2 cm
d_2  = 0\,, \hskip 2 cm
d_3  = 0\,.
\end{equation}
 In this case every remaining nonvanishing numerical coefficient
in front of a basis elements is $\pm 1$.  (Recall that
$N^{\rm{(i)}}_{2}$ absorbed the $1/3$ combinatorial factor mismatch
between diagram (i) and diagram (j).)  Of course this is not some
``best'' choice of the $d_i$, given that the amplitude
is unchanged for any other choice of $d_i$.

Once the coefficients in front of each basis numerator are determined,
we are left with the question of whether the basis numerators properly
capture all terms that are present in the amplitude.  To answer this
we turn to the method of maximal cuts~\cite{MaximalCuts}.  This is a
variation on the standard generalized unitarity method, but organized
by starting with maximal cuts and systematically checking cuts with
fewer and fewer propagators set on shell.  This method has
been described in considerable detail in Ref.~\cite{MaximalCuts}, so
we only mention a few points.

The overall power counting of the three-loop $\NeqFour$
sYM amplitude is such that it can be written with at
most two powers of loop momenta in the
numerator~\cite{GravityThreeLoop,BCJLoop}. This means that in
principle we can fully determine the amplitude using only
next-to-maximal cuts.  However, here we use a higher-power counting
representation with up to four powers of loop momenta in the numerator
corresponding to as many as two canceled propagators.  This implies that
to completely determine the amplitude using our representation we need
to check cuts down to the next-to-next-to-maximal level.  We have
explicitly checked all next-to-next-to-maximal cuts, proving that the
amplitudes obtained by inserting the numerators in \eqn{Solution} into
\eqns{LoopGauge}{BasisExpansion} gives the complete amplitude, and
that it is entirely equivalent to earlier representations of the
amplitude~\cite{GravityThreeLoop,Manifest3,BCJLoop}. Because each
numerator basis element is constructed such that each integrand has only
logarithmic singularities and no poles at infinity, this proves
that the full nonplanar three-loop four-point $\NeqFour$
sYM amplitude has these properties, as conjectured in Ref.~\cite{Log}.

%%%%%%%%%%%%%%%%%%%%%%%%%%%%%%%%%%%%%%%%%%%%
\subsection{Relation to rung rule}
\label{sec:Heuristic rules}

Is it possible to determine the coefficients of the basis integrands as
they appear in the $\NeqFour$ sYM amplitude from simple
heuristic rules? Such rules can be useful both because they offer a simple
way to cross-check derived results, and because they can often point to deeper
structures.  Here we show that the rung rule of Ref.~\cite{BRY} gives
at least some of the coefficients\footnote{We thank Lance Dixon for
  pointing out to us that the rung rule is helpful for identifying
  nonplanar integrals with uniform transcendentality, suggesting a
  match to our construction as well.}.

%%%%%%%%%%%%%%%%%%%%% FIGURE %%%%%%%%%%%%%%%
\begin{figure}[tb]
\begin{center}
\includegraphics{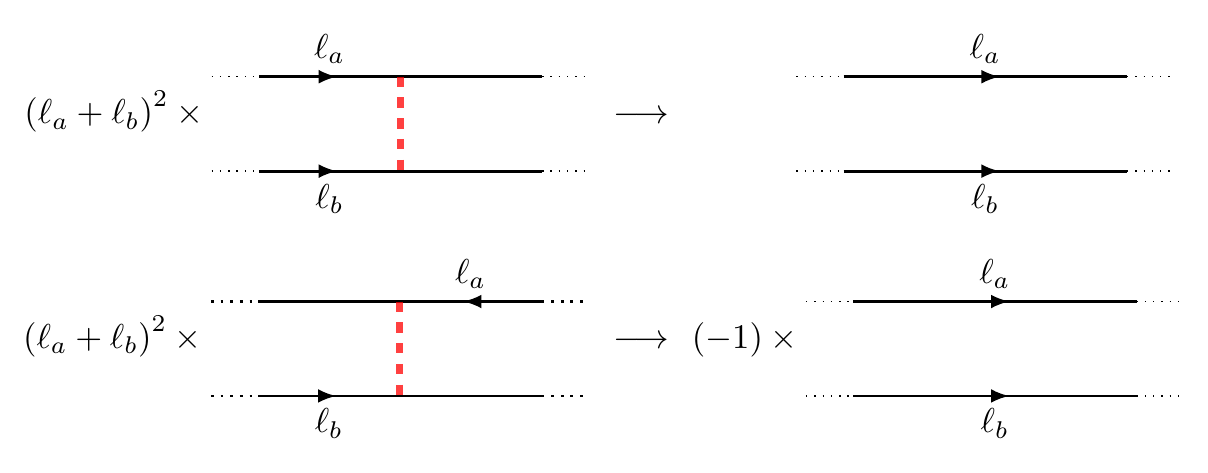}
\end{center}
\caption{The rung rule gives the relative coefficient between an $L$-loop diagram
  and an $(L-1)$-loop diagram. The dotted shaded (red) line represents the
  propagator at $L$ loops that is removed to obtain the $(L-1)$-loop
  diagram. As indicated on the second row, if one of the lines is twisted
  around, as can occur in nonplanar diagrams, there is an additional sign
  from the color antisymmetry.}
\label{fig:RungRule}
\end{figure}
%%%%%%%%%%%%%%%%%%%%%%%%%%%%%%%%%%%%%%%%%%%%

The rung rule was first introduced as a heuristic rule for generating
contributions with correct iterated-two-particle cuts in $\NeqFour$
sYM amplitudes~\cite{BRY}. It is also related to certain
soft collinear cuts. Today the rung rule is understood as a means for
generating contributions with simple properties under dual conformal
invariance.  In the planar case the rule applies even when the
contributions cannot be obtained from iterated two-particle
cuts~\cite{Bern:2006ew}.  However, the rung rule does not capture all
contributions.  It can also yield contributions that do not have the
desired properties, but differ by contact terms from desired
ones.
For this reason, the rule is most useful once we have a basis of integrands and are
interested in understanding the coefficients as they appear in amplitudes.

%%%%%%%%%%%%%%%%%%%%% FIGURE %%%%%%%%%%%%%%%
\begin{figure}[tb]
\begin{center}
\includegraphics{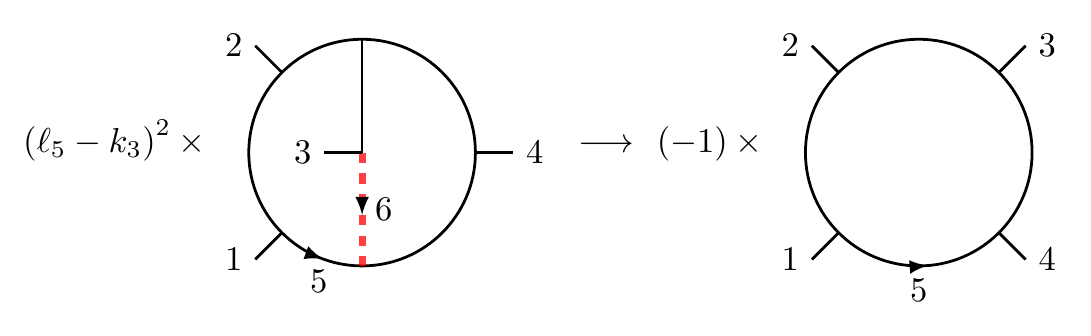}
\end{center}
\caption{
	The rung rule determines the relative sign between the
	two-loop nonplanar contribution and the one-loop box to be negative.
    		}
\label{fig:TwoLoopRung}
\end{figure}
%%%%%%%%%%%%%%%%%%%%%%%%%%%%%%%%%%%%%%%%%%%%

The rung rule was originally applied as a means for generating
new $L$-loop contributions from $(L-1)$-loop ones.
Here we use the rule in the opposite direction, 
going from an $L$-loop basis integrand to an $(L-1)$-loop contribution so
as to determine the coefficient of the $L$-loop contribution
to the amplitude.  As illustrated in
\fig{RungRule}, if we have a basis integrand containing a factor of
$(\ell_a + \ell_b)^2$ and a propagator indicated by a dotted
line, we can remove these to obtain a diagram with one fewer loop.
According to the rung rule, the overall coefficient of the diagram
obtained by removing a rung
matches that of the lower-loop diagram in the amplitude.
In the nonplanar case the diagrams can be twisted
around, as displayed on the second line in \fig{RungRule}, leading to
relative signs.  These relative signs can be thought of as coming
from color factors. 

Because we have already determined the three-loop
$\dlog$ numerators, we only need the rung rule to
determine the sign of the numerator in the amplitude.
 This allows us to slightly generalize the rung rule beyond its original form.
In the original version of the rung rule,
the rung carries an independent loop momentum
that becomes a new loop momentum in the diagram when the rung is added. 
The reverse of this means removing a rung from the diagram
requires also removing an independent loop momentum.  We will encounter
cases where removing a rung and its loop momentum
prevents the original version of the rung rule from
matching the desired $\dlog$ numerators.  We therefore slightly modify the rung rule by
allowing the factors to be matched in any order of removing a given
set of rungs or propagators. If we can match each factor in a numerator
in at least one order of rung removal, then we just read off the overall
sign as for other cases.

To illustrate how the rung rule determines a coefficient,
consider the two-loop four-point amplitude.
As discussed in \sect{TwoLoopSection}, after removing the overall
$\K$ from the amplitude, the only allowed numerator for the nonplanar double box in
\fig{TwoLoopParent}(b) with the desired properties is given in
\eqn{TwoLoopNumerators}.
The first step is to determine if a given numerator can be obtained
from the rung rule.  The first term, $(\ell_5 - k_3)^2$,
in the nonplanar numerator $N^{\rm (NP)}$ (\eqn{TwoLoopNumerators})
can be so determined. The rung corresponding
to the $(\ell_5 - k_3)^2$ term is displayed as the dotted (red) line on
the left side of \fig{TwoLoopRung}.  Removing this rung gives the
one-loop box diagram on the right side of \fig{TwoLoopRung},
which has coefficient $\K$.  However, we need to flip over leg 3 to obtain the
standard box from the diagram with the rung removed,
resulting in a relative minus sign between color factors.
This fixes $(\ell_5 - k_3)^2$ to enter the amplitude
with a negative sign, because the box enters the amplitude with
a positive sign.  This precisely matches the sign in \eqn{NonplanarDoubleBoxSign}
obtained from the maximal cut.

%%%%%%%%%%%%%%%%%%%%% FIGURE %%%%%%%%%%%%%%%
\begin{figure}[tb]
\begin{center}
\includegraphics{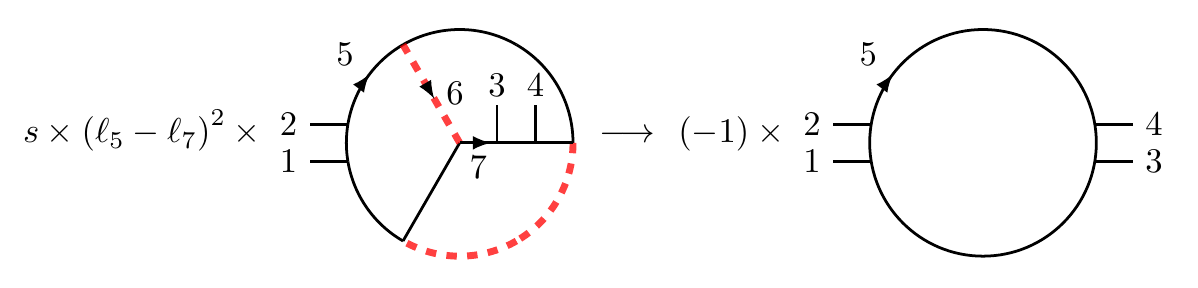}
\end{center}
\caption{The rung rule determines that the basis element $N^{\rm (c)}$
enters the amplitude with a relative minus sign.
}
\label{fig:ThreeLoopRungC}
\end{figure}
%%%%%%%%%%%%%%%%%%%%%%%%%%%%%%%%%%%%%%%%

At three loops the idea is the same.  Consider, for example diagram (c).
Examining the numerator basis element $N_1^{\rm (c)}$ from \tab{BasisTenProp},
we can identify the term $s (\ell_5-\ell_7)^2$
as a rung-rule factor. In \fig{ThreeLoopRungC}, the dotted (red) line in the
top part of the diagram corresponds to the factor $(\ell_5
-\ell_7)^2$.  After removing the top rung, the bottom rung is just a
factor of $s=(k_1+k_2)^2$.  An overall sign comes from the fact that
the first rung was twisted as in \fig{RungRule}.  This determines
the coefficient to be $-1$, and symmetry then fixes the second rung rule
numerator to have the same sign.  This matches the sign of the
numerator in \eqn{Solution} found via unitarity cuts.

%%%%%%%%%%%%%%%%%%%%% FIGURE %%%%%%%%%%%%%%%
\begin{figure}[tb]
\begin{center}
\includegraphics{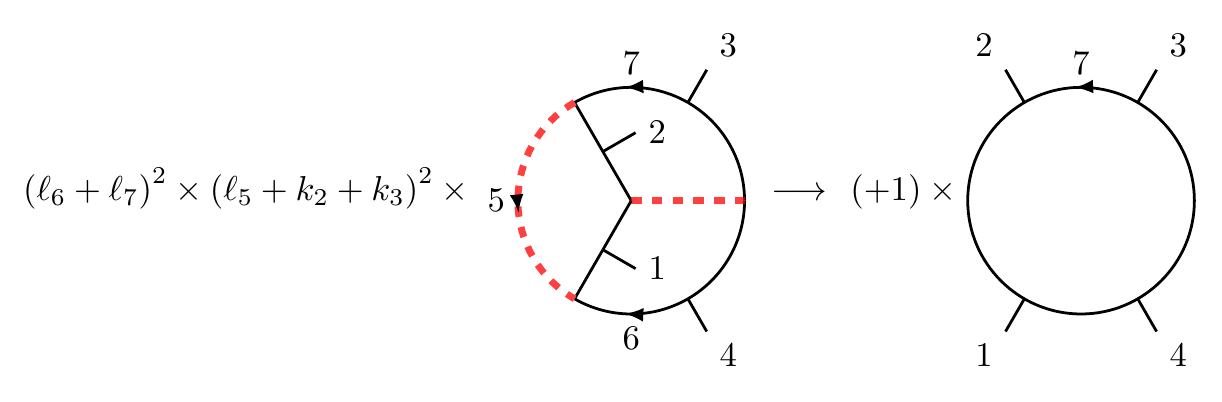}
\end{center}
\caption{ The rung rule determines that basis element
   $N^{\rm (h)}_{1}$ enters the amplitude with a relative
  plus sign.  
  }
\label{fig:ThreeLoopRungHOne}
\end{figure}
%%%%%%%%%%%%%%%%%%%%%%%%%%%%%%%%%%%%%%%%

Now consider the more complicated case of diagram (h).
In \tab{BasisTenProp}, the first term
in $N_1^{\rm (h)} = (\ell_6 + \ell_7)^2(\ell_5 + k_2 + k_3)^2 + \cdots $ is a more interesting example, because
the original rung rule does not apply.
Nevertheless, using the slightly modified rung rule described above, we can still
extract the desired coefficient in front of this term.
Examining \fig{ThreeLoopRungHOne}, notice that if we first remove the
left rung, the rung rule gives one factor of $N_1^{\rm (h)}$: $ (\ell_6 + \ell_7)^2 $,
while if we first remove the right rung, the rung rule gives the other factor $ (\ell_5 + k_2 + k_3)^2 $.
In both cases the rung rule sign is positive.  Furthermore, flipping
legs 1 and 2 to get the one-loop diagram on the right side of
\fig{ThreeLoopRungHOne} does not change the sign.  Thus the sign
is positive, in agreement with \eqn{Solution}.

The rung rule does not fix all coefficients of $\dlog$
numerators in the amplitude.  In particular, 
since the rule involves adding two propagators
per rung, it can never generate terms proportional
to propagators, such as those in
\tabs{BasisParentContactEightNine}{BasisEightandNine}. Nor is there any
guarantee that basis integrands without canceled propagators
can be identified as rung rule contributions.
One might be able to find various extensions of the rung rule that handle more of these cases.
Such an extension was discussed in Ref.~\cite{CachazoSkinner}, but for now we
do not pursue these ideas further.

%%%%%%%%%%%%%%%%
\section{Finding $\dlog$ forms}
\label{sec:DlogSection}

In the previous section we performed detailed checks showing that the
three-loop four-point $\NeqFour$ sYM amplitude has only logarithmic
singularities and no poles at infinity. The first of these conditions
is equivalent to being able to find $\dlog$ forms, so if we can find
such forms directly then we can bypass detailed analyses of the
singularity structure of integrands. There is no general procedure for
how to do this, so we have to rely on a case-by-case analysis.  We
build up technology at one and two loops, then apply that technology
to a few examples at three loops, relegating a detailed discussion to
the future. As expected, exactly the same Jacobians that lead to
double or higher poles in our analysis of the singularity structure
block us from finding $\dlog$ forms, unless the Jacobians are
appropriately canceled by numerator factors.

In this section, we use the terminology that an $L$-loop integrand form is a
$\dlog$ form if it can be written as a linear combination,
\begin{align}
d{\cal I} & = d^4\ell_5\dots d^4\ell_{4+L} \frac{N^{(x)}(\ell_r,k_s)
}{D^{(x)}(\ell_r,k_s)} 
= \sum_{j} c_j\,\,\dlog\,f^{(j)}_1 \wedge
\dlog\,f^{(j)}_2\wedge \cdots \wedge \dlog f^{(j)}_{4L} \,,
\label{eqn:sumdlog}
\end{align}
where $N^{(x)}(\ell_r,k_s)$ is a diagram numerator, the
denominator $D^{(x)}(\ell_r,k_s)$ is the usual product of propagators, $f_i^{(j)}=
f_i^{(j)}(\ell_r,k_s)$ is a function of loop and external momenta.
The coefficients $c_j$ are the leading singularities of $d\I$ on a $4L$
cut, where we take $f^{(j)}_1=f^{(j)}_2=\dots=f^{(j)}_{4L}=0$.  It is
still an open question whether
the smallest irreducible $\dlog$ forms 
may be expressed as a single form with unit leading singularity,
\begin{equation}
d {\cal I} \overset{?}{=} \dlog\,f_1 \wedge \dlog\,f_2 \wedge \dots\wedge \dlog\,f_{4L} \,.
\label{eqn:singdlog}
\end{equation}
We can determine on a case by case basis if the change of variables
(\ref{eqn:singdlog}) exists by checking if the integrand
form has: (i) only logarithmic singularities and (ii) only unit
leading singularities.  An integrand form has unit leading
singularities if the $4L$-cut of the integrand form is
\begin{equation}
\res{\ell_r=\ell^\ast_r} d{\cal I} = \pm 1,0 \,,
\end{equation}
where $\ell^\ast_r$ are positions for quadruple cuts for all loop
variables. In the $\dlog$ form it is a simple matter to extract the residues
via
\begin{equation}
\res{f_1 = 0} d {\cal I} = \dlog\,f_2 \wedge \dots\wedge \dlog\,f_{4L} \,,
\end{equation}
and the residues at the other $f_j=0$ may be obtained analogously.
In doing so there are signs from the wedge products which we do not track.
Clearly, it is better to find single-term
$\dlog$ forms as in \eqn{singdlog}, which we do in many examples. However 
the multiterm $\dlog$ form (\ref{eqn:sumdlog}) is sufficient for
our purposes because it makes manifest that the integrand form has only logarithmic 
singularities.

In the previous sections, we normalized the forms such that a factor
${\cal K}$, defined in \eqn{KappaDef}, was factored out.  In this
section, we restore this factor as
\begin{equation}
\K=stC_1\,,
\end{equation}
using the definitions from \eqns{KappaDef}{CoefficientDef}.  In some cases it is
best to use the symmetry (\ref{eqn:KappaSymmetry}) to write instead,
\begin{equation}
\K =suC_2\,, \hskip 1.5 cm \K =tuC_3\,.
\end{equation}
As noted in Sect.~\ref{subsec:AmpsandDlogs}, in general, we write the
integrand forms as linear combinations of $\dlog$ forms using the $C_i$
as prefactors, as we find below.

\subsection{One loop}
\label{subsec:dlogOneLoop}

At higher loops, a good starting point for finding $\dlog$ forms is to
express one-loop subdiagrams in $\dlog$ forms.  Following
standard integral decomposition methods, any one loop integrand form
with no poles at infinity can be decomposed in terms of box and
pentagon forms:
\begin{equation}
d{\cal I} = \sum_j a_j^{(5)}\,d{\cal I}_5^{(j)} + \sum_k b_k^{(4)}\,d{\cal I}_4^{(k)} \,.
\label{eqn:OneLoopExp}
\end{equation}
Triangle or bubble integrand forms do not appear in this decomposition
because they would introduce poles at infinity. 

The decomposition in \eqn{OneLoopExp} is valid beyond the usual
one-loop integrals. We can consider any integrand form with
$m$ generalized propagators which are at most quadratic in the momenta:
\begin{equation}
d{\cal I} = \frac{d^4\ell\,\,N_m}{F_1F_2\dots F_m}\,, \qquad
\mbox{where \quad $F_j=\alpha_j\ell^2 + (\ell\cdot Q_j) + P_j$} \,.
\label{eqn:GeneralizedPropagators}
\end{equation}
We can then use the same expansion (\ref{eqn:OneLoopExp}) for these
objects and express it in terms of {\it generalized boxes} and {\it
  generalized pentagons} which are integrals with four or five
generalized propagators, $F_j$. Unlike in the case of regular one-loop
integrals, there is no simple power-counting constraint on the
numerator such that $d{\cal I}$ is guaranteed not to have any poles at
infinity.  Instead one needs to check for poles at infinity case by case.

%\subsubsection*{\it Box integrands}

%%%%%%%%%%%%%%%%%%%%% FIGURE %%%%%%%%%%%%%%%
\begin{figure}[tb]
\begin{center}
\begin{tabular}
{
>{\centering\arraybackslash}m{0.30\textwidth}
>{\centering\arraybackslash}m{0.30\textwidth}
}
\includegraphics{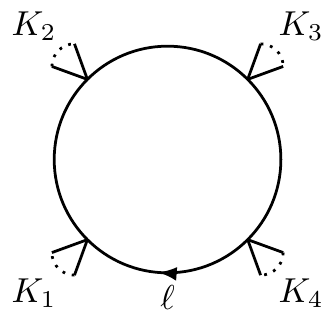} &
\includegraphics{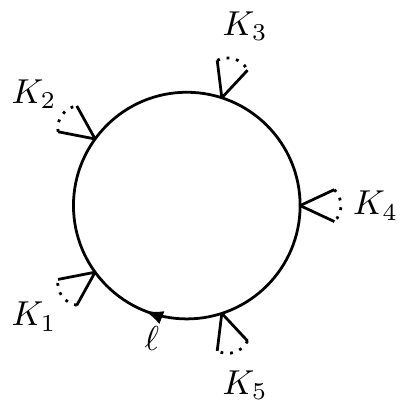} \\
(a) & (b) \\
\end{tabular}
\end{center}
\caption{One-loop box and pentagon diagrams.  }
\label{fig:BoxPentagon}
\end{figure}
%%%%%%%%%%%%%%%%%%%%%%%%%%%%%%%%%%%%%%%%

At one loop, \eqn{OneLoopExp} tells us that we need only consider
boxes and pentagons, since the higher-point cases can be reduced to
these.  First consider the standard box form with (off-shell) external
momenta $K_1,K_2,K_3,K_4$ shown in \fig{BoxPentagon}(a):
\begin{equation}
d{\cal I}_4 \left[ \begin{array}{cc} \ell^2 & (\ell-K_1-K_2)^2\\ (\ell-K_1)^2&(\ell+K_4)^2\end{array}\right]
 \equiv d^4\ell\,\frac{N_4}{\ell^2(\ell-K_1)^2(\ell-K_1-K_2)^2(\ell+K_4)^2} \,.
\end{equation}
Here on the left hand side we introduce a compact notation for the
$\dlog$ form that will be useful at higher loops. The actual positions
of the arguments do not matter, since swapping the locations of
arguments will only alter the overall sign of the form due to the
wedge products;  in amplitudes such
signs are fixed using unitarity.  The numerator
$N_4$ is just the Jacobian $J_\ell$ given in \eqn{GeneralJacobian},
using the labels of the box in \fig{BoxPentagon}(a).  With this
numerator $d{\cal I}_4$ has unit leading singularities, and we can
write it as a single-term $\dlog$ form,
\begin{align}
&d{\cal I}_4 \left[ \begin{array}{cc} \ell^2 & (\ell-K_1-K_2)^2\\ 
(\ell-K_1)^2&(\ell+K_4)^2\end{array}\right]\label{eqn:BoxDlog} \\
&\hspace{1cm}= \dlog\frac{\ell^2}{(\ell-\ell^\ast)^2}
\wedge \dlog\frac{(\ell-K_1)^2}{(\ell-\ell^\ast)^2}
\wedge \dlog\frac{(\ell-K_1-K_2)^2}{(\ell-\ell^\ast)^2}
\wedge \dlog\frac{(\ell+K_4)^2}{(\ell-\ell^\ast)^2}\,, \nonumber
\end{align}
as already noted in Sect.~\ref{subsec:LoopIntegralsPoleInfty}.
Here the $\dlog$ form depends on
$\ell^\ast$, which is a solution for $\ell$ on the quadruple cut
\begin{equation}
\ell^2 = (\ell - K_1)^2 = (\ell - K_1 - K_2)^2
  = (\ell + K_4)^2 = 0\,.
\end{equation}
There are two independent $\ell^\ast$ that satisfy these equations, and
we are free to choose either of them.  Both give the same results when
substituted into the $\dlog$ form. 

An important nontrivial property of a $\dlog$ form is that the
residue located at $(\ell-\ell^\ast)^2=0$ cancels.  If it were not to
cancel, then there would be an unphysical singularity which could feed
into our higher-loop discussion.  We illustrate the cancellation for
the massless box where $K_i=k_i$ with $k_i^2=0$ for all $i=1,\ldots,4$. In this case
$\ell^\ast=-\frac{[12]}{[24]}\lambda_1\widetilde{\lambda}_4$
and the
residue of $d{\cal I}_4$ in (\ref{eqn:BoxDlog}) on
$(\ell-\ell^\ast)^2=0$ is
\begin{align}
{\rm Res}\,d{\cal I}_4 =\null & \dlog(\ell^2) \wedge  \dlog(\ell-k_1)^2 \wedge \dlog(\ell-k_1-k_2)^2
\nonumber \\
& \null
 - \dlog(\ell^2) \wedge \dlog(\ell-k_1)^2 \wedge \dlog(\ell+k_4)^2 \nonumber\\
&\null
 + \dlog(\ell^2) \wedge \dlog(\ell-k_1-k_2)^2 \wedge \dlog (\ell+k_4)^2 \nonumber \\
&\null 
- \dlog(\ell-k_1)^2 \wedge \dlog (\ell-k_1-k_2)^2 \wedge \dlog(\ell+k_4)^2 \,.
\label{eqn:resDlog}
\end{align}
The simplest way to see the cancellation is that on the solution of $(\ell-\ell^\ast)^2=0$,
the following identity is satisfied
\begin{equation}
\ell^2(\ell-k_1-k_2)^2 = (\ell-k_1)^2(\ell+k_4)^2\,.
\end{equation}
Using this we can express, say, $\ell^2$ in terms of other inverse propagators
and substitute into \eqn{resDlog}. All terms in \eqn{resDlog} then cancel
pairwise because of the antisymmetry property of the wedge product. A
similar derivation can be carried out for the generic four-mass case,
but we refrain from doing so here.

The {\it generalized box}, in terms of which \eqn{GeneralizedPropagators}
can be expanded, is:
\begin{equation}
d{\cal I}_4\left[ \begin{array}{cc} F_1&F_2\\ F_3&F_4\end{array}\right]
 = \frac{d^4\ell \,\,N}{F_1F_2F_3F_4}
    = \dlog \frac{F_1}{F^\ast} \wedge \dlog\frac{F_2}{F^\ast} \wedge
              \dlog\frac{F_3}{F^\ast} \wedge \dlog\frac{F_4}{F^\ast} \,,
\label{eqn:GenBoxDlog}
\end{equation}
where the numerator $N$ is again a Jacobian
(\ref{eqn:GeneralJacobian}) of the solution to the system of equations
$P_i=0$ for $P=\{F_1,F_2,F_3,F_4\}$ and $F^\ast=(\ell-\ell^\ast)^2$.
Here $\ell^\ast$ is the solution for $\ell$ at $F_i=0$ for
$i=1,2,3,4$.  It is not automatic that
$d{\cal I}_4$ can be put into a $\dlog$ form for any set of $F_i$'s.
This depends on whether $d{\cal I}_4$ has only logarithmic
singularities.  If it has other types of singularities, then no change of
variables will give a $\dlog$ representation for $d{\cal I}_4$.
As a simple example of a form that cannot
be rewritten in $\dlog$ form consider the generalized box
\begin{equation}
d^4\ell \, \frac{N_4}{
\ell^2(\ell+k_1)^2(\ell+k_2)^2(\ell+k_4)^2}\,,
\end{equation}
where the numerator is independent of loop momentum $\ell$.
Using a parametrization of the type of \eqn{OneLoopParametrization},
we find that on the collinear cut $\ell^2=(\ell+k_1)^2=0$ where
$\ell=\alpha_1 k_1$, there is a double pole $d\alpha_1/\alpha_1^2$.
Therefore no $\dlog$ form exists. 
In any case, at higher loops we will
find the notion of a generalized box very useful for finding $\dlog$
forms.

%\subsubsection*{Pentagons}

Next we consider a generic one-loop pentagon form,
\begin{align}
d{\cal I}_5 = \frac{d^4\ell\, N_5}{\ell^2(\ell-K_1)^2
         (\ell-K_1-K_2)^2(\ell-K_1-K_2-K_3)^2(\ell+K_5)^2}\,,
\label{eqn:GenericPentagonForm}
\end{align}
with off-shell momenta $K_j$. The numerator $N_5$ is not fixed by the
normalization whereas it was in the case of the box. Also unlike in the case of the box,
there are multiple numerators $N_5$ that give unit leading
singularities.  The constraint of no poles at infinity 
constrains $N_5$ to be quadratic: $N_5
= g_1\ell^2 + g_2 (\ell\cdot Q) + g_3$, where the $g_k$ are some
constants and $Q$ is a constant vector.  The simplest way to
decompose the pentagon form (\ref{eqn:GenericPentagonForm}) is to
start with a reference pentagon form,
\begin{equation}
d\widetilde{{\cal I}_5} \equiv \dlog\frac{(\ell-K_1)^2}{\ell^2}\wedge
\dlog\frac{(\ell-K_1-K_2)^2}{\ell^2} \wedge
\dlog\frac{(\ell+K_4+K_5)^2}{\ell^2}\wedge
\dlog\frac{(\ell+K_5)^2}{\ell^2} \,,
\label{eqn:parityOddPentagon}
\end{equation}
in terms of which we express all other pentagons.  This reference
$\dlog$ form corresponds to a parity-odd integrand form and gives zero
when integrating over Minkowski space.  In \eqn{parityOddPentagon} we
single out $\ell^2$, but one can show that all five choices of
singling out one of the inverse propagators are equivalent.  We then
can decompose the generic pentagon form
(\ref{eqn:GenericPentagonForm}) into the reference pentagon form
(\ref{eqn:parityOddPentagon}) ${d{\widetilde{\cal I}_5}} $ plus box
forms,
\begin{equation}
d{\cal I}_5 = c_0  d\widetilde{{\cal I}_5} + \sum_{j=1}^5 
c_j d{\cal I}_4^{(j)}\,,
\label{eqn:PentExp}
\end{equation}
where $c_j$ are coefficients most easily determined by imposing cut
conditions on both sides of \eqn{PentExp} and matching.  While we can
express \eqn{parityOddPentagon} as a loop-integrand, its numerator
$\widetilde{N}_5$ is complicated, and it is better to use directly the
$\dlog$ form (\ref{eqn:parityOddPentagon}) for obtaining cuts.

The expansion (\ref{eqn:PentExp}) is always valid for up to two powers
of loop momentum in the numerator $N_5$, but in higher-loop
calculations it is often more convenient to use alternative
decompositions. It is also possible to define generalized pentagons
with propagators other than the standard ones. These will be useful in
subsequent discussion.

%%%%%%%%%%%%%%%%%%%%%%
\subsection{Two loops}

At two loops there are only two distinct integrand forms to consider:
the planar and nonplanar double boxes displayed in
\figs{TwoLoopParent}{TwoLoopDlogForm}. As shown in Ref.~\cite{Log}, we
can choose the numerators such that all integrals individually have
only logarithmic singularities and no poles at infinity.  As already
noted, in previous sections we suppressed a factor of $\K$ (defined in
\eqn{KappaDef}), that we now restore to make the connection to $\dlog$
forms and the leading singularity coefficients more straightforward.

%\subsubsection*{Planar double box}

We start with the planar double box of \fig{TwoLoopParent}.  It
appears in the amplitude as
\begin{equation}
  d{\cal I}^{\rm (P)}  = \frac{d^4\ell_5\,d^4\ell_6\,\,s^2t C_1}
  {\ell_5^2(\ell_5+k_1)^2(\ell_5-k_2)^2(\ell_5+\ell_6-k_2-k_3)^2
    \ell_6^2(\ell_6-k_3)^2(\ell_6+k_4)^2}\,,
\end{equation}
where $C_1$ is defined in \eqn{CoefficientDef}. 
It is straightforward to put this integrand form into a $\dlog$ form by
iterating the one-loop single-box case in \eqn{CoefficientDef}.
We immediately obtain a product of two one-loop box integrand forms,
\begin{align}
d{\cal I}^{\rm (P)} & = 
C_1 \left[\frac{d^4\ell_5\,\,s(\ell_6-k_2-k_3)^2}
                      {\ell_5^2(\ell_5+k_1)^2(\ell_5-k_2)^2(\ell_5+\ell_6-k_2-k_3)^2}\right] \nonumber \\
& \hskip 1 cm \times 
\left[\frac{d^4\ell_6\,\,st}
  {\ell_6^2(\ell_6-k_3)^2(\ell_6+k_4)^2(\ell_6-k_2-k_3)^2}\right] \nonumber\\
 & = C_1\, d{\cal I}_4\left[ 
\begin{array}{cc}
\ell_5^2&(\ell_5+k_1)^2\\
(\ell_5-k_2)^2&(\ell_5+\ell_6-k_2-k_3)^2\end{array}\right]
\wedge d{\cal I}_4\left[ \begin{array}{cc} \ell_6^2&(\ell_6-k_3)^2\\
                           (\ell_6+k_4)^2&(\ell_6-k_2-k_3)^2\end{array}\right].   \hskip .5 cm 
\label{eqn:TwoLoopDlogForm1}
\end{align}
Thus,
$d{\cal I}^{(P)}$ is a $\dlog$ eight-form given by the wedge product of two 
$d{\cal I}_4$ box four-forms, multiplied by the coefficient $C_1$.  By
symmetry, we can also reverse the order of iterating the one-loop box
forms to obtain instead
\begin{align}
d{\cal I}^{\rm (P)} 
 =C_1\, d{\cal I}_4\left[ \begin{array}{cc} \ell_5^2&(\ell_5+k_1)^2\\
(\ell_5-k_2)^2&(\ell_5-k_2-k_3)^2\end{array}\right]\wedge 
 d{\cal I}_4\left[ \begin{array}{cc} \ell_6^2&(\ell_6-k_3)^2\\
                           (\ell_6+k_4)^2&(\ell_5+\ell_6-k_2-k_3)^2\end{array}\right] .
\label{eqn:TwoLoopDlogForm2}
\end{align}
Despite the fact that the two $\dlog$ forms in
\eqns{TwoLoopDlogForm1}{TwoLoopDlogForm2} look different, they are
equal. This is another illustration that $\dlog$ forms are
not unique, and there are many different ways to write them.

\begin{figure}[tb]
\begin{center}
\begin{tabular}
{
>{\centering\arraybackslash}m{0.30\textwidth}
>{\centering\arraybackslash}m{0.30\textwidth}
}
\includegraphics{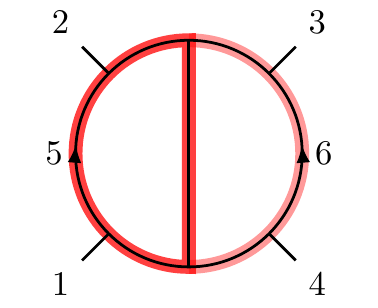} &
\includegraphics{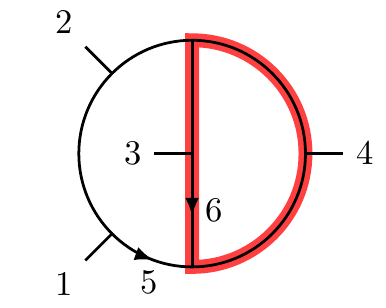} \\
(a) & (b) \\
\end{tabular}
\end{center}
\caption{The (a) planar and (b) nonplanar two-loop four-point parent diagrams.
  In each case one-loop box subdiagrams are shaded (red). In the planar diagram, the
￼￼￼￼￼￼￼￼￼￼Jacobian from the one-loop box subdiagram combines with the remaining three lightly shaded
         (light red) propagators to form a second box.
}
\label{fig:TwoLoopDlogForm}
\end{figure}
%%%%%%%%%%%%%%%%%%%%%%%%%%                                                                                              

The nonplanar double box in \fig{TwoLoopParent} is more complicated,
since it contains both box and pentagon subdiagrams. It is given by
\begin{equation}
d{\cal I}^{\rm (NP)} = \frac{d^4\ell_5\,d^4\ell_6\,\,\,C_1st[(\ell_5-k_4)^2+(\ell_5-k_3)^2]}
{\ell_5^2(\ell_5+k_1)^2(\ell_5+k_1+k_2)^2\ell_6^2(\ell_6+k_3)^2(\ell_6+\ell_5)^2(\ell_6+\ell_5-k_4)^2} \,.
\end{equation}
We start by writing the $\ell_6$ box subdiagram highlighted in
\fig{TwoLoopDlogForm} as a $\dlog$ form, so that
\begin{equation}
d{\cal I}^{\rm (NP)} = d{\cal I}_{\ell_6} \wedge d{\cal I}_{\ell_5} \,,
\end{equation}
where 
\begin{equation}
d{\cal I}_{\ell_6} = \frac{d^4\ell_6\,\,(\ell_5\cdot q)(\ell_5\cdot \overline{q})}
{\ell_6^2(\ell_6+k_3)^2(\ell_6+\ell_5)^2(\ell_6+\ell_5-k_4)^2} 
= d{\cal I}_4\left[ \begin{array}{cc} \ell_6^2&(\ell_6+\ell_5)^2\\ 
                   (\ell_6+k_3)^2 & (\ell_6+\ell_5-k_4)^2\end{array}\right] \,.
\label{NP1}
\end{equation}
The $d{\cal I}_{\ell_6}$ form is normalized with the Jacobian
numerator $(\ell_5\cdot q)(\ell_5\cdot\overline{q})$, where
$q=\lambda_3\widetilde{\lambda}_4$ and
$\overline{q}=\lambda_4\widetilde{\lambda}_3$.  This is just a
relabeling of the two-mass-easy normalization given in
\eqn{TwoMassEasyJacobian}. The remaining integral 
can then be divided into two parts,
\begin{equation}
d{\cal I}_{\ell_5} = C_1\, d{\cal I}_5^{\chi_1} + C_2\, d{\cal I}_5^{\chi_2} \,,
\end{equation}
where we have used $t C_1 = u C_2$ and
\begin{align}
d{\cal I}_5^{\chi_1} & \equiv \frac{d^4\ell_5\,\,st(\ell_5-k_4)^2}{\ell_5^2(\ell_5+k_1)^2(\ell_5+k_1+k_2)^2(\ell_5\cdot q)(\ell_5\cdot\overline{q})} \,, \nonumber \\
d{\cal I}_5^{\chi_2} & \equiv d{\cal I}_5^{\chi_1} \Bigr|_{k_3\leftrightarrow k_4} . 
\label{eqn:ChiDef}
\end{align}
These are exactly generalized pentagons, of the type we discussed in
the previous subsection. It is straightforward to check that there are
only logarithmic singularities and no poles at infinity. Because of
the two propagators linear in $\ell_5$, these two forms are not
canonical one-loop integrand forms.  Nevertheless, we can find a change
of variables for $d{\cal I}_5^{\chi_1}$ and $d{\cal I}_5^{\chi_2}$ so
that each is a single $\dlog$ form,
\begin{align}
d{\cal I}_{5}^{\chi_1} & = \dlog\frac{\ell_5^2}{(\ell_5\cdot \overline{q})}\wedge
\dlog\frac{(\ell_5+k_1)^2}{(\ell_5\cdot \overline{q})}\wedge
\dlog\frac{(\ell_5+k_1+k_2)^2}{(\ell_5\cdot q)}\wedge
\dlog\frac{(\ell_5-\ell_5^\ast)^2}{(\ell_5\cdot q)} \,,  %\hskip .6 cm 
\end{align}
where $\ell_5^\ast=\frac{\langle 34\rangle}{\langle
  31\rangle}\lambda_1\widetilde{\lambda}_4$ is the solution of cut
conditions $\ell_5^2=(\ell_5+k_1)^2=(\ell_5+k_1+k_2)^2=(\ell_5\cdot
q)=0$.  A similar result is obtained for $d{\cal I}_{\ell_5}^{\chi_2}$
by swapping $k_3$ and $k_4$.  The final result for the $\dlog$ form of
the nonplanar double box is
\begin{equation}
d{\cal I}^{\rm (NP)} = C_1\, d{\cal I}_{\ell_6} \wedge d{\cal I}_{\ell_5}^{\chi_1}
  + C_2 \, d{\cal I}_{\ell_6} \wedge d{\cal I}_{\ell_5}^{\chi_2}\,. 
\end{equation}
Since each term carries a different normalization, this expression
cannot be uniformly normalized to have unit leading singularities on
all cuts.  We choose a normalization such that $C_1$ or $C_2$ are the
leading singularities of the integrand form, depending on which
residue we take.  This construction is useful at three loops, as we
see below.

%%%%%%%%%%%%%%%%%%%%%%%%%%%
\subsection{Three loops}

%%%%%%%%% FIGURE %%%%%%%%%%%%%%%        
\begin{figure}[th]
\begin{center}
\begin{tabular}
{>{\centering\arraybackslash}m{0.30\textwidth}
>{\centering\arraybackslash}m{0.30\textwidth}
>{\centering\arraybackslash}m{0.30\textwidth}
}
\includegraphics{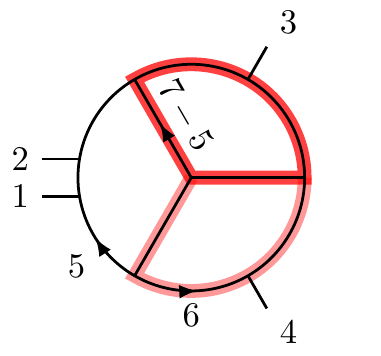} &
\includegraphics{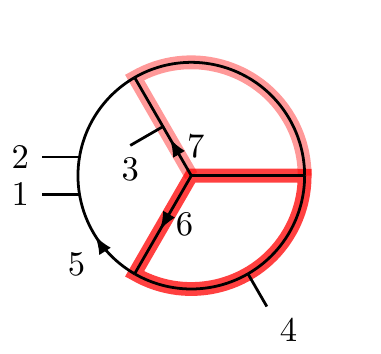} &
\includegraphics{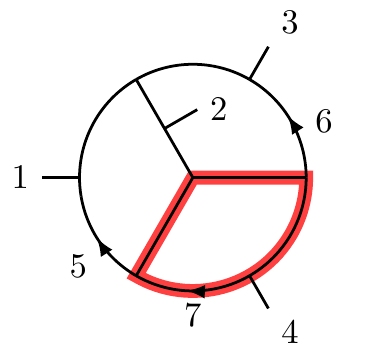} \\
(e) & (f) & (i)\\
\end{tabular}
\caption[]{
The three-loop diagrams with highlighted one-loop box subdiagrams used in the 
construction of $\dlog$ forms.  In diagram (e) we start with the top (red) box whose 
Jacobian generates the missing fourth propagator for the bottom (light red) box. 
In diagram (f) we start with the bottom (red) box whose 
Jacobian generates the missing fourth propagator for the top (light red) box. 
In diagram (i), there is only one box on the bottom (red).
     }
\label{fig:DLogThreeLoop}
\end{center}
\end{figure}
%%%%%%%%%%%%%%%%%%%%%%%%%%%%                                                                                              

We now turn to the main subject: constructing the three-loop
four-point nonplanar $\dlog$ forms.  Unfortunately, at present there
is no general procedure to rewrite high-loop order integrand forms
into $\dlog$ forms.  Nevertheless, we can proceed with our general
strategy: whenever there is a box subdiagram, we rewrite it in a
$\dlog$ form and then deal with the remaining forms by again looking
at subdiagrams. This tactic works well at three loops: We have worked
out $\dlog$ forms for all diagrams that have box subdiagrams. This
consists of all diagrams except for diagram (h), which is the most
complicated case because it has only pentagon subdiagrams.  Diagrams
(a) and (b) are simple because they are directly related to the one-
and two-loop cases. In this subsection we show explicit examples of
diagrams (e), (f) and (i), which are less trivial. Each example shows
how to overcome some new obstacle to constructing a $\dlog$ form.

We start with diagram (e) in \figs{ParentTenProp}{DLogThreeLoop}. 
The numerator is $\K N^{\rm (e)} = C_1 s^2 t (\ell_5 + k_4)^2$.  
This gives us the integrand form,
\begin{align}
d{\cal I}^{\rm (e)} = \null &
\frac{ d^4 \ell_5\, d^4 \ell_6\, d^4\ell_7 \,\,C_1s^2t (\ell_5 + k_4)^2}{ \ell_5^2
 (\ell_5 - k_1)^2 (\ell_5 - k_1 - k_2)^2 (\ell_5+\ell_6)^2 \ell_6^2 (\ell_6-k_4)^2} \nonumber \\
&\hskip 2 cm \null \times 
\frac{1}{
(\ell_7-\ell_5)^2(\ell_7+\ell_6)^2(\ell_7+k_4)^2(\ell_7-k_1-k_2)^2} \,,
\label{eqn:DiagramEDLogStart}
\end{align}
where $C_1$ is defined in \eqn{CoefficientDef}.  There are two box
subdiagrams in this case, both of which are highlighted in
\fig{DLogThreeLoop}(e).  We start with the top (red) box in \fig{DLogThreeLoop}, which carries loop momentum
$\ell_7$.  The $\dlog$ form for this box subdiagram is
\begin{equation}
d{\cal I}_{\ell_7}=d{\cal I}_4\left[ \begin{array}{cc} (\ell_7-\ell_5)^2 & 
(\ell_7+k_4)^2 \\ 
  (\ell_7+\ell_6)^2
&(\ell_7-k_1-k_2)^2\end{array}\right] .
\end{equation}
Using \eqn{CutBoxJacobian} and relabeling, we find that this box
carries a normalization factor
\begin{equation}
J_7 = (\ell_5+k_4)^2(\ell_6-k_3-k_4)^2-(\ell_5-k_1-k_2)^2(\ell_6-k_4)^2\,,
\end{equation}
which then goes into the denominator of the remaining $\ell_5,\ell_6$
forms. The $\ell_6$ integrand form is a generalized box formed from
the three bottom (light red) propagators in \fig{DLogThreeLoop} and a
generalized propagator $J_7$ . We can then rewrite the $\ell_6$
integrand form as a $\dlog$ form,
\begin{equation}
d{\cal I}_{\ell_6}=d{\cal I}_4\left[ \begin{array}{cc}
\ell_6^2 & (\ell_6+\ell_5)^2\\ 
 (\ell_6-k_4)^2& [(\ell_5+k_4)^2(\ell_6-k_3-k_4)^2-(\ell_5-k_1-k_2)^2(\ell_6-k_4)^2]
\end{array}\right].
\end{equation}
The normalization required by this generalized box can be computed from 
the generic Jacobian formula (\ref{eqn:GeneralJacobian}) and gives
\begin{equation}
J_6 = s[(\ell_5+k_4)^2]^2 \,,
\end{equation}
exactly matching \eqn{DiagramEJacobian} which was obtained by
searching for double poles.  This confirms that a factor
$(\ell_5+k_4)^2$ is needed in the numerator of
\eqn{DiagramEDLogStart}: it cancels the double pole in the remaining
$\ell_5$ form.  After canceling the double propagator against the
numerator factor, we then have the last box form,
\begin{equation}
d{\cal I}_{\ell_5}=d{\cal I}_4\left[ \begin{array}{cc} \ell_5^2 & 
  (\ell_5+k_4)^2\\ (\ell_5-k_1)^2&(\ell_5-k_1-k_2)^2\end{array}\right].
\end{equation}
The final result for the integrand form of \eqn{DiagramEDLogStart} is thus
\begin{equation}
d{\cal I}^{\rm (e)}  = C_1\, d{\cal I}_{\ell_5}\wedge d{\cal I}_{\ell_6}\wedge d{\cal I}_{\ell_7}\,.
\end{equation}
The derivation of a $\dlog$ form for this case is relatively straightforward,
because at each step we encounter only generalized box
forms.

As a less straightforward example, consider the nonplanar diagram (f) in 
\figs{ParentTenProp}{DLogThreeLoop}, using the
numerator $\K N_1^{\rm (f)}$ in \tab{BasisTenProp}. This integrand form is
\begin{align}
d{\cal I}^{(\rm f)} = \null & 
\frac{d^4 \ell_5\, d^4 \ell_6\, d^4\ell_7\,\,\,\,C_1st(\ell_{5}+k_{4})^{2}
    \left[(\ell_{5}+k_{3})^{2} + (\ell_{5}+k_{4})^{2} \right]}
{\ell_5^2\, (\ell_5 -k_1)^2 (\ell_5-k_1-k_2)^2
\ell_7^2  (\ell_7-k_3)^2 (\ell_5 + \ell_7+k_4)^2} \nonumber \\
& \hskip 4 cm \null \times
\frac{1}{\ell_6^2(\ell_6-\ell_5)^2(\ell_6-\ell_5-k_4)^2(\ell_6+\ell_7)^2}\,, 
\label{eqn:DefFForm}
\end{align}
where we include numerator $N_1^{\rm (f)}$ from \tab{BasisTenProp}.
As indicated in \fig{DLogThreeLoop} for diagram (f), there are two box
subdiagrams that can be put into $\dlog$ form. We write the $\ell_6$
and $\ell_7$ forms as box-forms straight away:
\begin{align}
d{\cal I}_{\ell_6} &= d{\cal I}_4 \left[ \begin{array}{cc} \ell_6^2 & (\ell_6-\ell_5-k_4)^2 \\
             (\ell_6+\ell_7)^2& (\ell_6-\ell_5)^2
\end{array}\right] , \nonumber \\
d{\cal I}_{\ell_7} &= d{\cal I}_4 \left[ \begin{array}{cc} \ell_7^2 & (\ell_7+\ell_5+k_4)^2\\ 
(\ell_7-k_3)^2&[(\ell_5+k_4)^2(\ell_5+\ell_7)^2-\ell_5^2(\ell_5+\ell_7+k_4)^2]\end{array}\right] .
\end{align}
The $\ell_6$ box introduced a Jacobian which is then used in the $\ell_7$
box as a new generalized propagator. The remaining $\ell_5$ form,
including also the Jacobian from the $\ell_7$ generalized box, is then a
generalized pentagon form,
\begin{equation}
d{\cal I}_{\ell_5} =  \frac{d^4\ell_5\,C_1 \, st\left[(\ell_{5}+k_{3})^{2} + (\ell_{5}+k_{4})^{2} \right]}
{\ell_5^2(\ell_5-k_1)^2(\ell_5-k_1-k_2)^2(\ell_5\cdot q)
(\ell_5\cdot \overline{q})} \,,
\end{equation}
where $q=\lambda_3\widetilde{\lambda}_4$,
$\overline{q}=\lambda_4\widetilde{\lambda}_3$.  This generalized pentagon 
form is a relabeling of the one we encountered for the two loop nonplanar 
double box so we can write, 
\begin{equation}
d{\cal I}_{\ell_5} = C_1 \, d{\cal I}_5^{\chi_1} 
  + C_2\, d{\cal I}_5^{\chi_2} \,,
\end{equation}
where the forms $d{\cal I}_5^{\chi_1}$ and $d{\cal I}_5^{\chi_2}$ are
defined in \eqn{ChiDef}. The final result for $d{\cal I}^{(\rm f)}$ in
\eqn{DefFForm} is then
\begin{equation}
d{\cal I}^{\rm (f)} = C_1\, d{\cal I}_{\ell_6}\wedge d{\cal
  I}_{\ell_7}\wedge d{\cal I}_5^{\chi_1} + C_2\, d{\cal
  I}_{\ell_6}\wedge d{\cal I}_{\ell_7}\wedge d{\cal I}_5^{\chi_2}\,.
\end{equation}

An even more complicated example is diagram (i) in
\figs{ParentTenProp}{DLogThreeLoop}.  Consider the first term in
numerator $N_1^{\rm (i)}$ in \tab{BasisTenProp} given by
$(\ell_6+k_4)^2(\ell_5-k_1-k_2)^2$. We will explicitly write the
$\dlog$ form for this part of the integrand but not the remaining
pieces for the sake of brevity.  Putting back the overall
normalization $C_1 st$, we have the integrand form
\begin{align}
d{\cal I}^{\rm (i)}_1 &= 
\frac{d^4 \ell_5 \, d^4 \ell_6 \, d^4 \ell_7\,\,\,C_1 st (\ell_6+k_4)^2(\ell_5-k_1-k_2)^2}
{\ell_5^2(\ell_5-k_1)^2\ell_6^2(\ell_6-k_3)^2(\ell_6+\ell_5-k_1-k_3)^2(\ell_6+\ell_5+k_4)^2}\nonumber \\
& \hskip 3 cm \null \times
\frac{1} {\ell_7^2(\ell_7+k_4)^2(\ell_7-\ell_5)^2(\ell_7+\ell_6+k_4)^2} \,.
\label{eqn:IntegrandFormI}
\end{align}
As in all other cases we start with a box subintegral.  Here
there is only a single choice, as highlighted in \fig{DLogThreeLoop}(i):
\begin{align}
d{\cal I}_{\ell_7} = \frac{d^4\ell_7\,\,\left[(\ell_5+k_4)^2(\ell_6+k_4)^2-\ell_5^2\ell_6^2\right]}
   {\ell_7^2(\ell_7+k_4)^2(\ell_7-\ell_5)^2(\ell_7+\ell_6+k_4)^2}
= d{\cal I}_4\left[ \begin{array}{cc} \ell_7^2 & (\ell_7-\ell_5)^2\\ 
(\ell_7+k_4)^2&(\ell_7+\ell_6+k_4)^2\end{array}\right] \,.
\end{align}
The $\ell_6$ integrand form is then a generalized pentagon,
\begin{equation}
d{\cal I}_{\ell_6} = \frac{d^4\ell_6\,\,st(\ell_6+k_4)^2}
{\ell_6^2(\ell_6-k_3)^2(\ell_6+\ell_5-k_1-k_3)^2(\ell_6+\ell_5+k_4)^2\left[(\ell_5+k_4)^2(\ell_6+k_4)^2-\ell_5^2\ell_6^2\right]}\,.
\end{equation}
In principle we could follow a general
pentagon decomposition procedure, but there is a simpler way to obtain 
the result. We can rewrite the numerator as
\begin{equation}
(\ell_6+k_4)^2 = \frac{1}{(\ell_5+k_4)^2}
  \left[(\ell_5+k_4)^2(\ell_6+k_4)^2-\ell_5^2\ell_6^2\right]
 + \frac{\ell_5^2}{(\ell_5+k_4)^2}\, \ell_6^2 \,.
\label{eqn:SimpleIdentity}
\end{equation}
After canceling factors in each term against denominator factors,
we get two generalized box integrand forms. The decomposition is 
\begin{equation}
d{\cal I}_1^{\rm (i)} =C_1\, d{\cal I}_{\ell_7} \wedge d{\cal I}_{\ell_6}^{(1)} \wedge d{\cal I}_{\ell_5}^{(1)} 
+ C_3\, d{\cal I}_{\ell_7} \wedge d{\cal I}_{\ell_6}^{(2)} \wedge d{\cal I}_{\ell_5}^{(2)}\,,
\label{eqn:DlogDiagi}
\end{equation}
where the $\ell_6$ integrand forms can be put directly into $\dlog$ forms:
\begin{align}
d{\cal I}_{\ell_6}^{(1)}&= \frac{d^4\ell_6\,\,[(\ell_5-k_1-k_2)^2(\ell_5-k_1-k_3)^2-(\ell_5+k_4)^2(\ell_5-k_1)^2]}{\ell_6^2(\ell_6-k_3)^2(\ell_6+\ell_5-k_1-k_3)^2(\ell_6+\ell_5+k_4)^2}\nonumber\\
&=d{\cal I}_4\left[ \begin{array}{cc} \ell_6^2 
& (\ell_6+\ell_5+k_4)^2\\ (\ell_6-k_3)^2&(\ell_6+\ell_5-k_1-k_3)^2\end{array}\right] , \nonumber\\
d{\cal I}_{\ell_6}^{(2)} &= \frac{d^4\ell_6\,\,(\ell_5\cdot q)(\ell_5\cdot
  \overline{q})(\ell_5-k_1-k_2)^2}{(\ell_6-k_3)^2
(\ell_6+\ell_5-k_1-k_3)^2(\ell_6+\ell_5+k_4)^2
\left[(\ell_5+k_4)^2(\ell_6+k_4)^2-\ell_5^2\ell_6^2\right]}\nonumber\\
&=d{\cal I}_4\left[ \begin{array}{cc} (\ell_6-k_3)^2 & (\ell_6+\ell_5-k_1-k_3)^2\\ (\ell_6+\ell_5+k_4)^2&(\ell_5+k_4)^2(\ell_6+k_4)^2-\ell_5^2\ell_6^2)\end{array}\right],
\label{eqn:TwodLogForms}
\end{align}
with $q=\lambda_2\widetilde{\lambda}_4$ and
$\bar q=\lambda_4\widetilde{\lambda}_2$.  Here we have normalized both
integrand forms properly to have unit leading singularities so that
they are $\dlog$ forms. As indicated in \eqn{DlogDiagi}, the remaining $\ell_5$
integrand forms are different for $d\I^{(1)}_{\ell_6}$ and
$d\I^{(2)}_{\ell_6}$.

Writing the integrand form for $\ell_5$ following from $d\I^{(1)}_{\ell_6}$, 
\begin{align}
d{\cal I}_{\ell_5}^{(1)} &= \frac{d^4\ell_5\,\,st(\ell_5-k_1-k_2)^2}{\ell_5^2(\ell_5-k_1)^2(\ell_5+k_4)^2
[(\ell_5-k_1-k_2)^2(\ell_5-k_1-k_3)^2-(\ell_5+k_4)^2(\ell_5-k_1)^2]}\nonumber\\
&=\frac{d^4\ell_5\,\,st(\ell_5-k_1-k_2)^2}
{\ell_5^2(\ell_5-k_1)^2(\ell_5+k_4)^2((\ell_5-k_1)\cdot q)((\ell_5-k_1)\cdot \overline{q})} ,
\end{align}
where $q=\lambda_2\widetilde{\lambda}_3$ and
$\overline{q}=\lambda_3\widetilde{\lambda}_2$.  In the last expression
we used the fact that the quartic expression was a two-mass-easy
Jacobian of the $\ell_6$ integrand, which factorizes into a product.
Up to relabeling, this is the same integrand as the first
nonplanar pentagon form in \eqn{ChiDef}, and we can write it as the $\dlog$ form
\begin{equation}
d{\cal I}_{\ell_5}^{(1)} = \dlog\frac{(\ell_5-k_1)^2}{((\ell_5-k_1)\cdot \overline{q})}\,\,
\dlog\frac{\ell_5^2}{((\ell_5-k_1)\cdot \overline{q})}\,\,
\dlog\frac{(\ell_5+k_4)^2}{((\ell_5-k_1)\cdot q)}\,\,
\dlog\frac{(\ell_5-\ell_5^\ast)^2}{((\ell_5-k_1)\cdot q)},
\end{equation}
where $q=\lambda_3\widetilde{\lambda}_2$,
 $\overline{q} = \lambda_2\widetilde{\lambda}_3$ and $\ell_5^\ast = \frac{\la 32\ra}{\la 31\ra}\lambda_1\widetilde{\lambda}_2$. 
For the second integrand form in \eqn{TwodLogForms}, the remaining
$\ell_5$ integral is (for
$q=\lambda_2\widetilde{\lambda}_4$ 
and $\overline{q}=\lambda_4\widetilde{\lambda}_2$)
just a generalized box and can be directly written as the $\dlog$
form,
\begin{equation}
d{\cal I}_{\ell_5}^{(2)} =
\frac{d^4\ell_5\,\,ut}{(\ell_5-k_1)^2(\ell_5+k_4)^2(\ell_5\cdot
  q)(\ell_5\cdot\overline{q})}= d{\cal I}_4\left[ \begin{array}{cc} (\ell_5-k_1)^2 & (\ell_5\cdot q)\\ (\ell_5+k_4)^2&(\ell_5\cdot\overline{q})\end{array}\right] \,.
 \label{eqn:DlogDiagi2}
\end{equation}
To obtain this, we used the relation $s C_1 = u C_3$ to write $C_3$ as
the overall factor of the second term in the assembled result 
given in \eqn{DlogDiagi}. 

We have carried out similar procedures on all contributions to the
three-loop four-point amplitude, except for the relatively complicated
case of diagram (h).  In all these cases we find explicit $\dlog$
forms.  These checks directly confirm that there are only logarithmic
singularities in the integrand, as we found in
\sect{ThreeLoopBasisSection}.  At relatively low loop orders, detailed
analysis of the cut structure, as carried out in
\sect{ThreeLoopBasisSection}, provides a straightforward proof of this
property. At higher loop orders, the space of all possible
singularities grows rapidly and finding dlog forms, as we did in the
present section, becomes a more practical way of showing that
there are only logarithmic singularities.  Even so, one cannot
completely avoid detailed checks of the singularity structure because,
in general, $\dlog$ forms do not necessarily make manifest that there are no
poles at infinity.

%%%%%%%%%%%%%%%%                                 
\section{Logarithmic singularities at higher loops}
\label{sec:HigherLoops}

Complete, unintegrated four-point $\NeqFour$ sYM
amplitudes, including their nonplanar parts, have been obtained at
four and five loops in Refs.~\cite{Bern:2010tq, ColorKinematics,
  FiveLoopNonPlanar}. In principle, we could repeat the same procedure
we did for three loops at higher loops to construct
$\dlog$ numerators.  However, the number of parent diagrams grows: at
four loops there are 85 diagrams and by five loops there are 410
diagrams. Many of the diagrams are simple generalizations of the
already analyzed three-loop diagrams, so their analysis is
straightforward. Some, however, are new topologies, for which an
exhaustive search for double or higher poles and
poles at infinity would be nontrivial. Such an analysis would require either
a more powerful means of identifying numerators with the desired
properties, or computer automation to sweep through all dangerous
kinematic regions of the integrands while looking for unwanted singularities.
This of course is an interesting problem for the future.

Here we take initial steps at higher loops, investigating sample four-
and five-loop cases to provide supporting evidence that only
logarithmic singularities are present in the nonplanar sector.  We do
so by showing compatibility between $\dlog$ numerators and 
known expressions for the amplitudes~\cite{Bern:2010tq,  FiveLoopNonPlanar}
on maximal cuts.

%\subsection*{Example 1: maximal cut at four loops}

%
%%%%%%%%%%%%%%%%%%%%% FIGURE %%%%%%%%%%%%%%%
\begin{figure}[tb]
\centering
\begin{tabular}{l}
\multicolumn{1}{c}{\includegraphics{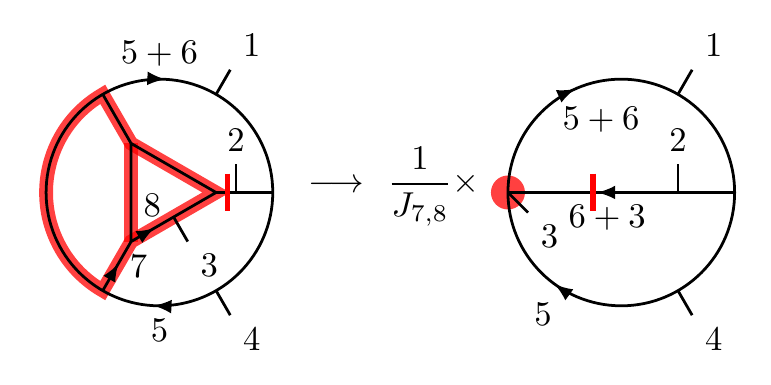}}
%\hskip 1.42 cm (4NP)
\end{tabular}
\caption{\label{fig:FourLoopNP} The left diagram contributes to the
  four-loop four-point $\NeqFour$ sYM amplitude~\cite{Bern:2010tq}.  The 
  shaded (red) lines indicate propagators that are replaced by
  on-shell conditions as given in \eqn{FourLoopCut}. These propagators
  are removed from the diagram and leave behind an inverse Jacobian, given in
  \eqn{doubleBoxJacobian4LNP}. The resulting simplified diagram is given on the right. 
   The vertical 
  shaded (red) line crossing the propagator carrying momentum $\ell_6 + k_3$
  indicates that it too is replaced with an on-shell condition
  at the start of this process.
 }
\end{figure}
%%%%%%%%%%%%%%%%%%%%%%%%%%%%%%%%%%%%%%%%
%

As a first example, consider the nonplanar four-loop diagram on the
left in \fig{FourLoopNP}. 
We wish to show that the maximal cuts are
compatible with a numerator that ensures only logarithmic
singularities and no poles at infinity.  Since this diagram has a
hexagon subdiagram carrying loop momentum $\ell_6$ and a pentagon
subdiagram carrying loop momentum $\ell_5$, the overall dimensionality
and asymptotic scaling constraints of \sect{StrategySection}
require $N^{\fourloop} \sim \mathcal{O}((\ell_6)^4 \,, (\ell_5)^2)$.

In order to derive the desired numerator for this diagram, we 
use the cut sequence
\begin{equation}
{\rm cuts} = \{ (\ell_6+k_3)^2 , B(\ell_8) , B(\ell_7,\ell_7-k_3) \}\,,
\label{eqn:FourLoopCut}
\end{equation}
where the notation is defined in Sect.~\ref{subsec:LoopIntegralsPoleInfty}.
The first cut setting $ (\ell_6+k_3)^2 = 0$  is indicated in \fig{FourLoopNP} by the 
vertical shaded (red) line crossing the corresponding propagator.  The
remaining cuts leave behind Jacobians; the propagators placed on-shell by these cuts
are indicated by the shaded (red) thick lines.
The resulting Jacobian is
\begin{equation}
\label{eqn:doubleBoxJacobian4LNP}
J_{7,8} = (\ell_5-k_3)^2\left[\ell^2_6\right]^2\,.
\end{equation}
Since the Jacobian appears in the denominator, this gives us an
unwanted double pole in the integrand when $\ell_6^2 = 0$.  Thus, to
remove it on the cuts (\ref{eqn:FourLoopCut}) we require the numerator
be proportional to $\ell^2_6$:
\begin{align}
{N^{\fourloop}(\ell_5,\ell_6)}\bigr|_{\rm cut} =
{\ell^2_6 \widetilde{N}^{\fourloop}(\ell_5,\ell_6)}\bigr|_{\rm cut} \,.
\end{align}

After canceling one factor of $\ell^2_6$ from the Jacobian in
\eqn{doubleBoxJacobian4LNP} against a factor in the numerator, we can
use the remaining $\ell_6^2$ or $(\ell_5-k_3)^2$ from the Jacobian
together with the remaining uncut propagators on the right of
\fig{FourLoopNP} to give two distinct relabelings of the two-loop
nonplanar diagram in \fig{TwoLoopParent}(b), if we also cancel
the other propagator factor coming from the Jacobian.  We then relabel
the $\dlog$ numerators for the two-loop nonplanar diagram in
\eqn{TwoLoopNumerators} to match the labels of the
simplified four-loop diagram on the right in \fig{FourLoopNP}.  Including factors to cancel
the double pole and unwanted Jacobian factor, we have two
different $\dlog$ numerators for the four-loop diagram of
\fig{FourLoopNP}:
\begin{align}
\cut{N^{\fourloop}_{1}(\ell_5,\ell_6)} = & \cut{\ell^2_6 (\ell_5-k_3)^2 \left[(\ell_6-k_4)^2 + (\ell_6-k_1)^2\right]} \,,
\nonumber \\
\cut{N^{\fourloop}_{2}(\ell_5,\ell_6)} = & \cut{[\ell^2_6]^2 \left[(\ell_5-k_2-k_3)^2  + (\ell_5-k_1-k_3)^2\right]} \,.
\label{eqn:FourLoopNumerators}
\end{align}
The integrands with these numerators then have only logarithmic
singularities and no poles at infinity in the kinematic region of the
cut, as inherited from the two-loop nonplanar double box.

Are these $\dlog$ numerators compatible with the known four-loop amplitude? 
Relabeling the numerator of the corresponding
diagram~32 in Fig.~23 of Ref.~\cite{Bern:2010tq} to match our labels,
we see that a valid numerator that matches the maximal cuts is
\begin{equation}
N^{\fourloop}_{\text{old}} = 
 \ell_6^2 (s \ell_6^2 - t (\ell_5 - k_3)^2 - s (\ell_6 +\ell_5)^2 ) \,.
\label{eqn:OldFourLoopDiag32}
\end{equation}
To check compatibility with our $\dlog$ numerators we take the
maximal cut, replacing all propagators with on-shell conditions.
This selects out a piece unique to this diagram.\footnote{Other
diagrams do not mix with the one under consideration if we use all
solutions to the maximal cut.}
On the maximal cut, \eqn{OldFourLoopDiag32} simplifies to
\begin{equation}
\maxcut{N^{\fourloop}_{\text{old}}} = \ell_6^2 (s \ell_6^2 + t (2 \ell_5 \cdot k_3) )
= \maxcut{\Bigl(N^{\fourloop}_{1} - N^{\fourloop}_{2}\Bigr)} \,.
\label{eqn:OldFourLoopDiag32MaxCut}
\end{equation}
This shows that the maximal cut of diagram 32 with the old numerator is a linear combination of the
maximal cut of diagram 32 with the two $\dlog$ numerators in
\eqn{FourLoopNumerators}.   We note that by following
through the modified rung rule of \sect{Heuristic rules},
we obtain the same coefficients as those determined from the maximal cuts.

%\subsection*{Example 2: Maximal cut at five loops}

%%%%%%%%%%%%%%%%%%%%% FIGURE %%%%%%%%%%%%%%%
\begin{figure}[tb]
  \centering % trim =   left bottom right top
\begin{tabular}{l}
\multicolumn{1}{c}{\includegraphics{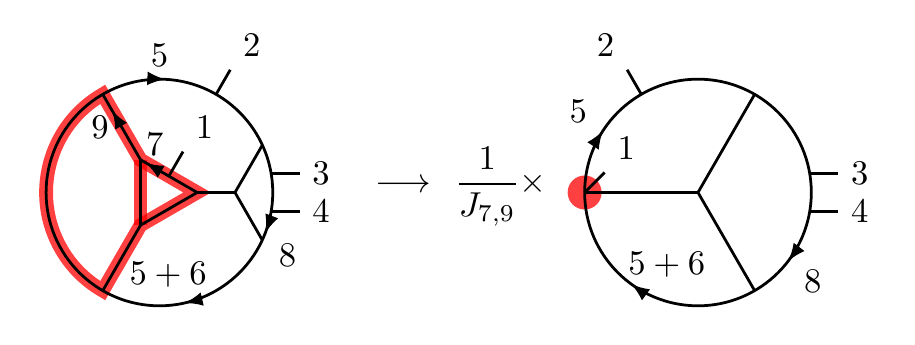}}
%\\
%\hskip 1.42 cm (5NP)
\end{tabular}
\caption{The left diagram contributes to the five-loop four-point
  $\NeqFour$ sYM amplitude~\cite{FiveLoopNonPlanar}.  The shaded 
  (red) lines indicate propagators that are replaced by
  on-shell conditions as given in \eqn{FourLoopCut}. These propagators are
  removed from the diagram and leave behind an inverse Jacobian, given in \eqn{J}.
  The resulting simplified diagram is given on the right.
  The factor $\ell_6^2$ in the Jacobian can be used to expand the shaded
  (red) region, resulting in a graph isomorphic to the three-loop
  diagram \fig{ParentTenProp}(g).
    }
 \label{fig:FiveLoopNP}
\end{figure}
%%%%%%%%%%%%%%%%%%%%%%%%%%%%%%%%%%%%%%%%

Next consider a five-loop example: the nonplanar five-loop diagram on the left of \fig{FiveLoopNP}.
As in the four-loop case, we identify potential double poles
by choosing a sequence of cuts that
uncovers a lower-loop embedding for which a $\dlog$ numerator is already known.
Our order of taking cuts is
\begin{equation}
{\rm cuts} = \{ B(\ell_7) , B(\ell_9,\ell_9+k_1) \} \,,
\end{equation}
resulting in the Jacobian
\begin{align}
 J_{7,9} = \ell^2_6 \left[\ell^2_6(\ell_5+k_1)^2-\ell^2_5(\ell_6-k_1)^2\right] .
\label{eqn:J}
\end{align}
Collecting the $\ell^2_6$-factor of this Jacobian with the remaining
uncut propagators reproduces a relabeling of three-loop diagram (g),
with numerator given in Eq.~(\ref{eqn:Solution}).  To ensure this
five-loop nonplanar integrand has a $\dlog$ numerator, we require the
numerator to cancel the Jacobian, as well as to contain a factor of the
three-loop numerator,
\begin{equation}
N^{\rm (g)} = -s (\ell_8-\ell_5)^2 
+ (\ell_5+k_1)^2 \left[(\ell_8-k_1)^2+(\ell_8-k_2)^2\right] ,
\end{equation}
obtained from \eqn{Solution} and relabeled to match 
\fig{FiveLoopNP}.  We have not included the last three terms in the
numerator given in \eqn{Solution}, because they vanish on maximal cuts,
which we impose below in our compatibility test. Here we are not
trying to find all $\dlog$ numerators, but just those that we
can use for testing compatibility with the known results.
Combining the Jacobian (\ref{eqn:J}) with the 
relabeled numerator $N^{(\rm g)}$ gives a valid $\dlog$ numerator, 
\begin{align}
\cut{N^{\fiveloop}} &= \cut{\left[\ell^2_6(\ell_5+k_1)^2
-\ell^2_5(\ell_6-k_1)^2 \right] N^{\rm (g)}} \,. 
\label{eq:N70logSing}
\end{align}

A straightforward exercise then shows that on the maximal cut of the 
five-loop diagram in \fig{FiveLoopNP}, $N^{\fiveloop}$ matches
the numerator from Ref.~\cite{FiveLoopNonPlanar}:
\begin{equation}
\maxcut{N^{\fiveloop}}
=
\maxcut{N^{\fiveloop}_{\rm old}}
=
-\frac{1}{2} s (\ell_8\cdot k_1) (\ell_6\cdot k_1) (\ell_5 \cdot k_1) \,.
\label{eq:N70Zvi}
\end{equation}
Here we have compared to diagram 70 of the ancillary file of
Ref.~\cite{FiveLoopNonPlanar} and shifted momentum labels to match
ours.  Again the modified rung rule matches the
$\ell_6^2(\ell_5+k_1)^2$ term, which is the only non-vanishing
contribution to $N^{\fiveloop}$ on the maximal cut.

We have also checked a variety of other four- and five-loop examples.
These provide higher-loop evidence that we should find
only logarithmic singularities and no poles at infinity.  We build on
this theme in the next section by considering the consequences of
$\dlog$ numerators at high loop-order in the planar sector.

%===========================================================================
\section{Back to the planar integrand}
\label{sec:PlanarIntegrand}
%==========================================================================

How powerful is the requirement that an expression has only
logarithmic singularities and no poles at infinity?  To answer this we
re-examine the planar sector of $\NeqFour$ sYM theory and argue that these
requirements on the singularity structure are even more restrictive than
dual conformal invariance. Specifically we make the following conjecture:
\begin{itemize}
\item Logarithmic singularities and absence of poles at infinity imply
  dual conformal invariance of local integrand forms to all loop
  orders in the planar sector.
\end{itemize}
To give supporting evidence for this conjecture, as well as to show
that the constraints on the singularities are even stronger than
implied by dual conformal symmetry, we work out a variety of
nontrivial examples.  In particular, we show in detail that at three-
and four-loops the singularity conditions exactly select the dual
conformal invariant integrand forms that appear in the amplitudes. We
also look at a variety of other examples through seven loops.  This
conjecture means that by focusing on the singularity structure we can
effectively carry over the key implications of dual conformal symmetry
to the nonplanar sector even if we do not know how to carry over the
symmetry itself.  This suggests that there may be some kind of
generalized version of dual conformal symmetry for the complete
four-point amplitudes in $\NeqFour$ sYM theory, including the
nonplanar sector.  At the integrated level dual conformal symmetry
leads to powerful anomalous Ward identities that constrain planar
amplitudes~\cite{DualConfWI}.  An interesting question is whether
anything analogous can be found for nonplanar amplitudes.  We put off
further speculation on these points until future work.

We also show that the conditions of no double poles are even
more constraining than dual conformal symmetry. In fact, we demonstrate that
the singularity constraints explain why certain dual conformal
numerators cannot appear in the $\NeqFour$ sYM integrand.  We 
describe simple rules for finding non-logarithmic poles in momentum
twistor space.  These rules follow the spirit of
Ref.~\cite{Drummond:2007aua} and allow us to restrict the set of dual
conformal numerators to a smaller subset of potential $\dlog$ numerators.  
While these rules do not fully eliminate all dual conformal numerators 
that lead to unwanted double poles, they offer a good starting point 
for finding a basis of $\dlog$ numerators.

Furthermore, we give explicit examples at five and six loops where the
pole constraints not only identify contributions with zero coefficient
but also explain nonvanishing relative coefficients between various dual
conformally invariant contributions.  From this perspective, requiring
only logarithmic singularities is a stronger constraint than requiring
dual conformal symmetry.

In our study we use the four-loop results from Ref.~\cite{Bern:2006ew}
and results through seven-loops from
Ref.~\cite{Bourjaily:2011hi}. Equivalent results at five and six loops
can be found in Refs.~\cite{MaximalCuts,Eden:2012tu,Bern:2012di}. 

\subsection{Brief review of dual conformal invariance}

%%%%%%%%% FIGURE %%%%%%%%%%%%%%%
\begin{figure}[tb]
\begin{center}
\begin{tabular}
{
>{\centering\arraybackslash}m{0.30\textwidth}
}
\includegraphics{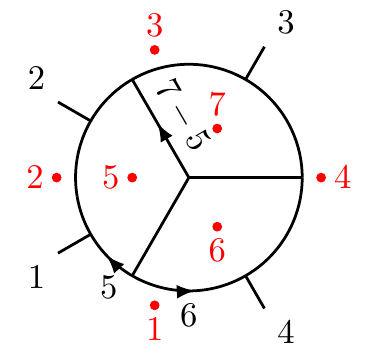} \\
(e)
\end{tabular}
\end{center}
\caption{The planar three-loop diagram (e),
including shaded (red) dots and labels to indicate the face or dual variables.}
\label{fig:DCIThreeLoopE}
\end{figure}
%%%%%%%%%%%%%%%%%%%%%%%%%%%%%%%%

Dual conformal symmetry~\cite{DualConformalMagic, Alday:2007hr,
  Drummond:2008vq} has been extensively studied for planar $\NeqFour$
sYM amplitudes. For a detailed review, see
Ref.~\cite{Elvang:2013cua,Henn:2014yza}.  Here we only require
the part useful for multiloop four point amplitudes, which we briefly
review.  Dual or region variables are the natural variables to make
dual conformal symmetry manifest.  To indicate the dual variables, we
draw graphs in momentum space with the corresponding dual faces marked
with a shaded (red) dot and labeled with a shaded (red) number. 
This is illustrated in \fig{DCIThreeLoopE}.

We define the relation between external momenta $k_i$ and external
dual variables (region momenta) $x_i$ as
\begin{equation}
k_i = x_{i+1} - x_{i} \,, \quad i = 1,2,3,4 \,,  \quad x_{5} \equiv x_1 \,.
\end{equation}
In term of dual variables, the Mandelstam invariants are
\begin{equation}
s = (k_1 + k_2)^2 \equiv x_{13}^2 \,,
\hskip 1cm
t = (k_2 + k_3)^2 \equiv x_{24}^2 \,.
\end{equation}
The internal faces are parametrized by additional $x_j$, with $j = 5, 6,\ldots, 4+L$
corresponding to loop momenta.
In terms of the dual coordinates, loop momenta are defined from the diagrams as:
\begin{equation}
\ell = x_{\rm right} - x_{\rm left} \,,
\end{equation}
where $x_{\rm right}$ is the dual coordinate to the right of $\ell$
when traveling in the direction of $\ell$, and $x_{\rm left}$ is the
dual coordinate to the left of $\ell$ when traveling in the direction of
$\ell$.  

The key symmetry property of the integrand forms is invariance 
under inversion, $x_i^\mu \rightarrow
x_i^\mu/x_i^2$ so that
\begin{equation}
x_{ij}^2 \rightarrow \frac{x_{ij}^2}{x_i^2 x_j^2} \,, \hskip 2 cm 
d^4x_i \rightarrow \frac{d^4 x_i}{x_i^2} \,.
\label{eqn:DualConformalWeight}
\end{equation}
We say that a four-point planar integrand form is dual conformally invariant 
if $\dI \rightarrow \dI$ under this transformation.

%					
%============================================
\subsection{Dual conformal invariance at three and four loops}
%============================================

First consider three loops. There are two planar diagrams, diagrams
(a) and (e) in \fig{ParentTenProp}. Diagram (a) is a bit trivial
because the numerator does not contain any loop momenta, so we do not
discuss it in any detail.  Diagram (e), together with its face
variables, is shown in \fig{DCIThreeLoopE}.  The only allowed 
$\dlog$ numerator for this diagram is given in \eqn{Solution} and
\tab{BasisTenProp}.  Written in dual coordinates, it is
\begin{equation}
N^{\rm (e)} =  s (\ell_5+k_4)^2 = x_{13}^2 x_{45}^2 \,.
\label{eqn:NdciE}
\end{equation}
This numerator exactly matches the known result~\cite{BRY,BDS} for the
three-loop planar amplitude consistent with dual conformal
symmetry~\cite{DualConformalMagic}, giving (somewhat trivial) evidence
for our conjecture. When counting the dual conformal weights via
\eqn{DualConformalWeight}, we need to account for the factor of $s t =
x_{13}^2 x_{24}^2$ in the prefactor $\K$ defined in \eqn{KappaDef}.
We note that the conditions of logarithmic singularities do not fix
the overall prefactor of $s$, but such loop momentum independent
factors are easily determined from maximal cut or leading singularity
constraints when expanding the amplitude.

%%%%%%%%% FIGURE %%%%%%%%%%%%%%%
\begin{figure}[tb]
\begin{center}
\begin{tabular}
{
>{\centering\arraybackslash}m{0.30\textwidth}
>{\centering\arraybackslash}m{0.30\textwidth}
>{\centering\arraybackslash}m{0.30\textwidth}
}
\includegraphics{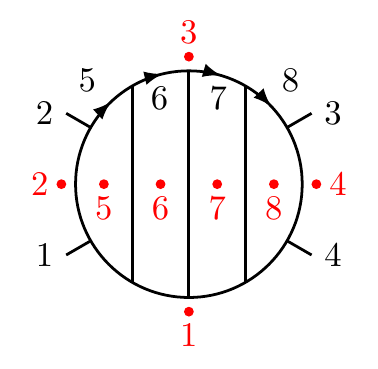} &
\includegraphics{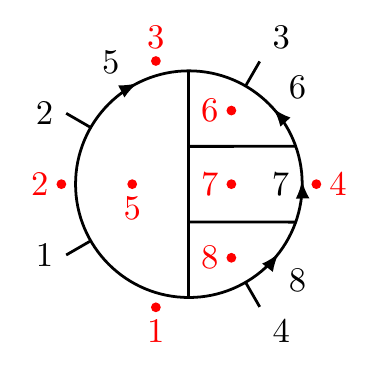} &
\includegraphics{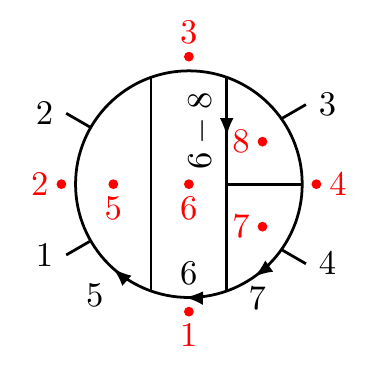} \\
(4a) & (4b) & (4c) \\
$\null$ \\
\includegraphics{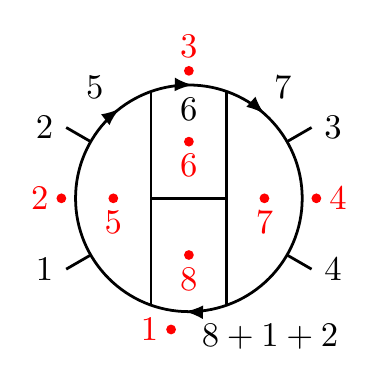} &
\includegraphics{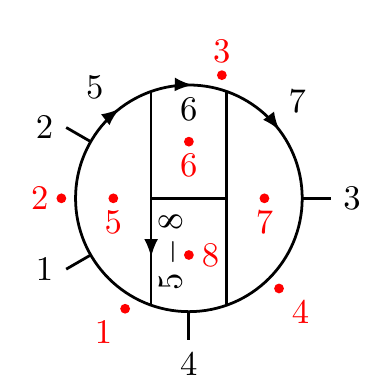} &
\includegraphics{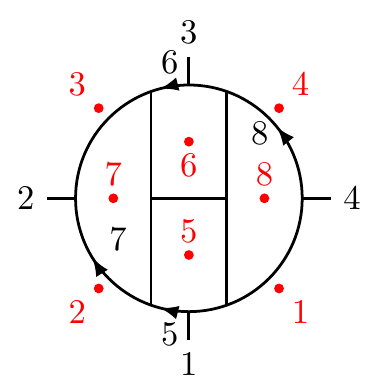} \\
(4d) & (4e) & (4f) \\
\end{tabular}
\end{center}
\caption{Parent diagrams contributing to the four-loop planar
  amplitude.  The shaded (red) dots indicate the face or dual labels
  of the planar graph. 
  }
\label{fig:DCIFourLoopParents}
\end{figure}
%%%%%%%%%%%%%%%%%%%%%%%%%%%%

A more interesting test of our conjecture starts at four loops.  We
construct a basis of $\dlog$-integrands for the planar amplitude
following the same techniques we used at three loops. We then compare
these to known results for the amplitude that manifest dual conformal
invariance~\cite{Bern:2006ew}.  Following the algorithm of
\sect{StrategySection}, we define trivalent parent diagrams.
These are given in \fig{DCIFourLoopParents}.

We have constructed all $\dlog$ numerators for the four-loop
four-point planar amplitude.  To illustrate this construction,
consider first diagram (4c) of \fig{DCIFourLoopParents}.
This is a particularly simple case, because it follows from taking diagram
(e) at three loops in \fig{DCIThreeLoopE} and forming an 
additional box by inserting an extra propagator between two external 
lines.   The extra box introduces only a factor of $s$ to the
three-loop numerator $N^{\rm (e)}$.  This then gives us the four loop
numerator
\begin{equation}
N^{\rm (4c)} = s \left. N^{\rm (e)} \right|_{\ell_5 \rightarrow \ell_6} 
 = (x_{13}^2)^2 x_{46}^2 \,,
\end{equation}
where the relabeling $\ell_5 \rightarrow \ell_6$ changes from the
three-loop diagram (e) labels of \fig{ParentTenProp} to the four-loop
labels of diagram (4c).

%%%%%%%%% FIGURE %%%%%%%%%%%%%%%
\begin{figure}[tb]
\begin{center}
\begin{tabular}
{
>{\centering\arraybackslash}m{0.30\textwidth}
%>{\centering\arraybackslash}m{0.30\textwidth}
>{\centering\arraybackslash}m{0.30\textwidth}
}
\includegraphics{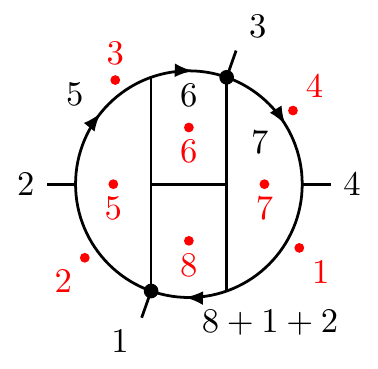} &
\includegraphics{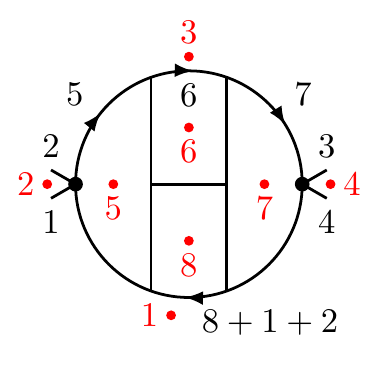} \\
(4d$_2$)
% & (4f$_2$)
 & (4d$'$)\\
\end{tabular}
\end{center}
\caption{
Diagram (4d$_2$) contributes to the planar amplitude at four-loops.
Diagram (4d$'$) does not. Shaded (red) dots represent dual coordinates.
Black dots represent contact terms. }
\label{fig:DCIFourLoopContact}
\end{figure}
%%%%%%%%%%%%%%%%%%%%%%%%%%%%

As a more complicated example, consider 
diagram (4d) of \fig{DCIFourLoopParents}. This contains two pentagon
subdiagrams parametrized by $\ell_5$ and $\ell_7$ and so has a
numerator scaling as $N^{\rm (4f)} \sim {\cal O}(\ell_5^2,\ell_7^2)$.
We skip the details here, and just list the two\footnote{There is a
  third numerator $s \ell^2_7(\ell_5+k_1+k_2)^2$ that is a relabeling
  of $N^{\rm (4d)}_2$ under automorphisms of diagram (4d).  Here and below
  we omit such relabelings.} independent numerators
that result from applying all $\dlog$-conditions:
\begin{align}
& N^{\rm (4d)}_{1} = s^2 (\ell_5-\ell_7)^2 = (x_{13}^2)^2 x_{57}^2 \,,
\label{eqn:N4d1DCINumerator}
\\
& N^{\rm (4d)}_{2} = s \ell^2_7(\ell_5+k_1+k_2)^2
= x_{13}^2 x_{37}^2 x_{15}^2
\longrightarrow N^{\rm (4d_2)} = x_{13}^2 \,.
\label{eqn:N4d2DCINumerator}
\end{align}
In \eqn{N4d2DCINumerator}, we have indicated that the numerator
$N^{\rm (4d)}_2$ cancels two propagators to produce
exactly \fig{DCIFourLoopContact} (4d$_2$), with numerator $N^{\rm (4d_2)}$.
The numerator $N^{\rm (4d)}_1$ is one of the known dual conformal numerators,
and the lower-propagator diagram \fig{DCIFourLoopContact}(4d$_2$) is
also a well-known dual conformal diagram. 

Interestingly, dual conformal invariance allows two additional numerators
\begin{align}
&  N^{\rm (4d)}_{3} = x_{13}^2 x_{27}^2 x_{45}^2 \,,
\label{eqn:N4d3DCINumerator} \\
& N^{\rm (4d)}_{4} = x_{13}^2 x_{25}^2 x_{47}^2 
\longrightarrow N^{\rm (4d')} = x_{13}^2 \,,
\label{eqn:N4d4DCINumerator}
\end{align}
where again $N^{\rm (4d)}_{4}$ reduces to diagram (4d$'$) in
\fig{DCIFourLoopContact} upon canceling propagators. These two
numerators do not meet the no double poles and no poles at infinity
constraints. This is not a coincidence and fits nicely with the fact that
these two numerators have zero coefficient in the
amplitude~\cite{Bern:2006ew,Drummond:2007aua}.

The other diagrams are similar, and we find that for all cases the
$\dlog$-requirement selects out the dual conformal planar integrands
that actually contribute to the amplitude and rejects those that do
not.  Our analysis also proves that, at least for this amplitude, each dual
conformally invariant term in the amplitudes, as given in
Ref.~\cite{Bern:2006ew}, is free of non-logarithmic singularities and
poles at infinity.

%%%%%%%%%%%%%%%%%%%%%%%%%%%%%%%%%%%%%%%%%%%%%%%%%%%%%%%%
%
\subsection{Simple rules for eliminating double poles}
\label{subsec:SimpleDoublePoleRules}
%
%%%%%%%%%%%%%%%%%%%%%%%%%%%%%%%%%%%%%%%%%%%%%%%%%%%%%%%%

In the previous subsection, we highlighted the relationship between
dual conformal invariance and the singularity structure of integrands. 
Here we go further and demonstrate that
the requirement of having no other singularities than logarithmic
ones puts tighter constraints on the integrand than 
dual conformal symmetry.

We start from the observation of Drummond, Korchemsky and Sokatchev
(DKS)~\cite{Drummond:2007aua} that certain integrands upon integration are not well
defined in the infrared, even with external off-shell legs.  They found that if any set of
four loop variables $\{x_{i_1},x_{i_2},x_{i_3}, x_{i_4} \}$ approach
the same external point $x_j$ such that $\rho^2 = x_{i_1 j}^2 + x_{i_2
  j}^2 +x_{i_3 j}^2 + x_{i_4 j}^2 \rightarrow 0$ and the integrand
scales as 
\begin{equation}
\dI = \frac{d^4x_{i_1}\cdots d^4x_{i_4}\,\,N(x_i)}{D(x_i)} \sim \frac{d\rho}{\rho} 
\label{eqn:IRprob} \,.
\end{equation}
The singularity  $\rho \rightarrow 0$ 
corresponds to an integrand double pole in our language, as we shall see below. 
The singularity (\ref{eqn:IRprob}) occurs in the region of integration 
and results in an infrared divergent integral even for off-shell external
momenta.  It is therefore not a sensible dual conformal
integral.  Such ill-defined
integrals should not contribute, as DKS confirmed in various examples, leading
to their all loop order conjecture~\cite{Drummond:2007aua}. A trivial generalization is
to group $l$ loop variables
at a time, $\rho^2 = x_{i_1 j}^2 + x_{i_2 j}^2 + \cdots + x_{i_l j}^2
\rightarrow 0$.  Again the requirement is that the integral not be
divergent with off-shell external momenta.   Of course, this rule 
was not meant to explain all vanishings of coefficients nor to 
explain why terms appear in certain linear combinations.  

Here we wish to extend this line of reasoning using our new insight 
into the singularity structure of amplitudes. For this exercise
it is convenient to switch to momentum twistor coordinates, for which the
problem of approaching certain dangerous on-shell kinematic regions becomes
purely geometric; see Ref.~\cite{Hodges, ArkaniHamed:2010gh} for a
discussion of momentum twistor geometry.  To facilitate comparisons to
existing statements in the literature, we translate the results back
to dual coordinates (region momenta) at the end.

%%%%%%%%% FIGURE %%%%%%%%%%%%%%%
\begin{figure}[tb]
\begin{center}
\begin{tabular}
{
>{\centering\arraybackslash}m{0.30\textwidth}
>{\centering\arraybackslash}m{0.30\textwidth}
>{\centering\arraybackslash}m{0.30\textwidth}
}
\includegraphics{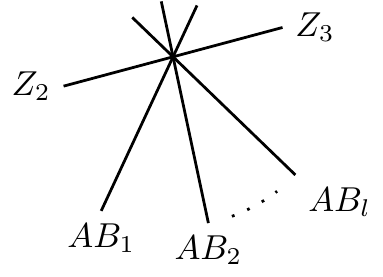} &
\includegraphics{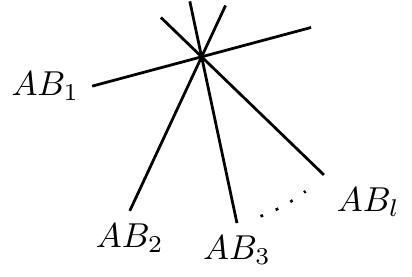} &
\includegraphics{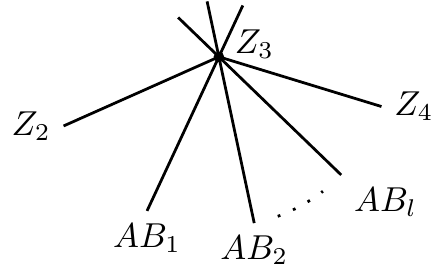} \\
(a) & (b) & (c) \\
\end{tabular}
\end{center}
  \caption{Cut configurations in momentum twistor geometry. Our type I conditions correspond to (b), 
  type II to (a) and type III to (c).
	}
  \label{fig:twistor1}
\end{figure}
%%%%%%%%%%%%%%%%%%%%%%%%%%%%%%

First we rewrite the DKS observation in momentum twistor variables.
To be concrete, we can take $x_j = x_3$ to be the designated 
external point, but in fact there is nothing special about that choice.
Consider the case of $l$ loop variables.
The $l$ loop variables $\{x_{i_1},\ldots,x_{i_l}\}$ correspond to
$l$ lines $(AB)_1,\dots,(AB)_l$ in momentum twistor space.
In our notation, the point $x_3$ in dual coordinates corresponds 
to the line $Z_2Z_3$ in momentum twistor space.
Geometrically, the condition $\rho^2\rightarrow0$ corresponds to 
a configuration in momentum twistor space for which all $l$ lines intersect 
the line $Z_2Z_3$ at the same point, as in \fig{twistor1}(a). 

If we parametrize
\begin{equation}
A_i = Z_2 + \rho_i Z_3 + \sigma_i Z_4 \,,
\end{equation}
where $\rho_i$, $\sigma_i$ are free parameters,
then setting $\rho_i=\rho^\ast$ and $\sigma_i=0$ results in the
desired configuration, where $\rho^\ast$ is arbitrary but the same for all $i$.
We use a collective coordinate: 
\begin{equation}
\rho_1 = \rho^\ast,\qquad \rho_j =
\rho^\ast + t\widetilde{\rho}_j \qquad \sigma_i = \widetilde{\sigma}_i t \,,
\end{equation}
for $j=2,\dots,l$ and $i=1,\dots,l$,  which sets all parameters to the desired configuration
in the limit $t\rightarrow0$.
In this limit, the measure scales as
\begin{equation}
\prod_{i=1,j=2}^l d\rho_j \, d\sigma_i \sim t^{2l-2}dt \,,
\end{equation}
and all propagators of the form $\la (AB)_i\,(AB)_j\ra$ and $\la
(AB)_i\,Z_2Z_3\ra$ scale like $t$. The result is that the integrand
behaves as
\begin{equation}
\dI = \prod_{i=1,j=2}^l d\rho_j \, d\sigma_i 
	\frac{N(\rho_j,\sigma_i)}{D(\rho_j,\sigma_i)}
	\sim dt\,\,t^{2l-2} \cdot \frac{t^N}{t^P} = \frac{dt}{t^{P-N-2l+2}}\,,
\end{equation}
where $N(t)\sim t^N$ is the behavior of the numerator in this limit
and $D(t)\sim t^P$ is the behavior of the denominator, meaning that
$P$ is the number of propagators that go to zero as $t\rightarrow0$.
To avoid unwanted double or higher poles, we demand that $P < N+2l$.
Note the shift by one in the counting rules with respect to \eqn{IRprob}.
That equation counts overall scaling in the integration,
while we study singularities in the
integrand in an inherently on-shell manner. Either way we arrive at the same conclusion.

As an example consider diagram (4d).  One of the numerators is
$N^{\rm(4d)}_1 = (x_{13})^2 x_{57}^2$ and so has $N=1$, while there
are $l=4$ loops, and there are a total of $P=8$ propagators of the
form $\la (AB)_i\,(AB)_j\ra$ and $\la (AB)_i\,Z_2Z_3\ra$.  In this
case
\begin{equation}
P = 8 < 9 = 1 + 2\cdot 4 = N + 2l \,,
\end{equation}
and so the numerator is allowed by this double pole constraint.
In fact, both numerators $N^{\rm (4d)}_1$ and $N^{\rm (4d)}_2$
from \eqns{N4d1DCINumerator}{N4d2DCINumerator}
have the same values of $P$, $l$, and $N$, and so each passes this double pole test
and has only single poles.
In contrast, the numerators $N^{\rm (4d)}_3$ and $N^{\rm (4d)}_4$
from \eqns{N4d3DCINumerator}{N4d4DCINumerator} have $N=0$ and
fail the inequality, so they have double poles and do not contribute to the amplitude.

Now we can generalize this and consider two similar cases:
all lines $(AB)_i$ intersect at a generic point as in \fig{twistor1}(b),
or all lines intersect at a given external point as in \fig{twistor1}(c).
The crux of the argument is the same as the first: 
if the integrand has a double pole we reject it.
The resulting inequalities to avoid these singularities follow analogously; 
the only difference with the first case is the geometric
issue of how many of the $l$ lines are made to intersect.
We summarize the results in terms of $N$, the
number of numerator factors that vanish, $P$,
the number of vanishing propagators,
and the subset of $l$ loop dual coordinates $\{x\}_L \equiv \{x_{i_1},\ldots,x_{i_l}\}$.
Corresponding to each of the diagrams in \fig{twistor1}, we obtain  three types of conditions:
\begin{itemize}
\item Type I (\fig{twistor1}(b)): 
\begin{equation}
P<N+2 l-2 \,, \label{ineq1}
\end{equation}
in the limit that loop dual coordinates are light-like separated from each other:
$x_{ij}^2=0$ for all $x_i,x_j\in \{x\}_L$.
\item Type II (\fig{twistor1}(a)):
\begin{equation}
P<N+2 l \,, \label{ineq2}
\end{equation}
in the limit that loop dual coordinates are light-like separated from each other and from one external point:
$x_{ij}^2 = x_{ki}^2 = 0$ for all $x_i,x_j\in \{x\}_L$, $k=1,2,3,4$
\item Type III (\fig{twistor1}(c)):
\begin{equation}
P<N+2 l+1 \,, \label{ineq3}
\end{equation}
in the limit that loop dual coordinates are light-like separated from each other and from two external points:
$x_{ij}^2 = x_{ki}^2 = x_{k'i}^2 =0$, for all $x_i,x_j\in \{x\}_L$,
$k,k'=1,2,3,4$.\footnote{The equations $x_{ki}^2=x_{k'i}^2=0$ have two solutions so
we have to choose the same solution for all $x_i$.}
\end{itemize}

These rules prevent certain classes of non-logarithmic singularities
from appearing.  In the four-loop case, these rules are sufficient to
reconstruct all dual conformal numerators, automatically precluding
those that do not contribute to the amplitude. Up to seven loops, we
used a computer code to systematically check that all contributions to
the amplitude pass the above rules. Furthermore, we were able to
explain all coefficient zero diagrams up to five loops and many
coefficient zero diagrams up to seven loops using these rules and the
available data provided in Ref.~\cite{Bourjaily:2011hi}.  In the next
subsection we give examples illustrating the above three conditions in
action, as well as examples of non-logarithmic singularities not
detected by these tests.  Not surprisingly, as the number of loops
increases there are additional types of nonlogarithmic singularities.
Indeed, at sufficiently high loop order we expect that cancellations
of unwanted singularities can involve multiple diagrams.

\subsection{Applications of three types of rules}

We now consider three examples to illustrate the rules.
First we examine a five-loop example where the rules 
forbid certain dual conformal numerators from 
contributing to the amplitude. We will see in that example 
that double poles beyond the scope of the above three rules
determine relative coefficients between integrands consistent 
with the reference data \cite{Bourjaily:2011hi,MaximalCuts}.
We then consider two different six-loop diagrams that have zero coefficient
in the expansion of the amplitude. In the first example, the three rules
are sufficient to determine that the numerator
has zero coefficient in the amplitude,
while the integrand in the second example has hidden double poles
not accounted for by the rules.

%%%%%%%%% FIGURE %%%%%%%%%
\begin{figure}[tb]
\begin{center}
\begin{tabular}
{
>{\centering\arraybackslash}m{0.30\textwidth}
}
\includegraphics{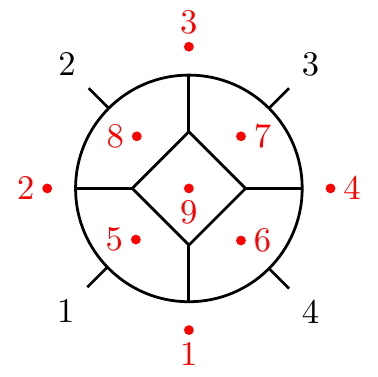}\\
(5a)\\
\end{tabular}
\caption{
A sample five-loop planar diagram.
Shaded (red) dots and labels represent dual coordinates. 
\label{fig:DCIFiveLoopParent} }
\end{center}
\end{figure}
%%%%%%%%%%%%%%%%%%%%%%%
%%%%%%%%% FIGURE %%%%%%%%%%%%%%%
\begin{figure}[tb]
\begin{center}
\begin{tabular}
{
>{\centering\arraybackslash}m{0.23\textwidth}
>{\centering\arraybackslash}m{0.23\textwidth}
>{\centering\arraybackslash}m{0.23\textwidth}
>{\centering\arraybackslash}m{0.23\textwidth}
}
\includegraphics{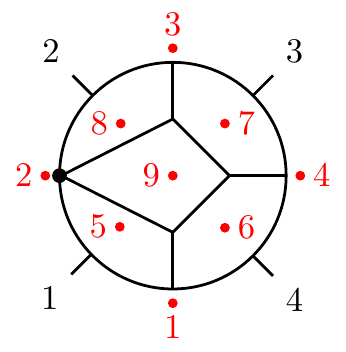} &
\includegraphics{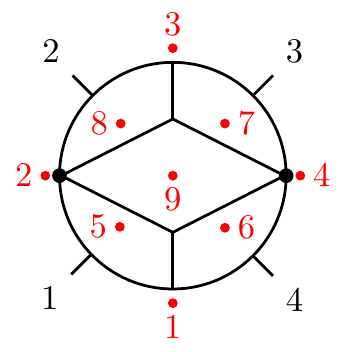} &
\includegraphics{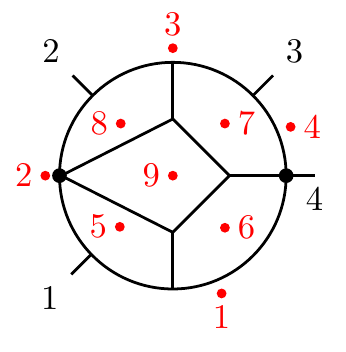} &
\includegraphics{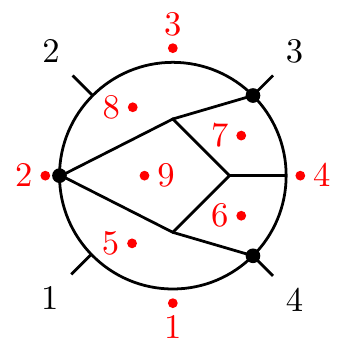}
\\
(5b) & (5c) & (5d) & (5e) \\
\end{tabular}
\end{center}
\caption{ Descendants of the five-loop planar diagram of
  \fig{DCIFiveLoopParent} with numerator coefficients determined to be
      {\it non-zero} by testing for non-logarithmic singularities.
 }
\label{fig:DCIFiveLoopDescendantsNonZero}
\end{figure}
%%%%%%%%%%%%%%%%%%%%%%%%%%%%%%
%%%%%%%%% FIGURE %%%%%%%%%%%%%%%
\begin{figure}[tb]
\begin{center}
\begin{tabular}
{
>{\centering\arraybackslash}m{0.23\textwidth}
>{\centering\arraybackslash}m{0.23\textwidth}
>{\centering\arraybackslash}m{0.23\textwidth}
>{\centering\arraybackslash}m{0.23\textwidth}
%>{\centering\arraybackslash}m{0.30\textwidth}
}
\includegraphics{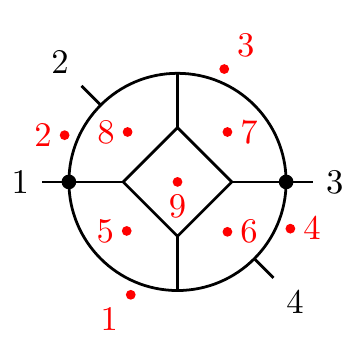} & 
\includegraphics{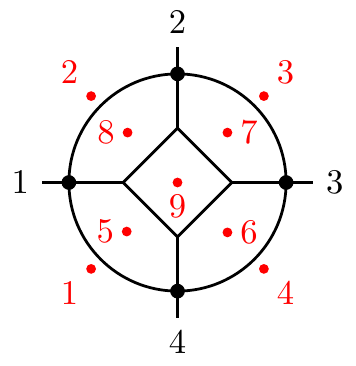} &
\includegraphics{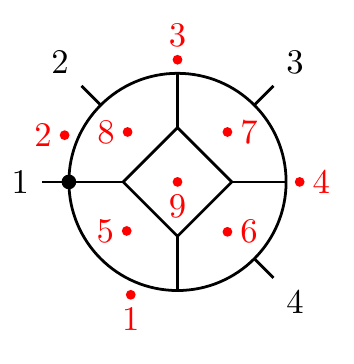} &
\includegraphics{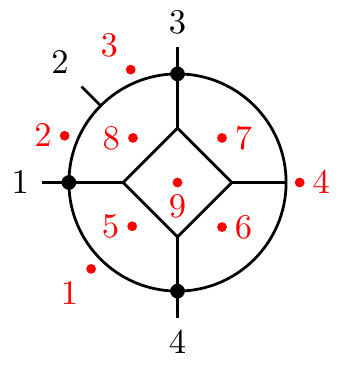} 
\\
(5f) & (5g) & (5h) & (5i)\\
\end{tabular}
\end{center}
\caption{
Descendants of the five-loop planar diagram of \fig{DCIFiveLoopParent} 
with numerator coefficients determined to be
{\it zero} by testing positive for non-logarithmic singularities.
}
\label{fig:DCIFiveLoopDescendantsZero}
\end{figure}
%%%%%%%%%%%%%%%%%%%%%%%%%%%%%%

We first consider the  diagram of \fig{DCIFiveLoopParent}.
We take a slightly different approach here than in previous subsections.
First we list the set of all dual conformal numerators allowed by power counting,
then eliminate numerators that do not pass the three rules of the previous subsection.

The dual conformal numerators that do not collapse any propagators in
\fig{DCIFiveLoopParent} are
\begin{align}
&  N^{\rm (5a)}_1 = x_{24}^2 x_{35}^2 x_{17}^2 x_{68}^2   \,,
&& N^{\rm (5a)}_2 = - x_{13}^2 x_{24}^2 x_{57}^2 x_{68}^2 \,,
\nonumber
\\
& N^{\rm (5a)}_3 = x_{18}^2 x_{27}^2 x_{36}^2 x_{45}^2 \,,
\label{eqn:N5aDCINumerators}
\end{align}
where we omit any dual conformal numerators that are relabelings of
these numerators under automorphisms of the diagram. These three 
numerators correspond to diagrams 21, 22 and 35, respectively,
of Ref.~\cite{MaximalCuts}. However, notice that in the notation used here an
overall factor of $st = x^2_{13}x^2_{24}$ has been stripped off. For
the three kinematic conditions of the rules, this diagram has three different values of
$P$:
\begin{equation}
P_{\rm I} = 8 \,, \quad P_{\rm II} = 10 \,, \quad {\rm and} \quad  P_{\rm III} = 12 \,,
\end{equation}
where the subscript denotes the kinematic case we consider.  The type
I kinematics is most constraining in this example, and for $l=5$
requires $N>0$.  Converting this back to a statement about the
numerator, we conclude that all $\dlog$ numerators for this diagram must have at least one
factor of the form $x_{l_1 l_2}$, for $x_{l_1}$ and $x_{l_2}$ in the
set of loop face variables. Only $N^{\rm (5a)}_1$ and $N^{\rm (5a)}_2$
have this correct loop dependence. So we conclude that both $N^{\rm
  (5a)}_1$ and $N^{\rm (5a)}_2$ can appear in the amplitude, while
$N^{\rm (5a)}_3$ yields an integrand with non-logarithmic poles, and
so has coefficient zero in the amplitude.

In addition to the numerators in \eqn{N5aDCINumerators},
there are other dual conformal numerators that cancel propagators
of the parent diagram, resulting in contact-term diagrams depicted in 
Figs.~\ref{fig:DCIFiveLoopDescendantsNonZero} and~\ref{fig:DCIFiveLoopDescendantsZero}.
If we consider only the contact terms that can be obtained from the diagram in 
\fig{DCIFiveLoopParent},
the numerators that pass the three types of checks are
\begin{align}
& N^{\rm (5b)}    = -x_{24}^2 x_{17}^2 x_{36}^2 \,, 
& & N^{\rm (5c)}  = x_{13}^2 x_{24}^2  \,, %x_{58}^2 x_{67}^2
\nonumber
\\
& N^{\rm (5d)}   = -x_{13}^2 x_{27}^2 \,,%x_{46}^2 x_{58}^2
& & N^{\rm (5e)}  = x_{24}^2  \,,%x_{16}^2 x_{37}^2 x_{58}^2
\label{eqn:N5bcdeNumerators}
\end{align}
where the four numerators respectively correspond to diagrams 31, 32,
33, and 34 in Ref.~\cite{MaximalCuts}.  Besides $N^{\rm (5a)}_3$,
there are four more numerators that display dual conformal invariance
at the integrand level, but are invalid by applying the type II rules,
which is equivalent to the DKS observation that they are ill defined:
\begin{align}
&  N^{\rm (5f)} = x_{18}^2 x_{36}^2 \,, 
&& N^{\rm (5g)} = 1 \,, 
\nonumber
\\
&  N^{\rm (5h)} = x^2_{17} x^2_{36} x^2_{48} \,,
&& N^{\rm (5i)} = x^2_{35} \,.
\label{eqn:N5fgNumerators}
\end{align}
These correspond to diagrams 36, 37, 38 and 39, respectively, of Ref.~\cite{MaximalCuts}.
The numerators listed in \eqn{N5bcdeNumerators} are numerators
for the lower-propagator topologies in \fig{DCIFiveLoopDescendantsNonZero},
and the numerators listed in \eqn{N5fgNumerators} are numerators
for the lower-propagator topologies in \fig{DCIFiveLoopDescendantsZero}.
We again omit the other dual conformal numerators that are relabelings 
of these numerators under automorphism of the diagram.

With this analysis, we have not proved that $N^{\rm (5a)}_{1,2}$ through $N^{\rm (5e)}$
ensure a $\dlog$ form; we have only argued that the corresponding 
integrands do not contain the types of non-logarithmic singularities detected by our three rules.
It is still possible for those integrands to have non-logarithmic poles 
buried in certain kinematic regimes deeper in the cut structure.  Indeed, 
under more careful scrutiny we find additional constraints from the requirements
of no double poles.  In particular, we find that only the following combinations
of integrands corresponding to \figs{DCIFiveLoopParent}{DCIFiveLoopDescendantsNonZero}
are free of double poles:
\begin{equation}
\I^{(A)} = \I^{\rm (5a)}_1 + \I^{\rm (5b)} + \I^{\rm (5e)} \,,
\hskip 1cm
\I^{(B)} = \I^{\rm (5a)}_{2} + \I^{\rm (5c)} \,,
\hskip 1cm
\I^{(D)} = \I^{\rm (5d)} \,.
\end{equation}
The notation is, for example, that the integrand $\I^{\rm (5a)}_1$ has
the propagators of diagram (5a) and the numerator $N^{\rm (5a)}_1$ in
\eqn{N5aDCINumerators}. Similarly, the corresponding numerators for
the integrands of diagrams (5b)--(5e) are given in
\eqn{N5bcdeNumerators}.  The integrand for diagram (5a) with numerator
$N^{\rm (5a)}_3$ is not present, because no contact terms can remove
all double poles of $\I^{\rm (5a)}_3$.  In this case, all
cancellations of double poles are between the parent and descendant
diagrams.  However, at higher loops the situation can very well be
more complicated: unwanted singularities can cancel between different
parent diagrams as well.

%\subsubsection*{Zero coefficients}

%%%%%%%%% FIGURE %%%%%%%%%%%%%%%
\begin{figure}[tb]
\begin{center}
\begin{tabular}
{
>{\centering\arraybackslash}m{0.30\textwidth}
>{\centering\arraybackslash}m{0.30\textwidth}
}
\includegraphics{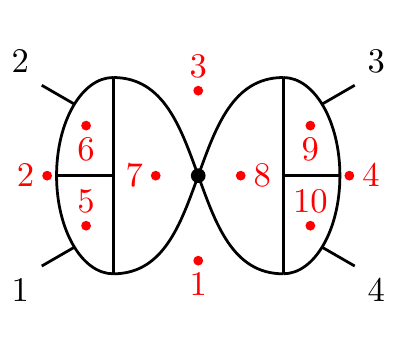} &
\includegraphics{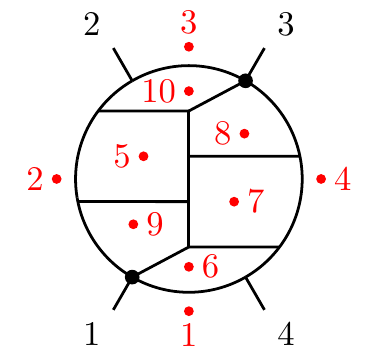}
\\
(6a) & (6b)
\end{tabular}
\end{center}
\caption{
Two six-loop diagrams that have coefficient zero in the amplitude
because they have non-logarithmic singularities.
Diagram (6a) has non-logarithmic poles detected 
by our rules. Diagram (6b) requires explicit checks to
locate double poles.
}
\label{fig:DCISixLoopExamples}
\end{figure}
%%%%%%%%%%%%%%%%%%%%%%%%%%%%%%

We now illustrate how pole constraints can explain
why some six-loop diagrams enter the planar amplitude with zero coefficient.
We choose two examples that both fall outside the type II classification 
of the effective rules of the previous subsection.
This means both numerators escape detection by the original DKS rule,
and so far could not be easily identified as coefficient-zero terms.
The two examples are the six-loop ``bowtie'' in \fig{DCISixLoopExamples}(6a)
and another six-loop diagram with two contact terms in \fig{DCISixLoopExamples}(6b).
The dual conformal numerators of these diagrams are~\cite{Bourjaily:2011hi}\footnote{These
diagrams and numerators can be found in the associated files of Ref.~\cite{Bourjaily:2011hi} 
in the list of six loop integrands that do not contribute to the amplitude. In our notation, 
we have again stripped off a factor of $st=x^2_{13}x^2_{24}$.}
\begin{align}
& N^{\rm (6a)} = x_{13}^3 x_{24} \,,
& & N^{\rm (6b)} = x_{24}^2 x_{27}^2 x_{45}^2 \,.
\end{align}
There are other dual conformal numerators for (6b), but they belong to lower-propagator 
diagrams, so we ignore them in this discussion.

We first consider diagram (6a). This integrand suffers from poles of type III.
We see this by cutting
\begin{equation}
x_{25}^2=x_{26}^2=x_{36}^2=x_{37}^2=x_{56}^2=x_{57}^2=x_{67}^2=0 \,.
\end{equation}
We are then looking at the $l=3$, $N=0$, $P=7$ case
and the corresponding inequality $P < N + 2l + 1$ is violated,
indicating a non-logarithmic pole. This means the non-logarithmic rules immediately
offer a reason why this diagram contributes to the amplitude with coefficient zero.
This agrees with Ref.~\cite{Bourjaily:2011hi}.

The six-loop example (6b) in \fig{DCISixLoopExamples} is more subtle,
since it is not ruled out by the three rules.  However, it does have a
double pole. We know from Ref.~\cite{Bourjaily:2011hi} that this
diagram with numerator $N^{\rm (6b)}$ does not enter the expansion of
the amplitude but has coefficient zero.  Presumably, the
double pole cannot cancel against other diagrams.

We also conducted a variety of checks at seven loops using the
integrand given in Ref.~\cite{Bourjaily:2011hi}.  We applied the three
rules to all 2329 potential contributions and found that all 456
contributions that failed the tests did not appear in the amplitude, as
expected.  We also checked dozens of examples that have vanishing
coefficients and we were able to identify problematic singularities. More
generally, as we saw at five loops, the double poles can cancel
nontrivially between different contributions.  We leave a detailed
study of the restrictions that logarithmic singularities and poles at
infinity place on higher-loop planar amplitudes to future work.  In
any case, the key implication is that we should be able to carry over
the key consequences of dual conformal symmetry to the nonplanar
sector, even though we do not know how to define the symmetry in this
sector.

%===========================================================================
\section{From gauge theory to gravity}
\label{sec:GravitySection}
%==========================================================================

Ref.~\cite{Log} noted that the two-loop four-point amplitude of
$\NeqEight$ supergravity has only logarithmic singularities and no
poles at infinity.  Does this remain true at higher loops?  
In this section we use BCJ duality to analyze this question.
Indeed, we make the following conjecture: 
\begin{itemize}
\item At finite locations, the four-point momentum-space integrand forms of
  ${\cal N} = 8$ supergravity have only logarithmic singularities.
\end{itemize}
However, we will prove that in $\NeqEight$ supergravity there are
poles at infinity whose degree grows with the loop order, as one might
have guessed from power counting.  This conjecture relies on two other
conjectures: the duality between color and kinematics~\cite{BCJLoop},
and the conjecture that nonplanar $\NeqFour$ sYM amplitudes have only
logarithmic singularities and are free of poles at
infinity~\cite{Log}.  Explicit local expressions for numerators that
satisfy the duality between color and kinematics are known at four
points through four loops~\cite{ColorKinematics}.  At higher loops the
duality is a conjecture and it may require nonlocal numerators for it
to hold, resulting in poles at finite points in momentum space for
supergravity amplitudes.  Our conjecture proposes that if this were to
happen it would introduce no worse than logarithmic singularities.
With modifications it should be possible to extend our conjecture
beyond four points, but for N$^k$MHV amplitudes with $k\ge 3$, the
second sYM conjecture that we rely on holds only in the Grassmannian
space and not momentum space, as noted earlier.  Given that all our
explicit studies are at four points, we leave our conjecture at this
level for now.

We note that our conjecture effectively states that one of the key
properties linked to dual conformal symmetry not only transfers to the
nonplanar sector of $\NeqFour$ sYM theory, but transfers to
$\NeqEight$ supergravity as well.  Because there are poles at
infinity, dual conformal symmetry is not quite present in
supergravity.  However, a strong echo remains in
$\NeqEight$ supergravity.

To gather evidence for our conjecture, we construct the complete
three-loop four-point amplitude of $\NeqEight$ supergravity, and do so
in a form that makes it obvious that the conjecture is true for this
case.  To demonstrate that there are poles at infinity, we analyze 
a certain easy-to-construct cut of the four-point amplitude
to all loop orders.  Using the duality between color and kinematics~\cite{BCJ,BCJLoop},
it is easy to obtain the complete three-loop four-point amplitude
of $\NeqEight$ supergravity in a format that makes the
singularity structure manifest. Here, we simply
quote a main result of the duality, and refer to
Ref.~\cite{HenrikJJReview} for a recent review.  According to the
duality conjecture, $\NeqEight$ supergravity numerators may be
constructed by replacing the color factors of each diagram of an
$\NeqFour$ sYM amplitude by kinematic numerators of a second copy,
constrained to the same algebraic relations as the color factors. 
Although the
general existence of numerators with the required property is
unproven, here we only need the three-loop case, for which such
numerators are explicitly known.  Whenever duality satisfying 
numerators are available we immediately have the $\NeqEight$ diagram
numerators in terms of gauge-theory ones:
\begin{equation}
N^{(x)}_{\rm \NeqEight} = N^{(x)} \, N^{(x)}_{\rm BCJ} \,,
\label{eqn:DoubleCopy}
\end{equation}
where $(x)$ labels the diagram.  The gauge-theory numerator $N^{(x)}$
is exactly one of the numerators in \eqn{Solution}, while $N^{(x)}_{\rm BCJ}$
is one of the $\NeqFour$ sYM BCJ numerators from Ref.~\cite{BCJLoop}.

To be concrete, we construct the $\NeqEight$ supergravity numerator for diagram (f) in
\fig{ParentTenProp}. Multiplying the sYM $\dlog$ numerator $N^{\rm{(f)}}$ in \eqn{Solution}
by the corresponding BCJ
numerator gives the $\NeqEight$ supergravity
numerator:
\begin{align}
N^{\rm (f)}_{\NeqEight} & = - \Bigl[(\ell_5 +k_4)^2 
     ((\ell_5 +k_3)^2 + (\ell_5 + k_4)^2)\Bigr] \nonumber\\
& \hskip 2.5 cm \null \times
   \Bigl[(s (-\tau_{35} + \tau_{45} + t) - t (\tau_{25} + \tau_{45}) +
           u (\tau_{25} - \tau_{35}) - s^2)/3 \Bigr] \,,
\end{align}
where $\tau_{ij} = 2 k_i \cdot \ell_j$.  As for the gauge-theory case,
we remove overall factors of $\K$ (defined in \eqn{KappaDef}).  The
construction of the complete three-loop supergravity amplitude is then trivial
using \eqn{DoubleCopy}, \eqn{Solution} and Table 1 of
Ref.~\cite{BCJLoop}.  This construction is designed
to give correct $\NeqEight$ supergravity unitarity cuts.

Based on the BCJ construction, we immediately learn some nontrivial
properties about $\NeqEight$ supergravity.  Since the supergravity and
sYM diagrams have identical propagators, and each numerator has a
factor of $ N^{(x)}$, all unwanted double poles located at finite
values are canceled.  However, in general the factor $ N^{(x)}_{\rm
  BCJ}$ in \eqn{DoubleCopy} carries additional powers of loop momenta.
These extra powers of loop momentum in the numerator compared to the
$\NeqFour$ sYM case generically lead to poles at infinity, as we prove
below.  However, because the three-loop BCJ numerators are at most
linear in loop momentum, only single poles, or equivalently
logarithmic singularities, can develop at infinity.  At higher loops,
the BCJ numerators contribute ever larger powers of loop momenta.
These additional loop momenta generate non-logarithmic singularities
as the orders of the poles at infinity grow.

%%%%%%%%%%%%%% FIGURE %%%%%%%%%%%%%  
\begin{figure}
\begin{center}
\begin{tabular}
{>{\centering\arraybackslash}m{0.30\textwidth}
>{\centering\arraybackslash}m{0.30\textwidth}
}
\includegraphics{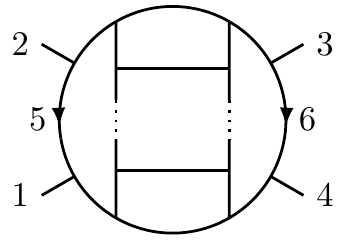} &
\includegraphics{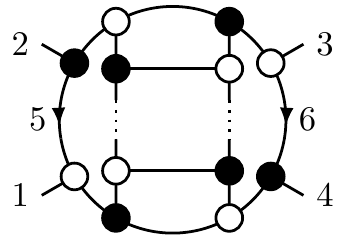} \\
(a) & (b) \\
\end{tabular}
\caption{At $L>3$ loops, diagram (a) contains a pole at infinity that
  cannot cancel against other diagrams.  By cutting all propagators in
  diagram (a) we obtain the corresponding on-shell diagram (b), which
  gives a residue of the amplitude on one of the solutions of the $L$-loop
  maximal cut. This is the only contribution.  }
  \label{fig:GravityTroubleMaker}
\end{center}
\end{figure}
%%%%%%%%%%%%%%%%%%%%%%%%%%%%%%%%%%%  

To analyze the poles at infinity, we turn to a particular set of cuts
chosen so that we can study poles at infinity at any loop
order.  While we do not yet know the four-point $\NeqEight$
supergravity amplitude at five or higher loops, we do have partial
information about the structure of the amplitude to all loop orders.
In particular, we know the value of the maximal cut of the diagram in
\fig{GravityTroubleMaker}(a) that is displayed in
\fig{GravityTroubleMaker}(b). One could evaluate the cut
directly in terms of amplitudes, using superspace
machinery~\cite{FreedmanSuperSums,SuperSums}.  However, it is much
simpler to use the rung rule~\cite{BRY}, which is equivalent to
evaluating iterated two-particle cuts.  This gives the value for the
numerator
\begin{equation}
N=\left[(\ell_5+\ell_6+k_2+k_3)^2\right]^{\delta (L-3)} \,,
\end{equation}
up to terms that vanish on the maximal cut. Here
 $\delta=1$ for $\NeqFour$ sYM theory and $\delta=2$ for $\NeqEight$
supergravity.  As usual factors of $\K$ have been removed.

We carefully\footnote{To avoid mixing in any additional solutions, we
  must first take a next-to-maximal cut, then make a final cut to hone
  in on the single solution in \eqn{GravityProblemSol}.} choose a set
of maximal cuts as encoded in \fig{GravityTroubleMaker}(b) so that
only a single diagram is selected. On this solution, the two loop
momenta labeled in \fig{GravityTroubleMaker} have solutions
\begin{equation}
\ell_5 = \alpha \lambda_1\widetilde{\lambda}_2\,, \hskip 1.5 cm \ell_6 =
\beta\lambda_3\widetilde{\lambda}_4\, .
\label{eqn:GravityProblemSol}
\end{equation}
The Jacobian for this cut is 
\begin{equation}
J = s^2\alpha\beta[(\ell_5+\ell_6+k_2+k_3)^2]^{L-2} F(\sigma_1,\dots \sigma_{L-3}) \,,
\end{equation}
where the function $F$ depends on the remaining $L-3$ parameters, $\sigma_i$,
of the cut solution, and not on $\alpha$ or $\beta$. On the cut, the
parametrization \eqn{GravityProblemSol} implies that
\begin{equation}
\left. (\ell_5+\ell_6+k_2+k_3)^2 \right|_{\rm cut} =
(\alpha\langle13\rangle + \langle23\rangle)(\beta[24]+[23]) \,.
\end{equation}
Then the residue in the sYM case is
\begin{equation}
\res{\rm cut} d \I_{\rm YM} \sim
\frac{d\alpha}{\alpha(\alpha-\alpha_0)}
\wedge \frac{d\beta}{\beta(\beta-\beta_0)}
\wedge \frac{d\sigma_1\dots d\sigma_{L-3}}{F(\sigma_1,\dots,\sigma_{L-3})} \,,
\label{eqn:YMform}
\end{equation}
with $\alpha_0= - \langle23\rangle / \langle13\rangle$, $\beta_0= - [23] / [24]$.
So the sYM
integrand has only logarithmic singularities and no pole at infinity in $\alpha$ or $\beta$.
On the other hand, in the supergravity case the residue is
\begin{equation}
\res{\rm cut} d \I_{\rm GR} \sim
\frac{d\alpha}{\alpha(\alpha - \alpha_0)^{4-L}}
\wedge \frac{d\beta}{\beta(\beta - \beta_0)^{4-L}}
\wedge \frac{d\sigma_1\dots d\sigma_{L-3}}{F(\sigma_1,\dots,\sigma_{L-3})} \,.
\label{eqn:GRform}
\end{equation}
We see that these forms have the same structure as
sYM for $L=3$, but for
$L>3$ they differ. In the latter case, the sYM expression in
\eqn{YMform} stays logarithmic with no poles at
$\alpha,\beta\rightarrow\infty$, while the supergravity residue
\eqn{GRform} loses the poles at $\alpha_0$ and $\beta_0$ for $L=4$ and
develops a logarithmic pole at infinity. However, for $L \ge 5 $ the
poles at infinity become non-logarithmic, and the degree grows linearly
with $L$.  Since the cut was carefully chosen so that no other
diagrams can mix with \fig{GravityTroubleMaker}(a), the poles at
infinity identified in \eqn{GRform} for $L \ge 4$ cannot cancel against other diagrams,
and so the ${\cal N}=8$ supergravity amplitudes indeed have poles
at infinity.  This can also be verified by the direct evaluation of
the on-shell diagram in \fig{GravityTroubleMaker}(b).  In fact,
at three loops there is another contribution (different from \fig{GravityTroubleMaker})
that leads to a pole at infinity as well. As it does not offer 
qualitatively new insights, we will not show this example here.

We conclude that in contrast to $\NeqFour$ sYM theory, $\NeqEight$
supergravity has poles at infinity with a degree that grows linearly
with the loop order.  An interesting question is what this might imply
about the ultraviolet properties of $\NeqEight$ supergravity. While it
is true that a lack of poles at infinity implies an amplitude is
ultraviolet finite, the converse argument that poles at infinity imply
divergences is not necessarily true.  There are a number of reasons to
believe that this converse fails in supergravity.  First, at three and
four loops the four-point $\NeqEight$ supergravity amplitudes have
exactly the same degree of divergence as the corresponding $\NeqFour$
sYM amplitudes~\cite{GravityThreeLoop,Manifest3,ColorKinematics}, even
though the supergravity amplitudes have poles at infinity.  Indeed,
when calculating supergravity divergences in critical dimensions where
the divergences first appear, they are proportional to
divergences in subleading-color parts of gauge-theory
amplitudes~\cite{ColorKinematics}. In addition, recent work in
$\NeqFour$ and $\NeqFive$ supergravity shows that nontrivial
cancellations, beyond those that have been understood by
standard-symmetry considerations, occur between the diagrams of any
covariant formulation~\cite{N4ThreeLoop,N5FourLoop}.  Furthermore,
suppose that under the rescaling $\ell_i \rightarrow t \ell_i$ with
$t\rightarrow\infty$ the supergravity integrand scales as $1/{t^m}$.
If $m\leq 4L$ where $L$ is the number of loops, we can interpret this
behavior as a pole at infinity. However, as we have demonstrated in
this paper, after applying cuts this pole can still be present or
disappear, and other poles at infinity can appear. Thus, the relation
between ultraviolet properties of integrated results and the presence
of poles at infinity is nontrivial.  It will be fascinating to study
this relation.

%=========================================================================
\section{Conclusion}
\label{sec:ConclusionsSection}

In this paper, we have studied in some detail the singularity structure of
integrands of $\NeqFour$ sYM theory, including nonplanar
contributions. These contributions were recently conjectured to have only
logarithmic singularities and no poles at infinity~\cite{Log}, just as
for the planar case~\cite{ArkaniHamed:2012nw}.  In this paper, besides
providing nontrivial evidence in favor of this conjecture, we made two
additional conjectures.  First, we conjectured that in the
planar sector of $\NeqFour$ sYM theory, constraints on the amplitudes
that follow from dual conformal symmetry can instead be obtained from
requirements on singularities. The significance of this conjecture
is that it implies that consequences of dual conformal symmetry on the
analytic structure of amplitudes carry over to the nonplanar sector.
We described evidence in favor of this conjecture through seven loops.  Our
second conjecture involves $\NeqEight$ supergravity.  While we proved
that the amplitudes of this theory have poles at infinity, we
conjectured that at finite locations, at least the four-point
amplitude should have only logarithmic singularities, matching the
$\NeqFour$ sYM behavior.

To carry out our checks we developed a procedure for
analyzing the singularity structure, which we then applied to the
three-loop four-point amplitude of $\NeqFour$ sYM theory.  Using this approach
we found an explicit representation of this amplitude, where the
desired properties hold term by term.  We also partially analyzed the
singularity structure of four-point amplitudes through seven loops.
We illustrated at three loops how to make the logarithmic
singularity property manifest by finding $\dlog$ forms.  

Our strategy for studying the nonplanar singularity structure required
subdividing the integrand into diagrams and assuming that we could
impose the desired properties on individual diagram integrands.
Unitarity constraints then allowed us to find the appropriate linear
combinations of integrands to build an integrand valid for the full
amplitude.  Interestingly, many coefficients of the basis integrands
follow a simple pattern dictated by the rung rule~\cite{BRY}. 

More generally, the study of planar $\NeqFour$ sYM amplitudes has
benefited greatly by identifying hidden symmetries.
Dual conformal symmetry, in particular, imposes an extremely powerful
constraint on planar $\NeqFour$ sYM amplitudes.  When combined with 
superconformal symmetry, it forms a Yangian
symmetry which is tied to the presumed integrability of the planar
theory. However, at present we do not know how to extend this
symmetry to the nonplanar sector.  Nevertheless, as we 
argued in this paper, the key analytic restrictions on the
amplitude do, in fact, carry over straightforwardly to the 
nonplanar sector.  This bodes well for future studies of full
amplitudes in $\NeqFour$ sYM theory.

Our basis integrands are closely related to the integrals used by Henn
et al.~\cite{Henn, HennSmirnov, HennSmirnov2, Henn:2014qga} to find a
simplified basis of master integrals determined from
integration-by-parts identities~\cite{IBP1,IBP2}.  In this simplified
basis, all master integrals have uniform transcendental weight, which
then leads to simple differential equations for the integrals.  This
basis overlaps with our construction, except that we include only
cases where the integrands do not have poles at infinity, since those
are the ones relevant for $\NeqFour$ sYM theory.  The $\dlog$ forms we
described are in some sense partway between the integrand and the
integrated expressions.

An interesting avenue of further exploration is to apply these ideas
to $\NeqEight$ supergravity.  Using BCJ duality~\cite{BCJ,BCJLoop}, we
converted the four-point three-loop $\NeqFour$ sYM integrand forms with into ones for
$\NeqEight$ supergravity.  We proved that the three-loop four-point
integrand form of $\NeqEight$ supergravity has only logarithmic
singularities.  However, there are singularities at infinity.  Indeed, we
proved that, to all loop orders, there are poles at infinity whose
degree grows with the loop order. A deeper understanding of these
poles might shed new light on the surprisingly tame ultraviolet properties
of supergravity amplitudes, and in particular on recently
uncovered~\cite{N5FourLoop} ``enhanced ultraviolet cancellations'', which 
are nontrivial cancellations that occur between diagrams.

In summary, by directly placing constraints on the singularity
structure of integrands in $\NeqFour$ sYM theory, we have a
means for carrying over the key consequences of dual conformal
symmetry and more to the nonplanar sector.  A key conclusion of our
study is that the nonplanar sector of $\NeqFour$ sYM theory is
more similar to the planar sector than arguments based on
symmetry considerations suggest.  Of course, one would like
to do better by finding a formulation that makes manifest the singularity
structure.  The explicit results presented in this paper should aid
that goal.

\subsection*{Acknowledgments}
We thank Nima Arkani-Hamed, Jacob Bourjaily, Scott Davies, Lance Dixon
and Josh Nohle for helpful discussions.  We
especially thank Johannes Henn for discussions and detailed
comparisons to unpublished results for various nonplanar master
integrals.  This work was supported in part by the US Department of
Energy under Award Numbers DE-{S}C0009937 and DE-SC0011632. J.~T. is
supported in part by the David and Ellen Lee Postdoctoral Scholarship.
E.~H. is supported in part by a Dominic Orr Graduate Fellowship.

%=========================================================================
%Bibliography - from "References.bib"
%=========================================================================
\bibliographystyle{JHEP}
\newpage
\phantomsection         %to get hyperlinks right for index
            %\addcontentsline{toc}{section}{\numberline{}References}
						% If you want to put a References section into the toc just comment in 
						% the previous line!
            \bibliography{References}
            %\printbibliography[prenote=preNote,postnote=postNote]
            \clearpage

\end{document}